\pdfoutput=1
\pdfoutput=1
\documentclass[11pt]{article}
\usepackage{compat-preamble}
\IfFileExists{content/lean_stats.tex}{% Auto-generated by scripts/build_papers.py. Do not edit manually.
\providecommand{\LeanTotalLines}{53935}
\providecommand{\LeanTotalTheorems}{2172}
\providecommand{\LeanTotalSorry}{0}

}{}

\title{Thermodynamic Limits of Proof}
\author{
  Tristan Simas\\
  McGill University\\
  \texttt{tristan.simas@mail.mcgill.ca}
}
\date{\today}

\makeatletter
\@ifundefined{ifdq@first}{\newif\ifdq@first}{}
\newcommand{\dqcommabreak}[1]{%
  \dq@firsttrue
  \@for\dq@item:=#1\do{%
    \ifdq@first
      \dq@item
      \dq@firstfalse
    \else
      , \allowbreak\dq@item
    \fi
  }%
}
\makeatother
\providecommand{\claimstamp}[2]{}
\renewcommand{\claimstamp}[2]{}
\providecommand{\LH}[1]{\hyperlink{lh:#1}{\texttt{#1}}}
\newcommand{\DQ}{\mathrm{DQ}}

\begin{document}
\maketitle

\begin{abstract}
Every irreversible recorded distinction has a positive thermodynamic work floor.  This positivity is the physical input that makes a finite-budget proof threshold exist; the arithmetic computes where the threshold falls.  Landauer's principle supplies the universal ideal floor $\varepsilon\ge k_B T\ln 2$ per irreversible bit, experimentally verified to $\pm 10\%$~\cite{berut2012experimental,jun2014high}.  Proof available to an agent is checkable information for that agent, so some substrate must produce, retain, and expose evidence that excludes answer-changing alternatives.  A finite detector array operating at temperature $T$ for finite time has finite signal-acquisition capacity.  Combining finite causal access, positive retained-record work cost, and exact lower bounds on required records gives the Physical Counting Impossibility Theorem: no fixed-budget physical substrate can provide universal exact proof once the retained-record lower bound exceeds the declared budget.  The theorem requires exactly $B<\infty$ and $\varepsilon>0$.

An answer reports a value; proof supplies checkable grounds for accepting it.  The two carry different retained-record obligations: a reversible device may compute an answer and erase its scratch history, while proof requires retained, inspectable records that exclude answer-changing alternatives.  A global answer register, oracle response, entanglement witness, finite survey catalog, or trusted device output supplies proof only under an interface that makes the relevant grounds available to the verifier.  A proposed interface must therefore identify the retained-record lower-bound family $R(n)$ it induces.  Sound operational claims about efficient solvability inherit the same finite-budget obstruction when their acceptance would license universal exact proof.  The result constrains proof-availability for agents.  Substrate-free derivability has proof status only when a physical verification event makes it available to an agent.  Parsing, checking, and retaining are physical operations with positive work cost per irreversible record.

\end{abstract}

\section{Introduction}\label{sec:introduction}

Proof is produced, retained, inspectable grounds that exclude alternatives.  Court, audit, peer review, mathematical verification, and safety certification instantiate this standard.  An answer without grounds is assertion, trust, oracle testimony, or answer production.  Information is physical: every irreversible recorded distinction has a positive work floor, and no signal reaches an agent faster than $c$.  A court that receives testimony without cross-examination, an auditor who accepts a figure without a receipt, and a reader who closes a book after checking a derivation all face the same retained-record requirement: the grounds must be available for inspection, and availability costs energy per retained distinction.  An agent is any system that acquires signals, retains records, and spends positive work per irreversible recorded distinction.  By those physical inputs, every agent in the theorem's scope is finite in budget and causal access.  Truth is free; proof costs energy per retained distinction.

\noindent\textbf{Evidence indistinguishability.} Proof of a proposition $P$ requires the verifier's retained accessible state to separate the in-scope alternatives that require different judgments about $P$.  If two possible worlds, preparations, histories, or transcripts leave the verifier in the same retained state while one supports $P$ and the other does not, the verifier's evidence is indistinguishable with respect to $P$.  The audit trail is the durable distinction-retaining object: it is the part of the physical state that remains available for local review after the process has produced its answer.  Weaker interfaces can be valuable when they prove their own declared claims, with resource accounting governed by their retained-record family, confidence threshold, or trust audit.\leanmeta{\LHrng{PROV}{75}{84}.}

\noindent\textbf{Physical counting preview.} For a declared proof interface, let $B<\infty$ be the work budget, let $\varepsilon>0$ be the cost floor for each retained evidence record, and let $R(n)$ be the retained-record lower-bound family for size $n$.  If $R(n)$ is unbounded, then some size satisfies $\varepsilon R(n)>B$, so universal exact proof is unavailable to that fixed-budget interface.  The inequality is elementary arithmetic once $\varepsilon>0$; that positivity is the physical input, and the retained-evidence bridge gives $R(n)$ physical content.  Special relativity bounds acquisition across a finite apparatus; thermodynamics supplies the positive per-record floor, with $k_B T\ln2$ as the Landauer-floor value for ordinary irreversible bits, experimentally verified to $\pm 10\%$~\cite{berut2012experimental,jun2014high}; query lower bounds specify how many distinctions must become available to exclude answer-changing alternatives.  Without the physical fact $\varepsilon>0$, no threshold exists.

Agent is used in the substrate-neutral sense fixed above: any system that acquires signals, retains records, and spends positive work per irreversible recorded distinction.  Exact proof is the finite-task instance of agent-available proof: a device may output an answer or follow a reversible trajectory correlated with one, but proof requires retained, locally inspectable evidence that answer-changing alternatives were excluded.  Resource-bounded computational capability has the same form: substrate-free complexity notation has no runtime or proof status for an agent until an evaluation rule supplies steps, time, records, and observer access.  Finite-interface bookkeeping, budget arithmetic, monotone threshold lemmas, and the proof-act bridge are mechanized in Lean 4; empirical physical inputs remain cited premises.\leanmeta{\LHrng{PROV}{37}{40}.}

The resource bound is physical counting.  A bounded region operating for finite time has only finitely many locally accessible acquisition events under the declared propagation and substrate assumptions.  If each retained event carries a positive work floor, then a fixed finite budget bounds the distinctions that can participate in proof.  The semantic lower-bound quantity is $D(f_n)$, the exact adaptive-query complexity of the finite task $f_n:X_n\to Y_n$.  The physical lower-bound quantity is $R(n)$, the retained-record demand induced by the declared proof interface.  In the separable audit interface, every required query distinction must leave a retained record available to the verifier, so $R(n)=D(f_n)$.  Compressed, quantum, probabilistic, interactive, or global-observable interfaces may induce a different $R(n)$.  The threshold uses the $R(n)$ supplied by that interface.

Standard query lower bounds provide the clean finite examples.  The first exhibit is separable parity: $D(\mathrm{PARITY}_n)=n$ while parity remains computationally easy, so audit-trail cost separates from algorithmic hardness.  The separation is combinatorial before it is thermodynamic.  A final parity bit is an answer record; retained proof under the separable audit interface must exclude every single-spin flip that would change the answer.  In Landauer-floor units, the two retained-record costs are different:
\[
\begin{array}{rcl}
\text{answer interface} &:& 1\text{ retained parity bit}\quad \leadsto\quad k_B T\ln 2,\\[2pt]
\text{audit-proof interface} &:& n\text{ retained spin records}\quad \leadsto\quad n k_B T\ln 2.
\end{array}
\]
The interface is not freely chosen: it is determined by what the claim asserts.  ``The parity is $1$'' licenses the global-observable interface.  ``Every spin was inspected and found to agree'' licenses the separable audit interface.  A claim and its proof interface are co-determined; changing one changes the other.

At $300\,\mathrm{K}$, retaining $n$ spin distinctions costs at least $n k_B T\ln 2$ joules in the Landauer-floor idealization.  The sign-query witness gives a SAT-shaped hidden-switch apparatus, and hidden truth-table proof gives a $2^n$ query lower bound when the object exposes $2^n$ locally distinguishable entries.

For a satisfiable SAT instance under the standard assignment-certificate interface, solving requires proving the answer.  The proof is the satisfying assignment together with the retained records needed to verify it: $n$ assignment bits give $R_{\mathrm{cert}}(n)\ge n$ when the assignment has $n$ variables.  A procedure that outputs an assignment without retaining those verification records is a guess, not a proof.  The instance is solved for the verifier only when the certificate bits and formula-evaluation records are available.  A claimed proof interface with smaller $R(n)$ must declare its retained evidence and soundness conditions; the counting theorem then applies to the retained-record family induced by that interface.\leanmeta{\LHrng{SATQ}{12}{14}, \LHrng{PROV}{101}{104}.}

Evidence indistinguishability is not special to parity.  A global-parity register may be a valid answer interface for the parity value itself while leaving the per-spin audit claim unresolved.  An entanglement witness can prove a declared witness threshold while leaving full-state alternatives unresolved.  A finite sky catalog can support a finite-confidence topology claim while exact global topology still requires a bridge from local records to global identification.  These examples carry the same accounting: changing the evidence object changes the proof interface and its lower-bound family.

\noindent The logical chain can be read left to right:
\[
\boxed{
f_n+\text{declared interface} \longrightarrow R(n) \longrightarrow \text{retained proof records}
\longrightarrow \varepsilon R(n) \longrightarrow B
\longrightarrow \varepsilon R(n)>B
}
\]

The chain separates truth from proof.  Whether a proposition holds is independent of any agent's budget.  Whether an agent can prove it is not.

This is a Carnot-class ideal limit theorem for proof.  Carnot upper-bounds extractable work from reservoirs; the counting theorem lower-bounds work required to retain proof records.  Both are elementary once the premises are in hand, and both set the ideal standard that real engines, or real proof systems, can only approach.  Carnot's formula is parameterized by reservoir temperatures; the counting threshold is parameterized by $R(n)$ and $\varepsilon$, where the task-specific and interface-specific content lives.

The accounting applies uniformly to reversible uncomputation, oracle testimony, compressed transcripts, substrate-free specifications, interactive proofs, zero-knowledge protocols, quantum procedures, and bounded-error checks.  Each chosen interface supplies its retained-record requirement, observer-access rule, cost accounting, and lower-bound family $R(n)$.  Bounded $R(n)$ falls outside the obstruction; unbounded evidence requirements inherit the finite-budget threshold.  Global-infinity cases require a local-budget bridge (Section~\ref{sec:global-infinity-local-budget}).

Abstraction can reduce the physical cost of proof by changing the representation, proof interface, or retained-record lower-bound family; retained records required by that interface still carry positive substrate-level cost.\leanmeta{\LH{PBC21}.}

Interface changes have precise effects.  Zero-temperature limits and Zeno-style measurements require replacement record-formation interfaces; reversible computation can produce answer-only output without retained audit records; and global infinitary structures alter the conclusion only when they supply a computation-local budget bridge.

\section{Physical Ingredients and Interface}\label{sec:foundations}

\subsection{Information Physicality and Availability}\label{sec:necessity-physical-availability}

Proof requires a physical state carrying checkable grounds for the claim, because every irreversible record has positive work cost and every acquisition event takes finite time.  A state that does not separate $P$ from an in-scope alternative on which $P$ is false does not carry proof of $P$ for that agent: it leaves alternatives indistinguishable when they require different judgments.  Confidence, testimony, answer receipt, and bounded-error acceptance prove only the claims declared by their own interfaces, with their own retained-record families and thresholds.  The retained separator condition is the physical consequence of asking what evidence remains available to the verifier.

\noindent\textbf{Physical verification events.} The reader parsing a proof is a physical system processing information.  Every proof verification event, including verification of the formal claims indexed here, proceeds through written, spoken, typed, stored, transmitted, parsed, remembered, or checked records.  Existing proof practice has the same structure: peer review, formal verification, reproducibility, certified computation, laboratory reporting, and audit trails all require physically instantiated, checkable evidence.

\subsection{Proof Versus Provability}\label{sec:proof-versus-provability}

The proof standard fixed in Section~\ref{sec:introduction} separates proof from provability through proof-acts.  Parsing, checking, and retaining a mathematical derivation are empirical events with the same accounting structure as examining a receipt, cross-examining a witness, or retaining a detector log.  A valid derivation in a formal system that no substrate has parsed, checked, or retained supplies a specification.  It has proof status for an agent only when a physical verification event makes it available.  The counting bound constrains that event of availability.\leanmeta{\LHrng{PROV}{75}{84}.}

Logical possibility has no work cost until an interface realizes it as a readable, checkable, retained state.  Once realized for an agent, the records that make the proof available enter the same physical accounting as any other retained information process.

\noindent\textbf{Proof-act and answer/proof distinction.}\label{par:proof-act-answer-distinction}
A proof-act is the event in which the separator condition is realized for an agent under a declared interface: the token is accessible, the declared verifier accepts it, and the records needed for local review are retained.  In physical terms, the proof-act removes the evidence indistinguishability relevant to the declared claim.  A formal derivation on its own is a specification for such an act.  It becomes proof for that agent only when some physical process parses, checks, and retains the relevant grounds.

Answer and proof separate at the interface boundary, and the interface is fixed by the epistemic content of the claim.  A device may compute a global predicate, witness statistic, or finite survey summary and then discard the transient evidence that produced it.  The final report can be correct.  It is not, by itself, a proof of a stronger audit claim.  A certified global readout needs the records that certify the readout; an entanglement witness needs the declared settings, outcomes, confidence threshold, and calibration trail; a finite-sky topology claim needs the catalog and bridge assumptions that connect the catalog to the global identification.  If the declared interface asks for separable proof over local coordinates, the unmeasured-coordinate argument gives the corresponding $R(n)$.  These are different claims with different proof costs, not the same claim under different accounting conventions.  Proof is counted through the declared evidence interface, not through answer production alone.

The claim that a computation was correct is itself a proposition requiring proof, with retained records carrying the same per-record floor.  An answer-only report, however correct, does not supply those records.  Otherwise the result belongs to assertion, trust, answer receipt, or bounded-confidence acceptance under a different interface.  Every interface that supplies checkable grounds has a retained-record family $R(n)$, and each retained irreversible distinction carries the positive floor.  Approximate and certificate-verification interfaces may have smaller $R(n)$ than separable exact audit, but they state different claims with their own cost structures.  ``I proved it, trust me,'' ``the experiment was correct,'' and ``the books are fine'' are not accepted as proof in mathematics, physics, law, or audit.  These practices require inspectable records.  The counting theorem identifies the thermodynamic floor beneath those requirements.

Correct oracle outputs, calibrated device readings, short certificates, and interactive transcripts are handled by the same accounting once their physical basis for verification is declared.  Sections~\ref{sec:physical-grounding}--\ref{sec:physical-impossibility} compute the retained-record family from that declared basis.

\leanmetapending{\LHrng{PROV}{37}{57}, \LH{PROV74}.}
\noindent\textbf{Physical Availability.}\label{scope:agent-relative-availability} The theorem constrains the event in which a specification (Section~\ref{sec:proof-versus-provability}) becomes available as proof through a physical verification process.  Every such event occurs in a physical substrate: a brain maintaining neural states by ATP-driven metabolism, a circuit switching transistors, a memory device refreshing cells, a proof assistant committing parser states to memory, or a detector exposing pointer records.  The per-record floor attaches to each substrate at its own declared cost.  When asserted as capabilities for agents, predicates such as ``polynomial-time'', ``provable'', and ``certifiable'' name realized processes or proof-acts.  A capability not realized in a physical substrate supplies a specification of what such a realization would require.

The proof-counting interface fixes three physical ingredients and one interface convention.  In the formal interface definitions (Appendix~\ref{app:formal-interface}), a finite proof task is a function $f_n:X_n\to Y_n$: the declared apparatus exposes finite admissible inputs and finite exact answers or proof statuses.  A query is one locally accessible measurement of one declared input entry, and the exact query complexity is
\[
D(f_n)=\min_T\max_{x\in X_n}\#\{\text{measurements made by }T\text{ on }x\}
\]
as in Definition~\ref{def:exact-query-complexity}.  Lower-bound proofs use this quantity directly: every exact protocol must have some admissible input on which at least $D(f_n)$ entries are measured.  Deterministic query complexity is used here as a lower-bound calculus for physical record acquisition, in the same way that combinatorial counting enters statistical-mechanical bounds.

The notation keeps two quantities separate.  $D(f_n)$ is the exact query lower bound for the declared finite task.  $R(n)$ is the retained-record lower-bound family induced by the proof interface: the number of retained distinctions needed to remove the evidence indistinguishability relevant to the declared claim.  In the separable audit interface, $R(n)=D(f_n)$ because each required query distinction is retained as evidence.  In compressed, quantum, probabilistic, interactive, or global-observable interfaces, $R(n)$ may differ because the evidence object differs.  The physical impossibility theorem applies to the $R(n)$ supplied by the declared interface.

\noindent\textbf{Running examples and physical transfer.}  An optical lattice holds $n$ two-level atoms, and the verifier must determine whether all spins are aligned along a declared axis.  The setup specifies which physical observables count as entries of $X_n$, which outcomes count as elements of $Y_n$, which records are retained, and what positive work floor attaches to each irreversible retained distinction.  Query $i$ may be the spin of atom $i$ along the declared axis; the retained record is the amplified pointer state; the per-record work floor is $\varepsilon\geq k_B T\ln 2$; and the total work budget is $B$.  In the optical-lattice examples above, the pointer starts in a prepared standard state and is set to the measurement outcome; this is the creation-event route.  Stable binary memory at $T>0$ has thermodynamic floor $k_B T\ln2$ per bit under the standard thermal-stability analysis, even though the operation is not strict erasure in Landauer's original sense.  A different axis, joint measurement, coherent oracle call, entanglement-witness interface, or finite-survey bridge is a different interface and must supply its own lower-bound family.  For large enough $n$, exact proof under any unbounded retained-record family exceeds $B$ because the required retained records exceed the declared budget.

\subsection{Record-Formation Accounting}\label{sec:no-physical-proof-exemption}

Any physical system that accepts a proof undergoes a state transition in a proof interface.  If the transition is coherently reversed or erased, its evidential trace is unavailable for later inspection.  If the transition remains as checkable evidence, it is a retained record and enters the irreversible record-formation branch of the cost accounting.  The formal transition interface records this exhaustion: erased reversible history is unavailable as retained evidence, and retained evidence is an irreversible record with positive declared cost.\leanmeta{\LHrng{PROV}{85}{89}.}

Symmetry proofs, null results, statistical excesses, topological guarantees, coherent quantum verifications, and working devices supply different evidence objects and induce different lower-bound families $R(n)$.  The accounting is the same; the interface inputs change.

\begin{definition}[Substrate-Neutral Agent]\label{def:substrate-neutral-agent}
An agent is any substrate that implements a declared proof interface: it acquires signals, retains records, and spends positive work per irreversible recorded distinction.  Every such substrate is finite in budget and causal access by the thermodynamic and relativistic inputs of Section~\ref{sec:physical-grounding}.  The interface contract specifies acquisition, retention, and cost rules rather than consciousness, intentionality, biological organization, or a particular computational architecture.  An optical lattice, a single atom with two addressable states, a macroscopic detector array, or a finite community of communicating subsystems counts as an agent when the interface conditions are satisfied.
\end{definition}

Report discipline follows the same interface definition.  A proof interface may abstain, report exactly, or report approximately.  Integrity means it never asserts an unsupported exact answer; competence additionally requires non-abstaining coverage within the declared resource bound.  In lower-bound-blocked regimes, integrity requires abstention, weakened guarantees, or changed assumptions; unsupported exact reports fail the discipline.

Exactness is a contract-level property.  If the declared task permits tolerance, bounded error, or noisy reports, the target relation and retained evidence rule change with it; the lower-bound family becomes the corresponding $R_{\mathrm{approx}}(n)$ or noisy-interface family.  The physical counting step is unchanged after that family is declared.  The ideal exact-proof inequality is $E_{\mathrm{proof}}\ge D(f_n)k_B T\ln2$ for an interface that retains the required query distinctions as irreversible evidence.  Real proof engines declare tolerances, sampling rules, failure probabilities, and retained summaries, and those declarations determine the lower-bound family that replaces the exact one.

For example, replace exact proof that all $n$ spins are aligned by the approximate claim that the fraction of misaligned spins is at most a declared tolerance with confidence $1-\delta$.  The evidence object is then the retained random seed, sampled indices, pointer outcomes, calibration records, and the statistical acceptance calculation.  Its retained-record family is the sample count required by the declared confidence rule, not the exact audit family $R(n)=n$.  At fixed tolerance and confidence this can be much smaller than $n$ because it certifies the stated finite-confidence sampling claim, not exact alignment of every spin.  Approximate proof with bounded $R(n)$ does not escape the theorem; it satisfies the theorem under a different contract.

The Lean development mechanically checks the arithmetic threshold and interface bookkeeping: finite-interface definitions, budget arithmetic, monotone and unbounded threshold lemmas, query-to-record bridges, proof-act bookkeeping, compressed and distributed audit accounting, assumption-boundary countermodels, and capability-interface clauses.  All physical content enters as explicit premises or cited empirical input.  The supplementary ledger maps each formal claim below to a Lean handle, and Appendix~\ref{app:formal-interface} gives the corresponding definitions.  Table~\ref{tab:lean-axiom-scope-auto} separates the machine-checked kernel footprint from empirical inputs: SR, Landauer cost, detector bandwidth, and record-formation contracts remain explicit premises or cited physics.  Reported \texttt{Classical.choice} uses occur in finite bookkeeping, including finite choices from finite types that are constructively available but not distinguished by Lean's axiom profiler, and are not physical existence premises.

\section{Physical Grounding of Proof Records}\label{sec:physical-grounding}

The counting gap becomes a substrate-side constraint through a declared finite-resolution proof interface: finite records, positive record cost, and exact query lower bounds compose into time and work lower bounds.

\subsection{Finite-Resolution Proof Interfaces}\label{sec:continuous-discrete}

Continuous mathematical models can describe underlying dynamics, limiting behavior, and symmetry structure.  Proof uses the finite-resolution records exposed by the declared interface; the continuous/discrete distinction concerns retained evidence at that interface.  The elementary counting contract is stated explicitly because later sections use its hypotheses separately.

\leanmetapending{\LH{BA10}, \LHrng{PBC}{8}{13}.}
\begin{remark}[Counting-Gap Contract and Budget Consequences]\label{rem:real-budget-floor-bound}\label{rem:universal-feasibility-bounds-record-growth}\label{rem:budget-monotonicity}\label{rem:cost-floor-monotonicity}
Let $\varepsilon, C \in \mathbb{N}$ with $\varepsilon > 0$ and $C > 0$. If each recorded-proof event consumes $\varepsilon$ discrete cost units, then for every $N \in \mathbb{N}$:
\[
  \varepsilon \cdot N \leq C \;\Rightarrow\; N \leq C.
\]
For real-valued physical budgets, $\varepsilon,B\in\mathbb{R}$ with $\varepsilon>0$ and integer record count $N$ satisfy
\[
  \varepsilon N\leq B \;\Rightarrow\; N\leq \left\lfloor B/\varepsilon\right\rfloor .
\]
Consequently, if every size is feasible under one fixed real budget $B$ and one fixed positive per-record cost $\varepsilon$, then $R(n)\leq\lfloor B/\varepsilon\rfloor$ for all $n$.  Increasing $B$ preserves feasibility, decreasing $B$ preserves infeasibility, and strengthening the cost floor preserves infeasibility.  These statements are arithmetic and use no physical premise.
\end{remark}

Finiteness plus positive retained-record cost supplies the incompatibility: fixed $B$, positive $\varepsilon$, and unbounded $R(n)$ cannot support universal feasibility.  Thermodynamics supplies the ordinary physical scale of $\varepsilon$, with Landauer giving the ideal irreversible-bit floor; SR supplies an independent acquisition ceiling through finite signal propagation; query lower bounds supply the task-specific growth of $R(n)$.

\subsection{Physical Scale and Positive Cost}\label{sec:positive-cost-derived}

Numerical scale comes from two empirical inputs for ordinary finite-resolution proof interfaces:

\begin{enumerate}
\item[\textbf{SR.}] \textbf{Special Relativity.}
  Signals and causal influence in ordinary relativistic substrates propagate no faster than $c$.
  Einstein established: $c = 2.998 \times 10^8$ m/s.
  \cite{einstein1905electrodynamics,minkowski1908space}

\item[\textbf{TD.}] \textbf{Thermodynamics (Landauer).}
  Each irreversible recorded distinction has a positive work floor.
  Landauer established the universal floor: $\varepsilon \geq k_B T \ln 2$ joules per irreversible bit. Theorem~\ref{thm:thermo-derived} and the thermodynamic-lift discussion below separate that universal floor from mismatch and residual overheads; when their premises hold, those overheads can only raise the per-record lower bound.\leanmeta{\LHrng{WC}{1}{3}, \LH{WM1}, \LHrng{WM}{3}{4}, \LHrng{WR}{1}{3}, \LH{WR6}.}
  \cite{landauer1961irreversibility,bennett1982thermodynamics,berut2012experimental,jun2014high,wolpert2024stochastic,manzano2024absolute,yadav2026minimal}
\end{enumerate}

The positive-cost premise can be supplied by two physical routes.  On the Landauer/Bennett route, the proof interface forms a durable irreversible distinction, and Landauer supplies the ideal record-creation floor $k_B T\ln2$ per bit.  On a thermal-maintenance or stabilization route, the interface instead declares the work needed to couple, amplify, protect, refresh, or hold the pointer state long enough for inspection.  The theorem uses only the common output of these routes: a positive per-record floor $\varepsilon>0$.  A dispute about whether a particular apparatus realizes the Landauer value is therefore a dispute about the numerical value of $\varepsilon$, not about the budget theorem itself.

\begin{table}[htbp]
\centering
\small
\setlength{\tabcolsep}{4pt}
\renewcommand{\arraystretch}{1.12}
\renewcommand{\tabularxcolumn}[1]{m{#1}}
\begin{tabularx}{\linewidth}{@{}>{\raggedright\arraybackslash}m{0.22\linewidth}>{\raggedright\arraybackslash}m{0.25\linewidth}>{\raggedright\arraybackslash}X@{}}
\toprule
\textbf{Quantity} & \textbf{Symbol} & \textbf{Physical meaning} \\
\midrule
Signal propagation bound & $c=2.998\times 10^8\,\mathrm{m/s}$ & Maximum signal speed in ordinary relativistic substrates \\
\midrule
Landauer floor per irreversible bit & $k_B T\ln 2\approx2.87\times 10^{-21}\,\mathrm{J}$ at $300\,\mathrm{K}$ & Universal ideal lower bound for one irreversible recorded distinction \\
\midrule
Total work budget & $B<\infty$ & Declared finite work available to the proof-producing substrate \\
\midrule
Effective per-record cost floor & $\varepsilon\ge k_B T\ln 2$ & Declared cost floor for each retained irreversible evidence record \\
\midrule
Maximum retained proof distinctions & $\lfloor B/\varepsilon\rfloor<\infty$ & Finite ceiling on retained records under the declared contract \\
\bottomrule
\end{tabularx}
\caption{The counting-gap contract requires exactly $\varepsilon>0$ and $B<\infty$; numerical values are illustrative.}
\label{tab:calibration-anchor}
\end{table}

Quantum mechanics supplies background measurement theory for quantum and atomic realizations of the interface, including outcome-state vocabulary and calibration relations such as $E=h\nu$~\cite{planck1901distribution,dirac1930principles}.  The counting theorem uses only the interface contract, the SR propagation bound, and the TD per-record work floor.

\noindent The counting-gap contract supplies the arithmetic structure; SR and TD supply the ordinary physical scale.  At $T>0$, an irreversible recorded bit has the Landauer floor $k_B T\ln 2$, so a separable-audit threshold for a budget $B$ sits near
\[
  n_{\max}(B,T)=\left\lfloor\frac{B}{k_B T\ln2}\right\rfloor
\]
in the Landauer-floor idealization.  As $T\to0^+$ this calibrated ceiling moves outward like $1/T$ and remains finite for each fixed positive temperature.

The zero-temperature limit is a boundary of the Landauer value, not an available free-record regime.  Nernst-style unattainability and finite-resource third-law formulations treat exact zero temperature as physically inaccessible by finite operations~\cite{nernst1906berechnung,masanes2017general}.  Quantum measurement also keeps costs in view: preparation, coupling, control, amplification, back-action management, and pointer retention remain physical operations, especially in Zeno-style interrogation~\cite{misra1977zeno,itano1990zeno,wiseman2010quantum,jacobs2006straightforward,clerk2010introduction}.  At exactly $T=0$, the Landauer expression no longer supplies the numerical value $k_B T\ln2$, so a usable proof interface must declare its replacement positive record-formation floor: coupling energy, amplification work, pointer-state stabilization, refresh cost, isolation overhead, or another physical cost that makes the record durable and inspectable.  Declaring $\varepsilon=0$ removes the positive-cost premise and changes the theorem's input rather than escaping its conclusion.

\noindent\textbf{Beyond the Landauer floor.}  Landauer gives the ideal reversible-limit floor for one irreversible recorded bit.  Stochastic thermodynamics refines that floor for constrained processes: finite-time driving, measurement feedback, distributional mismatch, residual dissipation, and absolute irreversibility add nonnegative terms under their corresponding premises~\cite{sagawa2012nonequilibrium,parrondo2015thermodynamics,wolpert2024stochastic,manzano2024absolute,yadav2026minimal}.  With per-record mismatch $m\ge0$ and residual overhead $r\ge0$, a proof retaining $N$ records satisfies
\[
  E \geq N(k_B T\ln2+m+r),
\]
under the corresponding constrained-process premises.  Non-equilibrium substrates change the effective $\varepsilon$ supplied by the declared interface.  Lower dissipation moves the proof horizon outward; finite-time, noisy, feedback, or absolute-irreversibility costs move it inward.  The impossibility theorem uses the weaker invariant common to these regimes: retained evidence has some positive per-record floor.

\begin{example}[Room-temperature Landauer scale]
At $T=300\,\mathrm{K}$, the ideal Landauer floor is
\[
k_B T\ln 2 \approx 2.87\times 10^{-21}\,\mathrm{J}
\]
per irreversible recorded bit.  A one-joule budget could therefore pay for at
most about $3.5\times 10^{20}$ irreversible recorded bits at the ideal floor.
The number is enormous, and real constrained processes generally dissipate
more.  A one-megawatt-year work budget
($\approx 3.2\times 10^{13}\,\mathrm{J}$) would pay for at most about
$1.1\times 10^{34}$ ideal irreversible recorded bits.  A proof requiring
$1000$ retained local measurement distinctions is far below that floor, while a hidden
truth-table task with $2^n$ retained distinctions exceeds the
one-megawatt-year ideal budget at $n=114$.
\end{example}

For ordinary laboratory-scale finite tasks, the ideal Landauer crossover is usually beyond the experimentally relevant regime; engineering dissipation, detector bandwidth, preparation error, and readout noise dominate first.  Exponential retained-record families can cross macroscopic ideal budgets at moderate formal size, as the hidden truth-table example does at $n=114$ for a one-megawatt-year budget.  Linear witnesses usually function as foundational scale examples.  Realistic approaches to the ideal floor occur in engineered low-dissipation memory or readout experiments where record stabilization is the optimized task; ordinary certification encounters nonideal readout, error correction, and calibration costs first.  The theorem fixes the ideal proof-record floor and identifies which premise must change when an interface claims universal exact proof.

\subsection{Continuity-Neutral Scope}\label{sec:continuity-neutral-scope}

The result is neutral about whether the universe is continuous or discrete
below the resolution of a declared proof interface. Whether finer
structure exists is irrelevant to proof unless that structure is made locally
accessible as recorded evidence within the declared budget.

\subsection{Bounded Information Acquisition Rate}\label{sec:bounded-acquisition}

In a bounded region $\mathcal{R}$ of diameter $d$ with signal speed $c$, a serialized acquisition whose proof depends on causal traversal across the region takes at least $d/c$ time.  A proof interface operating for time $T$ can therefore complete at most
\[
\text{acquisitions}(\mathcal{R}, T) \leq \frac{c \cdot T}{d}.
\]
This finite acquisition ceiling is independent of the thermodynamic bound and supplies a second physical limit on proof-record capacity.\leanmeta{\LHrng{BA}{1}{2}.}

\subsection{Recorded Acquisitions and Binary Distinctions}\label{sec:discrete-transitions}\label{sec:one-bit}

A declared finite-resolution proof interface has a finite state space, a transition rule, and a positive lower-bound cost for each irreversible recorded distinction.  Its recorded acquisitions are transition events in the associated finite transition system.  Each retained record embeds at least one binary before/after distinction: the interface was in state $A$ before the transition and state $B$ after it.  This distinction is the atomic unit of record-cost accounting.  Landauer/Bennett supplies the creation-event floor for irreversible bits; other interfaces may declare a stabilization or maintenance floor for the same retained distinction.  Larger registers may expose more labels, and unrecorded degrees of freedom remain outside the proof object.  Atomic measurement interfaces provide a standard physical instance of this bookkeeping.\footnote{For finite electronic configurations and transition vocabulary, conservation-law bookkeeping, and thermal scale, see~\cite{schrodinger1926undulatory,pauli1925zusammenhang,sakurai2017modern,noether1918invariante,boltzmann1877beziehung,kittel1980thermal}.}  Appendix~\ref{app:recorded-binary-proof} records the label-accounting refinement for larger registers.\leanmeta{\LHrng{CI}{1}{4}.}

\subsection{Quantum Verification Interfaces}\label{sec:quantum-verification-interfaces}

Quantum verification can keep evidence coherent during the computation.  Coherence changes where the evidence lives and which lower-bound family applies.  Acceptance still reaches a verifier-facing record.

A quantum interactive proof has a finite transcript, verifier challenges, measurements, and an accept/reject event.  A $\mathsf{QMA}$ verification has a quantum witness, verification circuit, and final acceptance record.  Mahadev-style classical verification replaces direct quantum access by a cryptographic interaction whose classical verifier retains keys, challenges, responses, and an acceptance transcript~\cite{watrous2009quantum,aharonov2002quantum,mahadev2018classical}.  In each case the interface declares the evidence object: transcript records, syndrome histories, outcomes, commitments, sufficient statistics, or final reports.

Complexity-theoretic proof systems show how large the change in $R(n)$ can be.  Classical interactive proofs characterize $\mathsf{PSPACE}$, quantum interactive proofs also characterize $\mathsf{PSPACE}$, and entangled multiprover protocols reach recursively enumerable languages under their mathematical idealizations~\cite{shamir1992ip,jain2010qip,ji2020mipstar}.  These results demonstrate that interaction, quantum witnesses, and entanglement can replace a large static certificate by a much smaller verifier transcript, witness check, challenge log, or acceptance record for some tasks.  The budget threshold is then computed from that interface.  If the declared quantum or interactive verifier has bounded, logarithmic, or polynomial retained-record family, the threshold moves accordingly; if the resulting retained-record family remains unbounded, every finite budget still has a proof horizon.

Exact quantum query tasks use exact quantum query lower bounds, while witness-verification tasks induce the retained transcript, syndrome, or measurement-family required by their declared verifier~\cite{beals2001quantum}.  Interactive and cryptographic protocols can reduce one verifier's transcript by moving evidence into prover state, commitments, challenge rounds, trap checks, or logs.  The compressed-audit theorem counts aggregate retained records under the declared multi-phase interface.\leanmeta{\LHrng{CI}{43}{54}.}

\begin{example}[Entanglement verification]
Entanglement verification illustrates the range of possible $R(n)$ values inside quantum information.  For a known target family, a stabilizer test, Bell statistic, or entanglement witness can retain a small list of measurement settings, outcomes, confidence parameters, and calibration records~\cite{horodecki2009quantum,guhne2009entanglement}.  That interface proves the declared witness or statistic: the retained records distinguish states that satisfy the witness contract from states that violate it.

The same retained state does not usually distinguish every other alternative in the Hilbert space.  Many distinct density matrices can share the same Bell statistic, stabilizer syndrome, or witness expectation.  If the claim is only ``this witness certifies entanglement above the declared threshold,'' the short record may be the right proof object.  If the claim is ``the laboratory prepared this particular unknown $n$-qubit state,'' the verifier must distinguish many more alternatives, and the retained-record family grows with the expectation values, calibration data, or density-matrix parameters needed for that stronger claim.  In the general case this approaches tomography, so the retained-record family grows with the state description, not one witness statistic.  Thus one laboratory setup can instantiate multiple proof contracts: a compact witness contract, a calibration-heavy device-certification contract, and a much larger state-identification contract.  The accounting follows the declared quantum verification interface in each case.
\end{example}

\begin{example}[Cosmological topology and finite sky records]
Cosmology gives the same structure at a different scale.  A model may assign a definite global spatial topology, while a situated observer receives only a finite record through a past light cone.  Matched-circle searches in the cosmic microwave background instantiate a proof interface with sky maps, masks, angular correlations, candidate circle pairs, calibration records, and model assumptions~\cite{cornish1998circles,cornish2004constraining}.  That finite catalog can support a declared finite-confidence topology claim.  Exact proof of the global topology is stronger: the retained records must distinguish all in-scope global geometries that induce the same observed catalog.  The retained-record family is therefore supplied by the declared bridge from finite sky records to the global identification being asserted.
\end{example}

Interpretations of quantum theory change the ontology assigned to the internal process, not the retained-record accounting used by a verifier.  Collapse accounts place the outcome in a post-measurement record; decoherence accounts identify stable pointer states and environmental records~\cite{zurek2003decoherence}; Everett-style accounts track records inside branches~\cite{everett1957relative}; relational and QBist accounts assign state claims relative to systems or agents~\cite{rovelli1996relational,fuchs2014introduction}.  Across these readings, operational proof requires a declared record that the verifier can inspect or use.  Bounded-error protocols prove their declared soundness/completeness claim with the retained records required by that protocol; exact proof uses the exact-proof interface of Section~\ref{sec:proof-versus-provability}.

The situated-agent version packages the same physical inputs directly: finite diameter $d$, finite lifetime $\tau$, finite signal speed $c$, finite budget $B$, and positive retained-record cost $\varepsilon$.  Any retained-record count feasible both thermodynamically and through the spacetime acquisition window is bounded by
\[
\min\!\left(\left\lfloor \frac{B}{\varepsilon}\right\rfloor,
\left\lfloor \frac{c\tau}{d}\right\rfloor\right),
\]
in the discrete interface units of the bounded-acquisition model.  A physically situated agent therefore has a finite lifetime proof-record capacity before any task-specific lower-bound family is inserted.\leanmeta{\LHrng{CI}{38}{42}.}

The binding constraint is the smaller term in this minimum after the task lower bound $R(n)$ is inserted.  Thermodynamics binds when $R(n)>\lfloor B/\varepsilon\rfloor$ before the serialized acquisition ceiling is reached; the SR traversal window binds when $R(n)>\lfloor c\tau/d\rfloor$ first.  Query complexity supplies the required-record growth curve, while SR and TD supply independent physical ceilings against which that curve is compared.

Finite scientific communities aggregate the same bound.  A finite collection of agents with budgets $B_i$ and positive retained-record costs $\varepsilon_i$ has aggregate retained-record capacity $\sum_i \lfloor B_i/\varepsilon_i\rfloor$ at any finite time.  The theorem concerns a finite communication graph and a finite-time retained transcript.  Spatial infinity needs a local-budget bridge that supplies the relevant records to the proof-producing computation.  Records transmitted along a finite communication graph become retained records at the receiving agent and are charged to that receiver's positive per-record floor.  Any feasible community transcript is bounded by this aggregate capacity, and every unbounded retained-record lower-bound family eventually exceeds it.\leanmeta{\LHrng{CI}{55}{59}.}

\subsection{Minimum Recorded Distinctions and Exact Query Complexity}\label{sec:query-physical}

\leanmetapending{\LHrng{CI}{3}{6}, \LHrng{CI}{27}{29}.}
\begin{theorem}[Exact Proof Requires Recorded Access]\label{thm:resolution-sufficient}
Let $f_n:X_n\to Y_n$ be an exact proof task with declared finite query interface.  Any evidence-retaining physical process that proves $f_n$ for all in-scope inputs must expose at least $D(f_n)$ locally accessible recorded distinctions in the worst case.  Computational complexity calls this number the exact adaptive query complexity.
\end{theorem}

\begin{proof}
If a purported proof-producing process exposes fewer than $D(f_n)$ recorded distinctions on every input, its physical access pattern defines an adaptive measurement protocol of depth $<D(f_n)$, contradicting the exact lower bound for the declared interface.  The bridge constrains substrate implementations: unaccessed variables, continuous amplitudes, and hidden internal correlations count as proof evidence only when the declared interface exposes those distinctions as locally accessible records.  When proof requires an audit trail, each evidence-bearing access must also be represented by at least one retained recorded distinction in the proof interface.
\end{proof}

The access-to-record step uses the separable pointer-record convention.  In the standard projective-measurement/classical-pointer interface, a measurement produces a retained classical pointer state that fixes the audit-accounting grain.  Joint, compressed, entangled, topological, or nonlocal evidence objects are handled by declaring the corresponding interface and lower-bound family $R_{\mathrm{joint}}(n)$.  The formal bridge represents an injection from accessed distinctions into retained pointer records; a proof object with fewer retained records is a different compressed evidence interface and must supply its own lower-bound family.\leanmeta{\LHrng{CI}{27}{29}.}

The proof-act distinction of Section~\ref{sec:proof-versus-provability} appears here in lower-bound form.  A joint record can answer a declared global predicate with one retained value, while retained-audit proof of an underlying coordinate-wise history is a stronger target with its own $R(n)$.  Thermodynamics enters after the declared interface retains the distinctions needed for local acceptance.\leanmeta{\LHrng{CI}{27}{33}, \LHrng{PROV}{58}{62}.}

Compressed audit trails redistribute this accounting across commitment, challenge, response, recomputation, and acceptance phases.  The aggregate protocol model defines sound retained-audit acceptance as an injection from underlying instance-bit distinctions into total retained protocol records; Merkle trees, polynomial commitments, and interactive arguments are special cases.  If aggregate retained records fall below the required distinctions, the delivered object belongs to the answer, trust, or bounded-confidence interface described in Section~\ref{sec:proof-versus-provability}, not to the declared retained-audit target.\leanmeta{\LHrng{CI}{30}{33}, \LHrng{CI}{43}{54}.}

\leanmetapending{\LHrng{CI}{3}{6}}
\begin{theorem}[Recorded Distinctions Set Information, Time, and Energy Bounds]\label{thm:query-physical}
For an exact proof task $f_n$ with $D(f_n)=k$:

\begin{enumerate}
\item \textbf{Information.}
  Proof requires at least $k$ locally accessible physical distinctions in the worst case.  Proof records those distinctions for the declared observer.

\item \textbf{Time.}
  Each locally accessed distinction requires an interface event of duration $\geq d/c$ under the declared propagation model.  Minimum serialized proof time has lower-bound scale $k\cdot d/c$ under that traversal-time model.

\item \textbf{Energy.}
  If proof retains each accessed distinction as irreversible evidence, each retained record carries at least the Landauer floor $k_B T\ln 2$ by \textbf{TD}.  Under that retained-evidence accounting, the irreversible-record energy lower bound is $E\geq k\,k_B T\ln 2$; stronger empirical lower bounds only raise this floor.
\end{enumerate}
\end{theorem}

\begin{proof}
The information clause is Theorem~\ref{thm:resolution-sufficient} with $k=D(f_n)$, where $D(f_n)$ names the minimum number of physically exposed distinctions for the declared finite interface.  The time clause uses the BA1 and BA2 serialized traversal formalization: a bounded region supplies a positive diameter and signal speed, and the acquisition count over time $T$ is bounded by the discrete tick budget $cT/d$.  The energy clause then applies the evidence-retention conversion law to the same worst-case access count.  Thus $k$ sets the serialized-time lower-bound scale and, when those distinctions are retained as proof evidence, the irreversible-record energy bound.
\end{proof}

\subsection{Thermodynamic Lower Bound as Physical Consequence}\label{sec:thermo-derived}

The additive bound uses the same separable-record convention as Definition~\ref{def:exact-physical-certification-contract}: each retained query distinction is counted as a separable irreversible interface event.  Joint or compressed proof objects can lower $R(n)$ only after the new interface declares its retained evidence and lower-bound family.

\leanmetapending{\LH{BA9}, \LHrng{CI}{5}{7}}
\begin{theorem}[Thermodynamic Cost of Evidence-Retaining Proof]\label{thm:thermo-derived}
For any declared finite-resolution proof interface that proves task $f_n$ at temperature $T > 0$ while retaining the required query distinctions as irreversible evidence records:
\[
E_{\mathrm{proof}} \geq D(f_n)\, k_B T \ln 2.
\]
\end{theorem}

\begin{proof}
Proof of $f_n$ requires at least $D(f_n)$ locally accessed physical distinctions in the worst case (Theorem~\ref{thm:resolution-sufficient}).  When those distinctions ground proof, each required access is retained as at least one separable irreversible before/after event in the proof interface.  By \textbf{TD}, each irreversible recorded bit costs $\geq k_B T\ln 2$.  Summing over those separable retained evidence records gives $E\geq D(f_n)\,k_B T\ln 2$.  A jointly compressed proof object changes the declared record family and must be counted through the lower-bound function for that compressed evidence interface.
\end{proof}

The constant $k_B T \ln 2$ is the universal Landauer floor for irreversible
bit erasure, experimentally verified to $\pm 10\%$
\cite{berut2012experimental,jun2014high}. It gives a floor; actual energetic costs for constrained computation generally require stronger process-specific accounting
\cite{wolpert2024stochastic,manzano2024absolute,yadav2026minimal}.

The impossibility theorem requires exactly a declared positive retained-record floor; the Landauer value supplies the universal irreversible-bit floor.  Stochastic-thermodynamic refinements add nonnegative mismatch, residual, and absolute-irreversibility terms under stronger constrained-process premises.  Appendix~\ref{app:stochastic-thermo-refinements} gives those branches.

\begin{remark}[Reversible Answer-Only Reports]
In thermodynamic form, the distinction of Section~\ref{sec:proof-versus-provability} says that work for an exact proof task family scales with its retained-record lower bound $R(n)$, while answer production is governed by its own declared evidence contract.

The $D(f_n)\,k_B T\ln 2$ bound applies when the proof interface retains the queried distinctions as checkable evidence.  The thermodynamic floor attaches when that interface creates an irreversible evidence record: the queried outcome, the certificate bit, the accepted report, or any durable environmental record that makes the exact claim checkable.  The formal accounting splits acquisition/computation cost from retention cost; the additive identity and zero-acquisition positive-retention theorem are separate from the query lower bound.\leanmeta{\LHrng{CI}{34}{37}.}
\end{remark}

Among nonconstant evidence-retaining exact proof tasks in the declared interface, tasks with $D(f_n)=1$ are the minimal positive case: $E_{\mathrm{proof}}\geq k_B T\ln 2$ per irreversible recorded query in the Landauer-floor idealization.  Tasks with $D(f_n)>1$ are above that floor by at least $(D(f_n)-1)k_B T\ln 2$ per worst-case proof, with stronger empirical per-record costs only enlarging the gap.

\subsection{Connection to the Formal Model}\label{sec:physical-to-formal}

The exact-query structure of Definition~\ref{def:decision-problem} is connected
to physical proof by the recorded-binary-distinction interface and Theorem~\ref{thm:query-physical}:

\begin{itemize}
\item The finite query entries exposed by the task correspond to the
  finite recorded distinctions exposed by the declared proof interface
  under the finite-resolution and serialized-acquisition bookkeeping above.

\item The Boolean query case is the elementary case:
  each queried bit is represented by one recorded before/after distinction
  at the proof interface.

\item $D(f_n)$ counts the minimum worst-case physical distinctions
  that must become recorded access events to prove $f_n$
  (Theorem~\ref{thm:query-physical}).

\item The thermodynamic lower bound (Theorem~\ref{thm:thermo-derived})
  is the physical content of the thermodynamic lift results.
\end{itemize}

The exact-query model is substrate-neutral (Section~\ref{sec:foundations}).
The physical grounding applies to finite-resolution
proof interfaces whose records are locally accessible, whose propagation
model is bounded, and whose irreversible records have a positive thermodynamic
work floor.
\leanmeta{\LH{BA9}, \LHrng{CI}{1}{7}, \LH{DS5}}

The physical domain is explicit: a substrate is in scope only when it declares the physical constants, observables, encoding map, exact-proof task contract, and boundary-condition statement required by Definition~\ref{def:physical-scope-gate} in Appendix~\ref{app:physical-scope-reporting}.

\section{Physical Lower-Bound Witnesses as Required-Record Families}\label{sec:complexity-infrastructure}

The physical impossibility theorem uses a lower-bound function $R(n)$ counting the locally accessible elementary distinctions that must be retained as evidence.  The witnesses are physical access problems first: spins that must be read, hidden switches that must be inspected, and table entries that must be exposed.  Standard deterministic query complexity supplies the finite bookkeeping.

The first witness is separable parity: it requires $n$ retained distinctions for exact proof while the answer remains computationally easy.  The anchored sign-query family gives a second linear witness and later receives a compact unit-clause SAT encoding.  Hidden truth-table proof supplies an exponential witness when a device exposes $2^n$ separately addressable entries.  Appendix~\ref{app:lower-bound-variants} records the formal lower-bound-family definition and the quantum, interactive, and approximate variants.

\begin{example}[Parity of $n$ Spins Gives an Unbounded Exact Query Lower Bound]
Let the input be $n$ two-state systems, such as spins, atoms, or any locally distinguishable binary physical variables with states $\uparrow$ and $\downarrow$.  The task is to prove the parity of the total number of $\uparrow$ states modulo $2$.  Exact proof must expose enough physical records to rule out a change in any one spin.  In query-complexity notation this retained-access lower bound is
\[
D(\mathrm{PARITY}_n)=n.
\]
Thus $R(n)=n$ is an unconditional lower-bound family for exact proof.  Evaluating the proof claim requires $n$ measurable spin distinctions to be retained for the verifier.  Parity remains polynomial-time and reversibly computable; the lower bound concerns the audit trail required to show that every answer-changing spin has been excluded.  Exact quantum query complexity does not remove the linear barrier for parity: $Q_E(\mathrm{PARITY}_n)=\Theta(n)$, so coherent access can change constants and implementation details without making separable exact parity proof bounded~\cite{beals2001quantum}.

\begin{proof}
Inspecting all $n$ systems and returning their parity gives $D(\mathrm{PARITY}_n)\le n$.  Conversely, suppose an exact adaptive protocol has an execution path on which some physical coordinate $i$ is not inspected.  Two configurations that agree on all inspected coordinates and differ only in coordinate $i$ produce the same retained evidence, so the protocol returns the same report on both.  Their parity values differ, contradicting exact correctness.  Hence every exact protocol has worst-case access depth at least $n$.  The lower bound already quantifies over adaptive strategies: adaptation can choose which coordinate is left uninspected on a path, but it cannot prove parity while any answer-changing coordinate remains unobserved.
\end{proof}
\end{example}

Three examples mark three different interface changes.  Separable parity gives the clean decision-tree witness: the separable audit interface has $R(n)=n$, while a certified global-observable interface can make $R(n)$ bounded for the parity value alone.  Entanglement verification separates a compact witness statistic from full-state verification.  Cosmological topology separates finite-sky evidence from exact global identification.  In each case the original proof target keeps its own lower-bound family until the declared evidence interface changes.

A finite-measurement analogue uses $n$ distinguishable two-state spin systems with declared axis and outcomes $\uparrow,\downarrow$.  Proving exactly that all spins are $\uparrow$ requires measuring every spin in the worst case by the same unmeasured-coordinate indistinguishability argument.

\begin{example}[Anchored Sign-Query Lower Bound]
An anchored sign-query interface exposes $n$ hidden two-state switches, each inspectable only by querying that switch, together with a public anchor setting.  The verifier must prove that no hidden switch changes the global constraint.  If switch $j$ is left uninspected, the all-agreeing configuration and the configuration with only $j$ flipped are indistinguishable while requiring different exact reports.  Hence the retained-record lower-bound family is linear and unbounded.

The same apparatus has a compact Boolean encoding.  Fix a public variable $x$ by an anchor clause requiring $x=\mathsf{true}$, and let each hidden switch choose whether an additional unit clause contains $x$ or $\neg x$.  The retained records in this witness are the hidden clause signs: an unqueried sign can flip satisfiability while preserving the retained view.  A claimed shortcut with fewer retained records must declare its proof interface and soundness conditions; the counting theorem applies to the $R(n)$ induced by that interface.\leanmeta{\LH{SATQ6}, \LH{SATQ8}, \LH{SATQ11}.}

\begin{proof}
The compact encoding preserves the physical indistinguishability pair: the all-agreeing configuration is satisfiable, while the flipped configuration encodes both $x$ and $\neg x$ and is unsatisfiable.  Appendix~\ref{app:lower-bound-variants} records the formal indistinguishability and unboundedness statements.\leanmeta{\LH{SATQ6}, \LH{SATQ8}, \LH{SATQ11}.}
\end{proof}
\end{example}

\begin{example}[Hidden Truth-Table Device Requires Full Support]
Let a device hide a Boolean table $g:\{0,1\}^n\to\{0,1\}$ behind separately addressable physical entries, and consider the exact proof task of deciding whether $g\equiv 0$.  Any exact retained-evidence procedure requires inspecting $2^n$ entries in the worst case.

\begin{proof}
If a procedure halts after inspecting a strict subset of $\{0,1\}^n$, choose an uninspected point $z$.  The all-zero table and the table that is zero everywhere except at $z$ give identical retained evidence while requiring different proof reports.  Exact correctness therefore requires full worst-case inspection.
\end{proof}
\end{example}

The hidden truth-table witness is a formal access witness rather than the most direct laboratory model: it assumes $2^n$ separately addressable entries.  The parity witness is the cleaner physical example because its $n$ local records can be realized as $n$ local two-state systems.

\begin{remark}[Protocol Variants and Standard Status]
Quantum fingerprints, $\mathsf{QMA}$-style witnesses, interactive proofs, probabilistic checks, zero-knowledge protocols, continuous-variable readouts, and approximate protocols change the proof game and therefore change $R(n)$.  Their retained evidence may be a transcript, commitment, challenge, measurement outcome, accept/reject record, sufficient statistic, or final report record.

If the resulting retained-record lower-bound family is unbounded, the same budget threshold applies.  If the protocol lowers the retained-record requirement, the thermodynamic bound scales with that family.  The deterministic witnesses above are ordinary decision-tree indistinguishability arguments~\cite{buhrman2002complexity}; parity is useful because it is unconditional, linear, and algorithmically easy, not because the accounting is parity-specific.  MAJSAT enters only as the finite counting predicate underlying $\PP$-complete stochastic comparisons.  Appendix~\ref{app:lower-bound-variants} records the protocol variants.
\end{remark}

\begin{table}[htbp]
\centering
\small
\setlength{\tabcolsep}{3pt}
\renewcommand{\arraystretch}{1.14}
\renewcommand{\tabularxcolumn}[1]{m{#1}}
\begin{tabularx}{\linewidth}{@{}>{\raggedright\arraybackslash}m{0.20\linewidth}>{\raggedright\arraybackslash}m{0.20\linewidth}>{\raggedright\arraybackslash}m{0.25\linewidth}>{\raggedright\arraybackslash}X@{}}
\toprule
\textbf{Declared interface} & \textbf{Natural task} & \textbf{Retained evidence} & \textbf{Resulting lower-bound family} \\
\midrule
Separable audit & Parity or alignment of $n$ local spins & One retained pointer record per answer-changing spin & $R(n)=n$ by the unqueried-coordinate argument \\
\midrule
Global observable & Calibrated collective parity readout & One global outcome plus retained calibration and apparatus records & Bounded in $n$ for the global-parity value interface, with calibration charged separately; the per-spin audit interface keeps $R(n)=n$ \\
\midrule
Merkle or commitment audit & Membership of one exposed leaf in an $n$-entry committed table & Commitment root and authentication path & $R(n)=O(\log n)$ for one membership check; universal exact table claims inherit the retained paths, challenges, leaves, or prover-side records declared by the protocol \\
\midrule
Certificate verification & SAT instance with a proposed satisfying assignment & Assignment bits and formula-evaluation trace & $R_{\mathrm{cert}}(n)\ge n$ when the assignment has $n$ bits, so the certificate interface remains unbounded at a lower threshold than exhaustive audit \\
\midrule
Interactive or probabilistic summary & Constant-query property test, or sum-check for an algebraic claim under a soundness gap & Random seed, challenge log, response transcript, and accept/reject record & Bounded $R(n)$ for fixed-query property tests at fixed confidence; sum-check retains $O(m)$ field challenges and responses for an $m$-variable claim, logarithmic in the explicit table size; exact universal proof uses the lower-bound family of the exact interface \\
\midrule
Quantum witness or coherent verifier & $\mathsf{QMA}$ witness check, quantum fingerprint, or verified quantum computation & Witness register access, measurement outcomes, trap checks, syndrome history, or verifier transcript & The witness or transcript can replace a large static audit; each task uses the quantum or interactive lower bound of the declared verifier \\
\midrule
Finite-sky/global-geometry bridge & Matched-circle topology claim or other finite cosmological survey claim & Sky map, mask, candidate correlations, calibration records, and model assumptions & Finite-confidence survey claims use the declared catalog; exact global topology requires the bridge family from finite records to global identification \\
\bottomrule
\end{tabularx}
\caption{Concrete ways compressed, interactive, and quantum interfaces change $R(n)$.  Entries with bounded $R(n)$ fall outside the impossibility theorem; entries with unbounded $R(n)$ inherit the finite-budget threshold.  The original separable-audit task retains its own lower-bound family.}
\label{tab:interface-rn-examples}
\end{table}

\section{Physical Impossibility of Universal Evidence-Retaining Proof}\label{sec:physical-impossibility}

The Physical Counting Impossibility Theorem derives an incompatibility from an explicit contract: encoded task, measurement protocol, observational setup, budget model, and retained-record rule.  No fixed-budget positive-cost substrate can provide universal exact proof when the required recorded distinctions exceed the available budget at some size.

Agent/report scope and physical-substrate scope coincide here.  Report admissibility governs asserted outputs.  Physical counting governs the work needed to make those outputs exact proof.

\begin{definition}[Evidence-Retaining Exact Physical Proof Contract]\label{def:exact-physical-certification-contract}
An evidence-retaining exact physical proof contract, equivalently a declared evidence-retaining measurement protocol for a task family, consists of:
\begin{enumerate}
\item a declared exact-proof task family $f_n:X_n\to Y_n$ with a finite query interface;
\item a measurement-to-query-interface map $E:\mathcal{P}\to\mathcal{F}$ satisfying the declared physical domain of Definition~\ref{def:physical-scope-gate} in Appendix~\ref{app:physical-scope-reporting};
\item a finite physical budget $B<\infty$ in the chosen work units;
\item a positive per-record cost floor $\varepsilon>0$ for irreversible evidence records;
\item a required-retained-record lower-bound family $R(n)$ for exact proof on the encoded family, for instance $R(n)=D(f_n)$ when each required query distinction is retained as evidence;
\item a record-separability convention for the standard audit interface: each retained record counted by $R(n)$ is a separable irreversible interface event.  A joint, nonlocal, topological, or compressed proof object is allowed only by declaring the retained-record interface whose own positive cost floor and lower-bound family are being used.
\end{enumerate}
\end{definition}

\subsection{From Detectors to Query Interfaces}\label{sec:detectors-to-query-interfaces}

A query is the coarse-grained interface event induced by a declared detector protocol.  Projective measurements, POVM bins, threshold crossings, detector clicks, time samples, finite-bandwidth readouts, weak measurements, and continuous traces instantiate finite proof interfaces after the observable, outcome partition, resolution, sampling window, tolerance convention, retained-record rule, and positive cost floor are declared~\cite{nielsen2010quantum,wiseman2010quantum,jacobs2006straightforward,clerk2010introduction,shannon1948mathematical1,shannon1948mathematical2}.

\begin{table}[htbp]
\centering
\small
\setlength{\tabcolsep}{4pt}
\renewcommand{\arraystretch}{1.12}
\renewcommand{\tabularxcolumn}[1]{m{#1}}
\begin{tabularx}{\linewidth}{@{}>{\raggedright\arraybackslash}m{0.21\linewidth}>{\raggedright\arraybackslash}m{0.22\linewidth}>{\raggedright\arraybackslash}m{0.24\linewidth}>{\raggedright\arraybackslash}X@{}}
\toprule
\textbf{Lab protocol} & \textbf{Query interface} & \textbf{Retained record} & \textbf{Premise set} \\
\midrule
Projective spin measurement & Boolean or finite-alphabet query & Pointer outcome record & Positive per-record cost \\
\midrule
POVM with finite outcome bins & Finite alphabet query & Recorded bin index & Lower-bound family for the chosen partition \\
\midrule
Continuous detector trace & Sampled or thresholded record stream & Time-sample or bin records & Bandwidth, sampling, and resolution budget \\
\midrule
Weak measurement & Stochastic query channel & Aggregated samples or sufficient statistics & Exactness and lower-bound premise \\
\midrule
Analog field readout & Finite-resolution digitized interface & Digitized amplitude or phase records & Declared resolution and scope gate \\
\midrule
Quantum witness or fingerprint & Declared quantum verification game & Measurement outcomes, accept/reject records, or retained sufficient statistics & Soundness, completeness, and retained-record rule \\
\bottomrule
\end{tabularx}
\caption{Detector protocols as declared query interfaces.}
\label{tab:detector-query-dictionary}
\end{table}

A declared finite-outcome detector induces a finite query interface and a finite proof interface with the same positive retained-record cost, and finite budget bounds the number of retained detector records.\leanmeta{\LHrng{CI}{12}{16}.}

\leanmetapending{\LHrng{PH}{11}{15}, \LHrng{PH}{26}{27}.}
\begin{theorem}[Physical Counting Impossibility Theorem]\label{thm:physical-impossibility-universal-exact}
For any exact-proof experiment declared by a contract in the sense of Definition~\ref{def:exact-physical-certification-contract}, finite work budget $B$, positive per-record physical cost floor $\varepsilon$, and unbounded retained-record lower-bound family $R(n)$ make universal exact proof physically unavailable within that declared interface.
\end{theorem}

\noindent\textbf{Physical reading.}  The obstruction is thermodynamic and evidential.  Once the experimental interface is fixed, budget, cost, and record arithmetic can block proof even when the detector can still produce the answer cheaply by reversible computation.

\begin{proof}
The declared physical domain (Definition~\ref{def:physical-scope-gate} in Appendix~\ref{app:physical-scope-reporting}) licenses applying the lower-bound premise only to encoded physical instances satisfying the boundary conditions, for any agent in the sense of Definition~\ref{def:substrate-neutral-agent}.  For size $n$, the contract requires at least $R(n)$ retained records; the positive floor converts these into work at least $\varepsilon R(n)$.  Since $R(n)$ is unbounded while $B$ is fixed and $\varepsilon>0$, the counting-gap contract gives a size $n$ with $\varepsilon R(n)>B$.  At that size, every within-budget system lacks the retained-record work required by the same encoding, cost floor, lower-bound family, and exact report target.
\end{proof}

\leanmetapending{\LH{PBC18}.}
\begin{corollary}[Monotone Lower Bounds Give Tail Failure]\label{cor:finite-budget-threshold-impossibility}
If the retained-record lower-bound family $R(n)$ is monotone and unbounded, then for every real budget $B>0$ and real per-record cost floor $\varepsilon>0$ there exists $n_0$ such that every $n\geq n_0$ requires work exceeding $B$.
\end{corollary}

A linear separable-audit family gives the simplest instance: parity proof has $D(\mathrm{PARITY}_n)=n$ as a monotone unbounded retained-record family, so every fixed positive-cost finite budget has a tail beyond which that exact audit proof exceeds the budget.  The same first-principles tail result applies to exponential lower-bound families such as $R(n)=2^n$ and to any declared quantum, interactive, or survey interface whose retained-record family is monotone and unbounded.\leanmeta{\LHrng{PBC}{14}{20}.}

\begin{remark}[Approximate Proof]
Allowing an $\eta$-approximate report at tolerance $\eta$ or a bounded error probability changes the lower-bound premise while leaving the physical counting step intact.  The contract must specify the approximate target relation, tolerance, confidence level, and evidence rule.  If that approximate contract has a required-record lower bound $R_{\mathrm{approx}}(n)$ that still outruns the local budget after cost conversion, the same counting theorem applies with $R_{\mathrm{approx}}$ in place of $R$.  If approximation admits a smaller or bounded lower bound, the impossibility conclusion weakens accordingly; the report supplies approximate evidence under the declared tolerance.
\end{remark}

\subsection{Global Infinity versus Local Computational Budget}\label{sec:global-infinity-local-budget}

The budget premise in Theorem~\ref{thm:physical-impossibility-universal-exact} is computation-local.  Global infinity matters when the declared interface supplies accessible operations to the computation being proved.  A spacetime, field theory, or many-particle limit may contain an infinitary object globally while each realizable proof-producing process still has a finite causal past, prepared support, or particle-number budget.  The local-budget bridge connects global infinity to the particular proof-producing computation.  Formally, ``the world has infinite resources'' and ``some computation has infinite budget'' are different predicates.

\begin{definition}[Local-Budget Bridge]\label{def:local-budget-bridge}
A local-budget bridge consists of a selected computation or observer worldline, an access relation from the global structure to locally available interface events, and a budget accounting rule that assigns those events to the proof-producing computation.  The bridge is infinite for a task family when, for every finite bound $K$, the interface supplies more than $K$ locally accessible events satisfying the task boundary conditions before the report is issued.
\end{definition}

In ordinary local quantum field theory practice, cutoffs, finite preparation energy, finite causal diamonds, horizons, and detector bandwidths implement this distinction.  Holographic and black-hole entropy bounds give stronger finite-budget candidates when their hypotheses apply~\cite{bekenstein1981universal,thooft1993dimensional,susskind1995world,bousso2002holographic}; each bounds the same local-budget premise.

\begin{example}[Spatially infinite field, finite detector]
A quantum field on a spatially infinite Minkowski slice has infinitely many formal modes, but a laboratory detector coupled to that field for finite time, aperture, bandwidth, memory, and readout energy accesses only the modes made available by that coupling.  Turning global mode infinity into proof budget requires infinitely many locally accessible retained records before the report is issued.  Without that bridge, the infinite field belongs to the background theory; the proof record contains detector-accessible modes.
\end{example}

\begin{example}[Cosmological topology, finite sky records]
A cosmological model may assign a definite global spatial topology, while a situated observer receives a finite record through a past light cone.  Matched-circle searches in the cosmic microwave background give a concrete interface: the retained evidence is a catalog of sky maps, angular correlations, candidate circle pairs, calibration records, and model assumptions~\cite{cornish1998circles,cornish2004constraining}.  A finite catalog supports the declared finite-confidence topology claim.  Its retained-record family grows with the number of independent sky pixels, correlation bins, and candidate identifications required by the search protocol.  Exact proof of a global topology requires the declared bridge from those finite records to the global identification being asserted, and its $R(n)$ is the retained-record family induced by that bridge.
\end{example}

\begin{example}[Vacuum structure, finite correlators]
A quantum field theory can specify vacuum sectors, order parameters, tunneling channels, or false-vacuum decay rates at the level of the formal model~\cite{coleman1977fate}.  A laboratory proof uses finite records: spectra, correlation functions, scattering data, lattice samples, or detector histories.  A finite set of correlators gives a finite proof task under the declared observable family and tolerance.  An exact global vacuum claim has the retained-record requirement supplied by the interface that connects those local records to the global sector.
\end{example}

The same pattern covers eternal inflation, spatial infinity, infinite-dimensional QFT Hilbert spaces, and anyonic infinite-particle limits: global structure matters for proof through the locally accessible records assigned to the selected computation.\leanmeta{\LHrng{GI}{3}{12}.}  Topological quantum computation is normally specified by finite anyon configurations and finite braid words~\cite{kitaev2003anyons,freedman2002modular,nayak2008nonabelian}; an infinite-particle limit changes the theorem when it supplies a corresponding local-budget bridge.

\leanmetapending{\LHrng{GI}{13}{16}.}
\begin{proposition}[Malament-Hogarth and CTC Interfaces Include the Missing Bridge]\label{prop:mh-ctc-bridge-boundary}
Malament-Hogarth and closed-timelike-curve interfaces occupy the bridge-present case.  In those interfaces, the structure explicitly supplies a distinguished computation or loop computation with infinite budget.  The formal implication holds because the local-budget bridge is an assumption of the interface.
\end{proposition}

These bridge-present cases give a selected worldline, loop, or causal arrangement computational access to resources that ordinary local finite-budget systems lack~\cite{hogarth1994non,etesi2002nonturing,nemeti2006general,andreka2009general}.  The counting theorem treats that access as a different interface premise.

\subsection{Physical Oracles and the Trust Budget}\label{sec:physical-oracles-trust-budget}

Oracle language becomes useful once it is typed physically.  Turing's oracle-machine abstraction leaves the oracle as an external answer source~\cite{turing1939systems}; later physical-computation discussions ask whether such answer sources could be realized by nonstandard mechanisms~\cite{copeland1997broad,aaronson2005np}.  A physical oracle is another agent with detector, memory, budget, record-formation rule, and proof interface.

If the oracle retains the audit trail and the agent receives only the answer bit, proof has moved to the oracle.  Local proof for the agent requires inspecting the oracle's retained evidence or retaining evidence about oracle reliability.  Black-box verification is the middle case where internal records exceed the consulting agent's budget.

\leanmeta{\LHrng{PROV}{66}{73}.}
\begin{proposition}[Physical Oracles Shift Audit Cost]\label{prop:physical-oracle-trust-budget}
A physical oracle consultation separates oracle-internal proof records, delivered answer records, and the consulting agent's trust-audit records.  Oracle-side proof availability is charged to the oracle's own positive-cost finite budget.  Delivered answer records below the declared proof requirement supply oracle testimony to the consulting agent.  If the agent audits trust in the oracle, the trust audit is itself a retained-record process under the agent's positive-cost finite-budget contract.  Consequently, every fixed positive finite trust-audit budget has a finite horizon for unbounded trust-audit lower-bound families.
\end{proposition}

\begin{proof}
The consultation structure separates oracle proof records, delivered answer records, trust-audit records, and the positive-cost budgets of the oracle and consulting agent.  Oracle-internal proof feasibility is charged to the oracle's budget; below-requirement answer transcripts remain delivered answers; trust-audit feasibility is charged to the consulting agent's budget.  The boundedness lemma used in Theorem~\ref{thm:physical-impossibility-universal-exact} applies again: if every trusted oracle answer required trust-audit records $T(n)$ under one fixed positive finite budget, then $T$ would be bounded by $\lfloor B/\varepsilon\rfloor$, contradicting unboundedness.
\end{proof}

Physical oracles redistribute retained evidence across agents.  Outsourcing verification can make an answer available to a smaller agent, but proof remains in the oracle or in the agent's trust audit; testimony and instrument certification have the same accounting structure.

\subsection{Assumption Boundary}

Each assumption is used: the threshold contradiction depends independently on finite budget, positive per-record cost, and a super-budget lower-bound family.
\leanmetapending{\LHrng{PH}{31}{33}.}
\begin{proposition}[Finite-Budget Necessity Countermodels]\label{prop:finite-budget-necessity-countermodels}
The formal development contains explicit countermodels for the premise boundary.  An unbounded budget removes the fixed threshold.  A zero per-record cost floor removes positive retained-record work.  A bounded retained-record lower-bound family can remain below the converted budget.
\end{proposition}

Appendix~\ref{app:assumption-countermodels} records the explicit countermodel forms.

\begin{remark}[Exact Truth Versus Exact Proof]
Exact proof concerns reports with retained grounds, not the ontology of exact facts.  A global vacuum property, an initial-condition parameter, or a nonlocal spacetime fact may have a determinate value in a chosen physical theory while remaining unprovable by any local finite-budget process.  Such cases are outside the universal exact-proof contract unless a declared interface supplies the required local evidence.
\end{remark}

\subsection{Capability Claims Require Evaluation Interfaces}

The evaluation-interface distinction matters physically because a substrate-free algorithm schema has no thermodynamic cost until an evaluation rule supplies steps, state transitions, observers, retained records, and costs.  A realized capability is a process available to agents with steps, time, accessible state, retained records, and observer access; bare specifications carry no runtime field, and polynomial time is a predicate on realized processes.\leanmeta{\LHrng{PROV}{41}{57}.}  The proof-act definition of Section~\ref{sec:proof-versus-provability} becomes a declared certification interface in the formal model: a proof-act exists for an agent exactly when a token is instantiated, accessible to that agent, and checkable under that interface.  If the scoped proof-act class is empty, there is no proof available to that agent, even when a substrate-free derivation exists as a specification.\leanmeta{\LHrng{PROV}{37}{40}, \LH{PROV74}.}

\leanmetapending{\LH{PROV31}, \LH{PROV36}, \LHrng{PROV}{58}{62}.}
\begin{proposition}[Tokenhood Does Not Supply Proof Extraction]\label{prop:tokenhood-insufficient-certifier-extraction}
Physical tokenhood, accessibility, and checkability underdetermine universal evidence-retaining proof.  There is a tokenhood-only operational scope gate with an available proof-act but no possible universal proof package for an impossible proof family; consequently no extraction from bare tokenhood to proof availability is licensed without separate operational-claim semantics.  Equivalently, an answer-only interface and an evidence-retaining proof interface are different objects: a trust-only system may accept both a correct and an incorrect claim, while a show-your-work system verifies claims by retained evidence whose inspection is sound.
\end{proposition}

The countermodels supply the boundary.  Tokenhood-only scope can make an operational proof-act available while no universal proof package exists.  Answer-only receipt can deliver correct outputs while failing show-your-work verification.  Under the declared contract, the evidence-retaining interface supplies proof; other contracts name token acceptance, witnessed certificate reports, answer production, or trust-audit processes with their own retained-record families.\leanmeta{\LHrng{PROV}{18}{36}, \LHrng{PROV}{58}{62}.}

Accepted tokens require licensed-content semantics: the accepted content fixes which proof processes the claim permits.  Certificate verification and finite meta-proof checking are both included in the same accounting.  A satisfying assignment with a retained evaluation trace proves the witnessed existential report for one satisfiable SAT instance, with its own $R_{\mathrm{cert}}(n)$; for SAT, $R_{\mathrm{cert}}(n)\ge n$ whenever the verifier must retain the full assignment.  A finite proof of a universal theorem has a finite, nonzero meta-proof cost governed by proof length and checker interface, while the instance-level $R(n)$ is determined by the proof acts the accepted content licenses.  Certificate verification supplies neither no-instance certification, solver-wide soundness, nor universal evidence-retaining proof for the full decision family; finite meta-audits likewise do not erase an unbounded object-level retained-record family.\leanmeta{\LHrng{CI}{17}{26}, \LHrng{PROV}{6}{35}, \LHrng{PROV}{63}{65}, \LHrng{PROV}{90}{104}.}

The counting theorem applies to an accepted operational token only after the token is accessible, the declared verifier accepts it, the accepted content licenses a universal evidence-retaining proof package rather than bare tokenhood or answer receipt, and that package supplies the retained-record family $R(n)$.  Succinct proof systems and distributed verification can reduce one verifier's local transcript by moving retained records into prover state, commitments, challenge rounds, recomputation, or logs; the aggregate work-conservation and finite-community theorems bound the full protocol transcript.\leanmeta{\LHrng{CI}{43}{59}.}

\leanmetapending{\LH{PH26}, \LHrng{CI}{17}{26}, \LHrng{PROV}{2}{3}, \LH{PROV5}, \LHrng{PROV}{7}{40}.}
\begin{corollary}[Thermodynamic Obstruction for Operational Proof-Acts]\label{cor:operational-provability}
Let an accepted operational proof-act, in the sense of Section~\ref{sec:proof-versus-provability}, have content that licenses a universal evidence-retaining proof package for a declared task family with unbounded retained-record lower-bound family $R(n)$.  By Theorem~\ref{thm:physical-impossibility-universal-exact}, no fixed-budget substrate satisfying the declared positive-cost retained-record interface can realize that package for all instance sizes.  The unavailable proof-act follows from retained-record thermodynamic cost exceeding every finite budget.  The direction is asymmetric: a proof-act whose content is the absence of such universal proof is consistent with the physical impossibility theorem.  For agents, the asymmetry between establishing unavailability and establishing universal availability is thermodynamic.
\end{corollary}

\begin{proof}
Tokenhood-like properties alone do not entail universal proof availability.\leanmeta{\LH{PROV31}.}  Sound operational proof acceptance adds verifier acceptance, soundness, and the semantic clause that accepted operational content licenses proof availability.  Once those clauses license a universal proof package with retained-record accounting, Theorem~\ref{thm:physical-impossibility-universal-exact} rules out that package under the declared finite-budget, positive-cost, unbounded-record interface.  The accepted-proof-act no-go follows by modus tollens.  The displayed handles instantiate the schema for the SAT-shaped operational capability case.\leanmeta{\LH{PH26}, \LHrng{CI}{17}{26}, \LHrng{PROV}{26}{35}.}
\end{proof}

\begin{remark}[Operational $\Pclass=\NP$ Instance]\label{rem:operational-pnp-instance}
Universal mathematical claims split into two proof tasks.  A finite derivation, when parsed and checked, has the retained-record cost of that proof audit.  A physical demonstration of the claim across all instances has the retained-record family licensed by the instance-level evidence interface, which can be unbounded.  This separation applies to any universal claim.  It becomes operationally sharp for $\Pclass=\NP$ because the point of the claim is the existence of an efficient SAT solver that works for all instances.  For a satisfiable SAT instance under the standard assignment-certificate interface, solving requires proving the answer.  The proof is the satisfying assignment together with the retained records needed to verify it: $n$ assignment bits give $R_{\mathrm{cert}}(n)\ge n$ when the assignment has $n$ variables.  A polynomial-time algorithm that finds an assignment without retaining the declared verification records is a guess, not a proof.  The instance is solved for the verifier only when the certificate bits and formula-evaluation records are available.  A claim that a smaller proof interface exists must declare that interface and its soundness conditions; the counting theorem applies to the retained-record family induced by that declared interface.  Proving that a shortcut works makes a universal claim over all instances, and that universal claim has an unbounded retained-record family under any declared exact proof interface.  Standard prize-problem formulations ask for proof, not bare truth: truth is free; proof is not.  Even if a polynomial-time SAT algorithm exists, no fixed-budget agent can certify its universal correctness by retained exact evidence across all instance sizes.  The obstruction is not $\Pclass\ne\NP$; it is the thermodynamic unavailability of universal exact certification for the solver capability.\leanmeta{\LH{PH26}, \LHrng{SATQ}{12}{14}, \LHrng{PROV}{26}{35}, \LHrng{PROV}{101}{104}.}
\end{remark}

\begin{remark}[Operational Scope of Short Certificates]\label{rem:scalable-certification-content}
Long-standing complexity-theoretic evidence concerning $\Pclass$ versus $\NP$ remains mathematical background~\cite{baker1975relativizations,razborov1997natural,aaronson2009algebrization,aaronson2005np}.  Corollary~\ref{cor:operational-provability} has five closure points: bare tokenhood does not yield proof extraction; certificate-verifiable answers have their own retained-record family and do not supply universal decision proof; compressed audit protocols conserve retained records; finite communities have additive retained-record capacity; and bare algorithm specifications have no runtime until a substrate-evaluation rule realizes them.\leanmeta{\LHrng{PROV}{36}{57}, \LHrng{PROV}{90}{100}, \LHrng{CI}{43}{59}.}
\end{remark}

\section{Related Work}\label{sec:related}

Physical counting for exact proof sits at the intersection of physical limits of computation, thermodynamic accounting, query lower bounds, and proof-carrying computation.  Aaronson frames physical solution claims for hard problems as claims about physics as well as algorithms~\cite{aaronson2005np,aaronson2008limits}; the Physical Counting Impossibility Theorem adds a proof-act layer and derives a contract-based obstruction for retained evidence.  Lloyd gives a complementary finite-resource perspective through ultimate operation-rate and information-capacity limits~\cite{lloyd2000ultimate}; Bekenstein entropy bounds give a stronger finite-information ceiling when their geometric hypotheses apply~\cite{bekenstein1981universal,bousso2002holographic}.  Lloyd-style and Bekenstein-style bounds constrain total physical capacity from above; the retained-record theorem constrains required proof capacity from below, starting from a lower-bound family $R(n)$ for inspectable evidence records, a finite local budget, and a positive per-record cost floor.  Gandy's locality conditions, Deutsch's quantum Church--Turing principle, and Piccinini's mechanistic account all treat physical computation as requiring explicit locality, state-preparation, implementation, and interpretation assumptions~\cite{gandy1980church,deutsch1985quantum,piccinini2015physical}.  The measurement-to-query-interface map and the specification-to-process typing supply the corresponding discipline for capability claims under resource constraints.

Aaronson's survey-style arguments against hypercomputation are abductive: they examine candidate mechanisms and find each wanting, so the conclusion rests on the completeness of the survey.  His formal barrier results, including relativization, algebrization, and natural proofs, are deductive in a different sense: they show that certain proof techniques cannot resolve $\Pclass$ versus $\NP$ within specified oracle, algebraic, or proof-technique settings~\cite{baker1975relativizations,razborov1997natural,aaronson2009algebrization}.  The retained-record obstruction is deductive in the present sense: it identifies premises and derives impossibility from them, with the arithmetic and interface bookkeeping represented in Lean.  The logical force is therefore different: survey arguments conclude that no known mechanism works, barrier results conclude that certain techniques cannot succeed, and the counting theorem concludes that no mechanism can work under the stated conditions without changing a premise.

Oracle-machine language enters in the same disciplined way.  Turing's o-machines and later broad-computation discussions mark an answer source outside the base machine's ordinary derivation rules~\cite{turing1939systems,copeland1997broad}.  The physical-oracle section asks what retained records, budgets, and trust audits a proposed physical oracle supplies when its answer is used as proof for an agent.

Proof-centered traditions provide adjacent vocabulary.  Constructivist mathematics treats proof as central to mathematical assertion~\cite{bishop1967foundations}; Wittgenstein's rule-following discussion emphasizes practices of use and checking~\cite{wittgenstein1953philosophical}; Dummett's verificationist semantics connects meaning with conditions of verification~\cite{dummett1978truth}.  These traditions differ over meaning, rule-following, and warranted assertion, but they share the operational fact used here: proof reaches a participant through checkable activity.  The result used here is narrower: once that activity is physically realized, the question is what thermodynamic accounting it entails.

Szilard, Landauer, Bennett, and Zurek supply the thermodynamic-information lineage behind the retained-record accounting~\cite{szilard1929entropieverminderung,landauer1961irreversibility,bennett1982thermodynamics,bennett2003notes,bennett1988logical,zurek1989thermodynamic}.  Szilard's engine makes measurement thermodynamic; Landauer supplies the erasure floor; Bennett separates reversible computation from record disposal and connects physical complexity with preserved computational history; Zurek connects algorithmic information, measurement, and entropy.  Experiments and stochastic-thermodynamic refinements support using Landauer as an ideal floor; device-level equality requires stronger process-specific assumptions.  Measurement, feedback, mismatch, residual dissipation, finite-time operation, and absolute irreversibility can add costs above the ideal bound~\cite{berut2012experimental,jun2014high,sagawa2012nonequilibrium,parrondo2015thermodynamics,wolpert2024stochastic,manzano2024absolute,yadav2026minimal}.  Much of the thermodynamics-of-computation literature asks how much work is required to run or erase a computational process.  The proof-act application asks a narrower question: how much irreversible record formation is required to make acceptance checkable by a local observer.  Quantum query complexity supplies the analogous lower-bound language for coherent query interfaces~\cite{beals2001quantum}.

Algorithmic information theory and resource-bounded Kolmogorov complexity measure shortest descriptions, shortest proofs, or shortest programs relative to a universal machine or resource bound~\cite{li2008introduction,buhrman2001resource}.  A shorter proof lowers the retained audit cost of checking that proof, but it does not determine the retained-record family for the licensed task: $R(n)$ is set by the declared evidence interface and the alternatives that the verifier's retained state must separate.

Relativistic and global-infinity issues enter through access conditions.  Interface discreteness can coexist with Lorentz-compatible structure; Snyder gives a classic example, while Einstein and Minkowski supply the relativistic structure~\cite{snyder1947quantized,einstein1905electrodynamics,minkowski1908space}.  Malament-Hogarth spacetimes and closed-timelike-curve computation are bridge-present hypercomputational models~\cite{hogarth1994non,earman1995bangs,etesi2002nonturing,nemeti2006relativistic,nemeti2006general,andreka2009general,deutsch1991quantum,bacon2004quantum,aaronson2009closed}; bare spatial infinity, eternal inflation, or infinite-dimensional state spaces supply local computational access when a local-budget bridge is present.  Holographic and black-hole entropy bounds provide additional finite-budget structure when their hypotheses apply~\cite{bekenstein1981universal,thooft1993dimensional,susskind1995world,bousso2002holographic}.

Certifying algorithms and proof-carrying code separate producing an answer from providing checkable evidence for it~\cite{mcconnell2010certifying,necula1997proof}.  The same distinction has physical content: an exact report is meaningful as a proof only when the system supplies the recorded evidence needed for local verification under the resource conditions.

Lean 4 and mathlib provide the proof-assistant environment behind the handle ledger~\cite{Lean2015,demoura2021lean4,mathlib2020}, alongside recent physics-facing Lean developments such as HepLean~\cite{toobysmith2024heplean} and the Lean formalization of the Generalized Quantum Stein's Lemma~\cite{meiburg2025quantumstein}.  The handles track proof bookkeeping, finite-budget arithmetic, monotone/unbounded threshold lemmas, and the decomposed proof-act/verifier/semantics bridge; empirical physical inputs remain cited premises.

\section{Conclusion}\label{sec:conclusion}

Proof is checkable information: produced, retained, inspectable grounds that exclude alternatives.  Court, audit, mathematical verification, and experimental verification instantiate the same retained-record structure.  The Physical Counting Impossibility Theorem identifies the thermodynamic floor beneath that structure.  Finite local budget, positive per-record work cost, an unbounded retained-record lower-bound family $R(n)$, and universal exact proof cannot all hold together.  Exact verification, operational capability claims, and proof-acts inherit the constraint once a declared interface supplies observables, records, observers, and cost accounting.

The physical object being counted is evidence indistinguishability removal.  A retained verifier state must distinguish the in-scope alternatives that require different judgments; otherwise the evidence remains indistinguishable with respect to the claim.  Which alternatives require different judgments, and therefore which distinctions must be retained, is fixed by the epistemic content of the claim.  The theorem says that for claims whose proof requires more retained distinctions than any fixed finite positive-cost substrate can retain, exact proof is physically unavailable to agents.

The examples show the range of the same mechanism.  Separable parity and spin-alignment tasks expose the local-coordinate audit cost.  Entanglement verification separates a compact witness record from the stronger claim of full-state identification.  Cosmological topology separates a finite survey catalog from exact proof of a global spatial fact.  Each case turns on the same question: which alternatives does the retained verifier state actually distinguish?

The proof-act definition keeps the boundary local.  Answer-only computation, oracle testimony, and trusted tokens can be valuable under their own interfaces; they become proof for the situated agent only when the declared interface makes the relevant grounds available as retained, checkable evidence.  Formal substrate-free truth values remain separate specifications.

The conceptual boundary is local: global infinitary structure gives a proof-producing computation infinitely many accessible operations when it supplies an explicit local-budget bridge.  Malament-Hogarth and closed-timelike-curve interfaces are bridge-present cases because their formal interfaces supply that bridge.

For a situated agent, finitude is a proof horizon.  The operational question turns on whether the retained-record requirement exceeds the agent's finite lifetime certification capacity; metaphysical infinity is secondary.  A larger agent or longer-lived community may move the horizon, as a stronger formal system may prove claims unavailable to a weaker one.  If the required retained records exceed the situated agent's joint thermodynamic and spacetime capacity, no local audit trail can both cover the requirement and fit the agent's budget window.\leanmeta{\LH{CI60}.}

The consequence for fundamental physics is a local verification bound internal to physical theory.  A theory may assign exact values to global states, vacuum properties, or boundary-condition facts while every finite subsystem lacks the causal access, retained-record capacity, or work budget needed to make those facts available as proof.  Exact facts and exact proof for a situated subsystem separate whenever the theory supplies truth without a local evidence bridge.

The accounting is uniform across mathematical and empirical proof, short and long derivations, human and machine verifiers.  It assigns each retained proof-act its cost under the declared interface.  The impossibility theorem identifies where that cost exceeds any fixed budget; the Landauer floor identifies where it begins.  The accounting reaches as far as physical information: wherever an agent retains a checkable record, the accounting applies.

The constructive direction changes the evidence interface while preserving the floor.  Certified global observables, compressed audits, interactive checks, quantum witnesses, and distributed transcripts can lower the retained-record family $R(n)$ when they supply their own sound retained evidence.  The theorem then gives the design target: make the durable separator smaller, or state exactly which physical premise supplies the additional budget.  For SAT yes-instances under the standard assignment-certificate interface, solving requires proving the answer.  The solution for a verifier requires retained certificate records, and retained records cost energy per irreversible distinction.  A faster procedure without retained records is a guess, not a proof.  A claim that a smaller proof interface exists is itself a proposition requiring proof under a declared interface.  Proving that a shortcut is sound makes a universal claim with its own retained-record family.  For $\Pclass=\NP$-style shortcut claims, the universal content ``this shortcut works for all instances'' has unbounded retained-record demand: exact certification across all instances is thermodynamically unavailable whether or not the mathematical statement is true.  The same accounting applies to objections: disputing the $n$-bit floor is a proof-act with positive thermodynamic cost.  The floor was always there.  It was never zero.

\appendix
\section{Lean Axiom Scan}\label{app:lean-axiom-scan}

\IfFileExists{content/axiom_scope_auto.tex}{% Auto-generated by scripts/build_papers.py --axiom-check. Do not edit manually.
% Generated: 2026-06-05T01:30:17.916150
% Paper: paper4_toc
\begin{table}[htbp]
\centering
\small
\setlength{\tabcolsep}{4pt}
\renewcommand{\arraystretch}{1.12}
\begin{tabularx}{\linewidth}{@{}X r X@{}}
\toprule
\textbf{Axiom-scan category} & \textbf{Count} & \textbf{Scope reading} \\
\midrule
Discovered theorem/lemma declarations & 121 & Candidate declarations found in the selected Lean proof tree. \\
\midrule
Successfully checked by Lean & 121 & Declarations for which \texttt{\#print axioms} returned a parsed result. \\
\midrule
No Lean kernel axioms reported & 3 & Fully constructive relative to imported definitions and theorem hypotheses. \\
\midrule
Reported Lean axiom \texttt{Classical.choice} & 114 & Distinct axiom name reported by \texttt{\#print axioms}; the count is declarations depending on this axiom. \\
\midrule
Reported Lean axiom \texttt{Quot.sound} & 114 & Distinct axiom name reported by \texttt{\#print axioms}; the count is declarations depending on this axiom. \\
\midrule
Reported Lean axiom \texttt{propext} & 118 & Distinct axiom name reported by \texttt{\#print axioms}; the count is declarations depending on this axiom. \\
\midrule
Exact axiom signature & 4 & \{\texttt{propext}\}; declarations whose reported axiom set is exactly this set. \\
\midrule
Exact axiom signature & 114 & \{\texttt{Classical.choice}, \texttt{Quot.sound}, \texttt{propext}\}; declarations whose reported axiom set is exactly this set. \\
\midrule
Private, timeout, or unresolved declarations & 0 & Not counted as hidden physical assumptions; inspect the raw scan if nonzero. \\
\bottomrule
\end{tabularx}
\caption{Lean axiom-scope summary generated from the build-system \texttt{\#print axioms} scan. Physical premises such as SR, Landauer cost, detector calibration, and retained-record contracts enter as explicit hypotheses or cited empirical inputs, not as hidden Lean kernel axioms.}
\label{tab:lean-axiom-scope-auto}
\end{table}
}{}

\section{Formal Interface Definitions}\label{app:formal-interface}

\begin{definition}[Exact Proof Task]\label{def:decision-problem}
An exact proof task at size $n$ is a finite function
\[
f_n:X_n\to Y_n,
\]
where $X_n$ is the finite set of admissible inputs exposed by the declared interface and $Y_n$ is the finite set of exact answers or exact report statuses.
\end{definition}

Physically, $X_n$ is the finite menu of configurations that the apparatus exposes at its declared resolution, and $Y_n$ is the finite menu of reports the verifier is allowed to accept.  The definition fixes the proof target before any budget arithmetic is applied.

\begin{definition}[Query Interface]\label{def:query-interface}
A query interface for $X_n$ declares a finite index set $I_n$ and, for each input $x\in X_n$, locally accessible entries $x_i$ for $i\in I_n$.  A query reads one entry $x_i$ and returns its declared finite value.
\end{definition}

Physically, a query is one calibrated detector channel, switch setting, spin readout, memory-cell read, or coarse-grained measurement outcome.  The finite index set records which local distinctions the proof-producing substrate can actually access.

\begin{definition}[Exact Adaptive Measurement Protocol]\label{def:exact-query-procedure}
An exact adaptive measurement protocol for $f_n$ is a sequential measurement procedure.  At each step it chooses a query index in $I_n$, possibly as a deterministic function of previous outcomes, and reads the corresponding entry.  It then outputs a value in $Y_n$.  It is correct when every admissible input $x$ leads to the value $f_n(x)$.

In decision-tree terminology, such a deterministic adaptive protocol has internal nodes given by measurements, edges given by measurement outcomes, and leaves given by accepted outputs.
\end{definition}

Physically, this is a laboratory control rule: each next measurement may depend on the retained outcomes already produced, and exactness means the final report is right for every admissible preparation in the declared task.

\begin{definition}[Exact Query Complexity]\label{def:exact-query-complexity}
The exact deterministic query complexity $D(f_n)$ is the minimum worst-case number of elementary measurements used by any exact adaptive measurement protocol for $f_n$:
\[
D(f_n)=\min_T\max_{x\in X_n}\#\{\text{queries made by }T\text{ on }x\},
\]
where $T$ ranges over exact adaptive measurement protocols for $f_n$.
\end{definition}

Physically, $D(f_n)$ is the smallest worst-case number of local detector events that any exact proof-producing process must expose before the report is accepted as proof.  This quantity supplies the record-demand input to the thermodynamic budget bound.

\section{Physical Scope and Reporting Definitions}\label{app:physical-scope-reporting}\label{sec:physical-transport}\label{sec:interpretive-foundations}

\begin{definition}[Measurement-to-Query-Interface Map]\label{def:physical-core-encoding}
For a physical instance class $\mathcal{P}$ and a core task class $\mathcal{F}$, a measurement-to-query-interface map is a declared map
\[
E:\mathcal{P}\to\mathcal{F}.
\]
It records how physical observables are assigned to the finite input entries and answers used by the exact proof task.
\end{definition}

Physically, $E$ is the apparatus dictionary: it says which observable, detector bin, pointer state, or register state realizes each abstract input entry and report value.

\begin{definition}[Declared Physical Domain of Validity]\label{def:physical-scope-gate}
A physical exact-proof claim is in scope only when it declares the physical state class, observable entries, the map $E$, measurement units for reported quantities, the objective/intervention class, and the boundary conditions relating physical assumptions to task assumptions.
\end{definition}

Physically, this scope gate prevents the theorem from being applied to an undeclared device.  A detector protocol must state what it measures, what counts as a record, what boundary conditions hold, and which physical reports are being supported as proof.

\begin{definition}[Proof Interface Integrity]\label{def:solver-integrity}
A proof interface has integrity for a declared relation when every asserted non-abstaining output is accepted by its checker and every checker-accepted output is true for that relation.  Abstention, or a report that the answer is unknown, makes no assertion about truth.
\end{definition}

Physically, integrity is report discipline: a device may abstain, while an asserted exact report requires the declared checkable evidence.

\begin{definition}[Competence Under a Regime]\label{def:competence-regime}
A proof interface is competent on a declared regime when it has integrity, covers the in-scope instance family without abstaining, and satisfies the declared resource bound.
\end{definition}

Physically, competence means the apparatus can issue exact reports with proof records throughout the declared family while staying inside its declared time, work, access, and retention limits.

\section{Recorded Binary Distinction: Proof Detail}\label{app:recorded-binary-proof}

A recorded interface transition is an exposed before/after event in the finite transition system described in Section~\ref{sec:discrete-transitions}.  The event supplies two endpoints to the record medium: the pre-transition state $A$ and the post-transition state $B$.  The map $0\mapsto A$ and $1\mapsto B$ gives the Boolean-indexed before/after pair used in the proof audit, so the recorded event carries at least the binary distinction needed for Landauer/Bennett-style energetic accounting~\cite{landauer1961irreversibility,bennett1982thermodynamics,sagawa2012nonequilibrium,parrondo2015thermodynamics}.

The count concerns the recorded before/after distinction used as evidence.  The observer-side description length needed to identify which transition occurred inside a larger $N$-state register is separate.  An external observer may require $\log_2 N$ bits to name the transition class.  If the same proof interface also stores the global labels of those states, those labels are additional retained records; the one-bit floor for the transition distinction remains.

For a joint-record interface the counted unit changes and the contract moves to the joint evidence game.  The interface must specify the joint record's distinguishable report states, its retention cost floor, and the proof lower-bound family.  A single topological, witness, or global-observable record can be enough for an answer whose declared target is only that global value.  Proof of the underlying coordinate-wise configuration additionally requires retained syndrome or local-access evidence that excludes all answer-changing alternatives in the original coordinate interface.

\section{Stochastic-Thermodynamic Refinements}\label{app:stochastic-thermo-refinements}

The universal Landauer floor is sufficient for Theorem~\ref{thm:thermo-derived}.  Standard stochastic-thermodynamic refinements strengthen the floor when additional constrained-process premises are declared: the constrained-process interface permits overhead above Landauer, the Wolpert decomposition splits that overhead into mismatch and residual terms, finite KL mismatch is nonnegative and strictly positive under an explicit finite-distribution mismatch witness, and the finite residual branch gives either a positive pairwise KL witness or an irreversible one-way transition.\leanmeta{\LHrng{WC}{1}{5}, \LHrng{WP}{1}{2}, \LHrng{WM}{1}{6}, \LHrng{WR}{1}{10}.}  Concretely, replacing the ideal floor by an effective per-record cost
\[
  \varepsilon_{\mathrm{eff}} = k_B T\ln2 + m + r
\]
with mismatch term $m\ge0$ and residual term $r\ge0$ changes only the numerical threshold: $N$ retained records cost at least $N\varepsilon_{\mathrm{eff}}$, and any strict positive overhead lowers the maximum feasible record count below the ideal Landauer estimate.  The broader stopping-time / absolute-irreversibility regime remains a cited premise.\leanmeta{\LH{WP5}.}  Any one positive branch is enough to force strict separation above the Landauer floor; query resources lower-bounded by mismatch further tighten the estimates, and the physical grounding bundle survives under this refined decomposition.\leanmeta{\LHrng{WP}{6}{9}.}  Exact equality is retained only as an optional idealized specialization.

\section{Lower-Bound Families and Interface Variants}\label{app:lower-bound-variants}

\begin{definition}[Exact Proof Lower-Bound Family]\label{def:exact-certification-lower-bound-family}
An exact proof lower-bound family is a function $R:\mathbb{N}\to\mathbb{N}$ such that every exact proof for the declared task family at size $n$ must realize at least $R(n)$ elementary accessible operations or distinctions.  The operations or distinctions that ground verifier acceptance must be represented by retained records available to the verifier.
\end{definition}

Physically, $R(n)$ is the required audit-trail size for the declared interface.  It counts the local distinctions that must become available as retained evidence.  Internal reversible steps erased before observer inspection belong to answer production; retained proof evidence is the stored audit record.

For a quantum query interface, $D(f_n)$ is replaced by the exact quantum query lower bound $Q_E(f_n)$, or more generally by a retained-record family $R_{\mathrm{quantum}}(n)$ for the declared quantum proof protocol.  Parity remains linear in the exact quantum query model: $Q_E(\mathrm{PARITY}_n)=\Theta(n)$~\cite{beals2001quantum}.  Section~\ref{sec:quantum-verification-interfaces} gives the main-text treatment of quantum interactive proofs, $\mathsf{QMA}$ witnesses, Mahadev-style classical verification, bounded-error reporting, and interpretation-neutral retained records.  Formally, quantum, weak, continuous, and generalized-probabilistic interfaces change the evidence object and therefore the lower-bound family.  Once finite-resolution observables, admissible measurements, exact-report records, and retained-record rules are declared, the same finite-budget threshold applies to the resulting $R_{\mathrm{quantum}}(n)$ or protocol-specific $R(n)$.

The anchored sign-query subfamily is a hidden-switch apparatus with a mechanized indistinguishable-pair obstruction: if switch $j$ is unqueried, the all-agreeing and $j$-flipped physical configurations have identical queried transcript but require different exact reports.  Unit-clause SAT is the compact Boolean encoding of this already declared physical interface; in that encoding the paired configurations have opposite satisfiability.  The all-signs lower bound and unboundedness of this sign-query lower-bound family are indexed by the corresponding handles.\leanmeta{\LH{SATQ6}, \LH{SATQ8}, \LH{SATQ11}.}

\section{Compressed and Distributed Audit Interfaces}\label{app:compressed-distributed-audit}

The compressed-audit interface is a multi-agent, multi-phase retained-record protocol.  Each phase is assigned to an agent and contributes a finite retained-record count; total retained records are the sum over phases.  Sound retained-audit acceptance for an $n$-bit instance is represented by an injection from the $n$ underlying instance-bit distinctions into the aggregate retained-record slots.  Finite cardinality gives the conservation law: total retained records are at least $n$, and multiplying by the positive per-record floor gives the retained-work lower bound.  Merkle-style audit trails (hash-tree summaries), polynomial-commitment audit trails (short commitments to large algebraic data), and interactive-argument audit trails (challenge-response evidence accepted under soundness assumptions) are specializations of this interface; their cryptographic details affect where records are held, not the aggregate conservation statement.\leanmeta{\LHrng{CI}{43}{54}.}

The distributed-community interface is a finite family of agents with per-agent budgets, positive per-record costs, and a finite communication graph.  Each agent's retained records are bounded by its budget divided by its retained-record cost; summing these bounds gives the aggregate community capacity.  Records received over a communication edge are included in the receiver's retained records, so communication redistributes evidence under receiver-side accounting.  Consequently, every unbounded retained-record lower-bound family eventually exceeds the aggregate capacity of any finite community at any finite time.\leanmeta{\LHrng{CI}{55}{59}.}

\section{Assumption-Boundary Countermodels}\label{app:assumption-countermodels}

The finite-budget, positive-cost, and unbounded-lower-bound premises are independently necessary for the threshold contradiction.

\begin{itemize}
\item \textbf{Unbounded budget.}  A size-dependent budget can satisfy $B_n\geq \varepsilon R(n)$ for every finite retained-record count $R(n)$.
\item \textbf{Zero per-record floor.}  If $\varepsilon=0$, then $\varepsilon R(n)=0$ for every retained-record count.  Record growth carries no positive work lower bound.
\item \textbf{No super-budget lower-bound family.}  A retained-record lower-bound family bounded below the converted budget stays feasible under the threshold test.
\end{itemize}

These countermodels remove one premise at a time and leave the remaining arithmetic intact.\leanmeta{\LHrng{PH}{31}{33}.}

\section{Proof-Act Formalization}\label{app:proof-act-formalization}

A proof-act is a proof made available to an agent through a token whose acceptance is backed by retained, locally inspectable grounds.  The scoped tokenhood-only countermodel has no universal proof package and no extraction function from bare tokenhood to an impossible proof family.  A finite answer-only countermodel separates trust-only acceptance from proof inspection: the trust-only system accepts both a correct and an incorrect claim, while the show-your-work system verifies claims through retained evidence whose inspection is sound.  Operational $\Pclass=\NP$ acceptance adds verifier acceptance, soundness, and the semantic clause that accepted operational content licenses universal evidence-retaining proof.\leanmeta{\LH{PROV31}, \LH{PROV36}, \LHrng{PROV}{58}{62}.}

The operational proof-act bridge represents physically available proof as a record-bearing object: proof content plus token, physical instantiation, agent accessibility, checkability, and retained audit records sufficient to substantiate acceptance.  The decoder layer makes accepted proof a physical process, and the parser specialization models the token as a finite string over a declared finite alphabet processed by a finite-state transducer with positive retained-record cost per consumed-symbol record.  Proof licensing is decomposed into operational $\Pclass=\NP$ acceptance with explicit decoder acceptance, general verifier soundness, and operational-claim semantics.  Accepted operational $\Pclass=\NP$ proof-acts produce a universal proof package with retained-record accounting; if that audit requirement is unbounded, the same finite-budget threshold applies.\leanmeta{\LHrng{CI}{17}{26}, \LHrng{PROV}{6}{35}.}

The typed gate states the operational domain: operational provability concerns proof objects that can be produced, stored, transmitted, checked, or used to convince an agent.  A token with no retained grounds for acceptance counts as an oracle answer, guess, or trusted black box for that verifier.

\section*{Statements and Declarations}

\paragraph*{Data and Code Availability.}
No empirical datasets were generated or analyzed for this manuscript. The Lean 4 formalization and supplementary artifact supporting the theorem-level claims are archived at \url{https://doi.org/10.5281/zenodo.19457896}; the supplementary ledger maps manuscript claims to their formal identifiers.

\paragraph*{Competing Interests.}
The author declares no competing interests.

\paragraph*{Generative AI Disclosure.}
Generative AI tools, including Codex, Claude Code, Augment, Kilo, and OpenCode, were used during manuscript preparation for prose refinement, notation consistency, \LaTeX{} formatting, structure cleanup, translation of the author's informal proof sketches into candidate Lean and \LaTeX{} artifacts, and adversarial reviewer-style critique prompts used to identify clarity gaps.

The author selected the problem, formulated the theorem statements and assumptions, determined the novelty framing, made all editorial decisions, and set the acceptance criteria for formal and prose claims. No theorem statement, assumption, citation, or technical claim was adopted from AI output without independent author derivation and validation. Formal claims reported as machine-verified were admitted only after Lean verification and direct author review; Lean was used as an integrity gate for responsible AI-assisted research. The author is solely responsible for all statements, citations, and conclusions.

\begingroup\small
\bibliographystyle{plain}
\bibliography{references}
\endgroup

\end{document}

\title{Supplementary Material: \PaperTitleAuto}
\author{Tristan Simas}
\date{\today}
\maketitle

\section{Full Lean Handle Ledger}

This supplement provides the Lean handle ledger from the supporting formal development.
The subset cited in the manuscript appears in the claim-to-handle mapping below.
The ledger includes handle identifiers, declaration names, and source module paths.

\IfFileExists{content/lean_handle_ids_auto.tex}{%
  % Auto-generated by scripts/build_papers.py. Do not edit manually.
% Generated: 2026-06-05T15:34:28.449874
\begingroup
\scriptsize
\setlength{\tabcolsep}{4pt}
\renewcommand{\arraystretch}{1.12}
\setlength{\LTpre}{2pt}
\setlength{\LTpost}{2pt}
\setlength{\emergencystretch}{3em}
\sloppy
\urlstyle{tt}
\makeatletter
\if@twocolumn
\begin{list}{}{\leftmargin=0pt\itemindent=0pt\itemsep=4pt\parsep=0pt\topsep=4pt}
\item \textbf{\nolinkurl{AB1}}\hypertarget{lh:AB1}{}\enspace{\ttfamily\nolinkurl{DecisionProblem.not_preservesOpt_iff_erasesDecisionRelevantDistinction}} {\tiny\ttfamily DecisionQuotient/\allowbreak AbstractionCollapse.lean}
\item \textbf{\nolinkurl{AB2}}\hypertarget{lh:AB2}{}\enspace{\ttfamily\nolinkurl{DecisionProblem.surjective_abstraction_factors_or_erases}} {\tiny\ttfamily DecisionQuotient/\allowbreak AbstractionCollapse.lean}
\item \textbf{\nolinkurl{AB3}}\hypertarget{lh:AB3}{}\enspace{\ttfamily\nolinkurl{DecisionProblem.collapseBeyondQuotient_physically_impossible}} {\tiny\ttfamily DecisionQuotient/\allowbreak AbstractionCollapse.lean}
\item \textbf{\nolinkurl{AB4}}\hypertarget{lh:AB4}{}\enspace{\ttfamily\nolinkurl{DecisionProblem.surjective_abstraction_with_feasible_collapse_map_factors}} {\tiny\ttfamily DecisionQuotient/\allowbreak AbstractionCollapse.lean}
\item \textbf{\nolinkurl{AN3}}\hypertarget{lh:AN3}{}\enspace{\ttfamily\nolinkurl{Physics.AssumptionNecessity.physical_claim_requires_physical_assumption}} {\tiny\ttfamily DecisionQuotient/\allowbreak Physics/\allowbreak AssumptionNecessity.lean}
\item \textbf{\nolinkurl{AN4}}\hypertarget{lh:AN4}{}\enspace{\ttfamily\nolinkurl{Physics.AssumptionNecessity.physical_claim_requires_empirically_justified_physical_assumption}} {\tiny\ttfamily DecisionQuotient/\allowbreak Physics/\allowbreak AssumptionNecessity.lean}
\item \textbf{\nolinkurl{AQ1}}\hypertarget{lh:AQ1}{}\enspace{\ttfamily\nolinkurl{ClaimClosure.AQ1}} {\tiny\ttfamily DecisionQuotient/\allowbreak ClaimClosure.lean}
\item \textbf{\nolinkurl{AQ2}}\hypertarget{lh:AQ2}{}\enspace{\ttfamily\nolinkurl{ClaimClosure.AQ2}} {\tiny\ttfamily DecisionQuotient/\allowbreak ClaimClosure.lean}
\item \textbf{\nolinkurl{AQ3}}\hypertarget{lh:AQ3}{}\enspace{\ttfamily\nolinkurl{ClaimClosure.AQ3}} {\tiny\ttfamily DecisionQuotient/\allowbreak ClaimClosure.lean}
\item \textbf{\nolinkurl{AQ4}}\hypertarget{lh:AQ4}{}\enspace{\ttfamily\nolinkurl{ClaimClosure.AQ4}} {\tiny\ttfamily DecisionQuotient/\allowbreak ClaimClosure.lean}
\item \textbf{\nolinkurl{AQ5}}\hypertarget{lh:AQ5}{}\enspace{\ttfamily\nolinkurl{ClaimClosure.AQ5}} {\tiny\ttfamily DecisionQuotient/\allowbreak ClaimClosure.lean}
\item \textbf{\nolinkurl{AQ6}}\hypertarget{lh:AQ6}{}\enspace{\ttfamily\nolinkurl{ClaimClosure.AQ6}} {\tiny\ttfamily DecisionQuotient/\allowbreak ClaimClosure.lean}
\item \textbf{\nolinkurl{AQ7}}\hypertarget{lh:AQ7}{}\enspace{\ttfamily\nolinkurl{ClaimClosure.AQ7}} {\tiny\ttfamily DecisionQuotient/\allowbreak ClaimClosure.lean}
\item \textbf{\nolinkurl{AQ8}}\hypertarget{lh:AQ8}{}\enspace{\ttfamily\nolinkurl{ClaimClosure.AQ8}} {\tiny\ttfamily DecisionQuotient/\allowbreak ClaimClosure.lean}
\item \textbf{\nolinkurl{BA1}}\hypertarget{lh:BA1}{}\enspace{\ttfamily\nolinkurl{Physics.BoundedAcquisition.BoundedRegion}} {\tiny\ttfamily DecisionQuotient/\allowbreak Physics/\allowbreak BoundedAcquisition.lean}
\item \textbf{\nolinkurl{BA2}}\hypertarget{lh:BA2}{}\enspace{\ttfamily\nolinkurl{Physics.BoundedAcquisition.acquisition_rate_bound}} {\tiny\ttfamily DecisionQuotient/\allowbreak Physics/\allowbreak BoundedAcquisition.lean}
\item \textbf{\nolinkurl{BA9}}\hypertarget{lh:BA9}{}\enspace{\ttfamily\nolinkurl{Physics.BoundedAcquisition.physical_grounding_bundle}} {\tiny\ttfamily DecisionQuotient/\allowbreak Physics/\allowbreak BoundedAcquisition.lean}
\item \textbf{\nolinkurl{BA10}}\hypertarget{lh:BA10}{}\enspace{\ttfamily\nolinkurl{Physics.BoundedAcquisition.counting_gap_theorem}} {\tiny\ttfamily DecisionQuotient/\allowbreak Physics/\allowbreak BoundedAcquisition.lean}
\item \textbf{\nolinkurl{BC1}}\hypertarget{lh:BC1}{}\enspace{\ttfamily\nolinkurl{Foundations.counting_nonneg}} {\tiny\ttfamily DecisionQuotient/\allowbreak BayesFoundations.lean}
\item \textbf{\nolinkurl{BC2}}\hypertarget{lh:BC2}{}\enspace{\ttfamily\nolinkurl{Foundations.counting_total}} {\tiny\ttfamily DecisionQuotient/\allowbreak BayesFoundations.lean}
\item \textbf{\nolinkurl{BC3}}\hypertarget{lh:BC3}{}\enspace{\ttfamily\nolinkurl{Foundations.counting_additive}} {\tiny\ttfamily DecisionQuotient/\allowbreak BayesFoundations.lean}
\item \textbf{\nolinkurl{BC4}}\hypertarget{lh:BC4}{}\enspace{\ttfamily\nolinkurl{Foundations.bayes_from_conditional}} {\tiny\ttfamily DecisionQuotient/\allowbreak BayesFoundations.lean}
\item \textbf{\nolinkurl{BC5}}\hypertarget{lh:BC5}{}\enspace{\ttfamily\nolinkurl{Foundations.entropy_contraction}} {\tiny\ttfamily DecisionQuotient/\allowbreak BayesFoundations.lean}
\item \textbf{\nolinkurl{CC1}}\hypertarget{lh:CC1}{}\enspace{\ttfamily\nolinkurl{DecisionQuotient.ClaimClosure.RegimeSimulation}} {\tiny\ttfamily DecisionQuotient/\allowbreak ClaimClosure.lean}
\item \textbf{\nolinkurl{CC3}}\hypertarget{lh:CC3}{}\enspace{\ttfamily\nolinkurl{DecisionQuotient.ClaimClosure.system_transfer_licensed_iff_snapshot}} {\tiny\ttfamily DecisionQuotient/\allowbreak ClaimClosure.lean}
\item \textbf{\nolinkurl{CC8}}\hypertarget{lh:CC8}{}\enspace{\ttfamily\nolinkurl{DecisionQuotient.ClaimClosure.boundaryCharacterized_iff_exists_sufficient_subset}} {\tiny\ttfamily DecisionQuotient/\allowbreak ClaimClosure.lean}
\item \textbf{\nolinkurl{CC9}}\hypertarget{lh:CC9}{}\enspace{\ttfamily\nolinkurl{DecisionQuotient.ClaimClosure.bounded_actions_detectable}} {\tiny\ttfamily DecisionQuotient/\allowbreak ClaimClosure.lean}
\item \textbf{\nolinkurl{CC10}}\hypertarget{lh:CC10}{}\enspace{\ttfamily\nolinkurl{DecisionQuotient.ClaimClosure.bridge_boundary_represented_family}} {\tiny\ttfamily DecisionQuotient/\allowbreak ClaimClosure.lean}
\item \textbf{\nolinkurl{CC11}}\hypertarget{lh:CC11}{}\enspace{\ttfamily\nolinkurl{DecisionQuotient.ClaimClosure.bridge_failure_witness_non_one_step}} {\tiny\ttfamily DecisionQuotient/\allowbreak ClaimClosure.lean}
\item \textbf{\nolinkurl{CC13}}\hypertarget{lh:CC13}{}\enspace{\ttfamily\nolinkurl{DecisionQuotient.ClaimClosure.certified_total_bits_split_core}} {\tiny\ttfamily DecisionQuotient/\allowbreak ClaimClosure.lean}
\item \textbf{\nolinkurl{CC14}}\hypertarget{lh:CC14}{}\enspace{\ttfamily\nolinkurl{DecisionQuotient.ClaimClosure.cost_asymmetry_eth_conditional}} {\tiny\ttfamily DecisionQuotient/\allowbreak ClaimClosure.lean}
\item \textbf{\nolinkurl{CC15}}\hypertarget{lh:CC15}{}\enspace{\ttfamily\nolinkurl{DecisionQuotient.ClaimClosure.declaredBudgetSlice}} {\tiny\ttfamily DecisionQuotient/\allowbreak ClaimClosure.lean}
\item \textbf{\nolinkurl{CC16}}\hypertarget{lh:CC16}{}\enspace{\ttfamily\nolinkurl{DecisionQuotient.ClaimClosure.declaredRegimeFamily_complete}} {\tiny\ttfamily DecisionQuotient/\allowbreak ClaimClosure.lean}
\item \textbf{\nolinkurl{CC17}}\hypertarget{lh:CC17}{}\enspace{\ttfamily\nolinkurl{DecisionQuotient.ClaimClosure.declared_physics_no_universal_exact_certifier_core}} {\tiny\ttfamily DecisionQuotient/\allowbreak ClaimClosure.lean}
\item \textbf{\nolinkurl{CC18}}\hypertarget{lh:CC18}{}\enspace{\ttfamily\nolinkurl{DecisionQuotient.ClaimClosure.dichotomy_conditional}} {\tiny\ttfamily DecisionQuotient/\allowbreak ClaimClosure.lean}
\item \textbf{\nolinkurl{CC19}}\hypertarget{lh:CC19}{}\enspace{\ttfamily\nolinkurl{DecisionQuotient.ClaimClosure.epsilon_admissible_iff_raw_lt_certified_total_core}} {\tiny\ttfamily DecisionQuotient/\allowbreak ClaimClosure.lean}
\item \textbf{\nolinkurl{CC20}}\hypertarget{lh:CC20}{}\enspace{\ttfamily\nolinkurl{DecisionQuotient.ClaimClosure.exact_admissible_iff_raw_lt_certified_total_core}} {\tiny\ttfamily DecisionQuotient/\allowbreak ClaimClosure.lean}
\item \textbf{\nolinkurl{CC21}}\hypertarget{lh:CC21}{}\enspace{\ttfamily\nolinkurl{DecisionQuotient.ClaimClosure.exact_certainty_inflation_under_hardness_core}} {\tiny\ttfamily DecisionQuotient/\allowbreak ClaimClosure.lean}
\item \textbf{\nolinkurl{CC22}}\hypertarget{lh:CC22}{}\enspace{\ttfamily\nolinkurl{DecisionQuotient.ClaimClosure.exact_raw_eq_certified_iff_certainty_inflation_core}} {\tiny\ttfamily DecisionQuotient/\allowbreak ClaimClosure.lean}
\item \textbf{\nolinkurl{CC23}}\hypertarget{lh:CC23}{}\enspace{\ttfamily\nolinkurl{DecisionQuotient.ClaimClosure.exact_raw_only_of_no_exact_admissible_core}} {\tiny\ttfamily DecisionQuotient/\allowbreak ClaimClosure.lean}
\item \textbf{\nolinkurl{CC24}}\hypertarget{lh:CC24}{}\enspace{\ttfamily\nolinkurl{DecisionQuotient.ClaimClosure.explicit_assumptions_required_of_not_excused_core}} {\tiny\ttfamily DecisionQuotient/\allowbreak ClaimClosure.lean}
\item \textbf{\nolinkurl{CC25}}\hypertarget{lh:CC25}{}\enspace{\ttfamily\nolinkurl{DecisionQuotient.ClaimClosure.explicit_state_upper_core}} {\tiny\ttfamily DecisionQuotient/\allowbreak ClaimClosure.lean}
\item \textbf{\nolinkurl{CC26}}\hypertarget{lh:CC26}{}\enspace{\ttfamily\nolinkurl{DecisionQuotient.ClaimClosure.hard_family_all_coords_core}} {\tiny\ttfamily DecisionQuotient/\allowbreak ClaimClosure.lean}
\item \textbf{\nolinkurl{CC27}}\hypertarget{lh:CC27}{}\enspace{\ttfamily\nolinkurl{DecisionQuotient.ClaimClosure.horizonTwoWitness_immediate_empty_sufficient}} {\tiny\ttfamily DecisionQuotient/\allowbreak ClaimClosure.lean}
\item \textbf{\nolinkurl{CC28}}\hypertarget{lh:CC28}{}\enspace{\ttfamily\nolinkurl{DecisionQuotient.ClaimClosure.horizon_gt_one_bridge_can_fail_on_sufficiency}} {\tiny\ttfamily DecisionQuotient/\allowbreak ClaimClosure.lean}
\item \textbf{\nolinkurl{CC29}}\hypertarget{lh:CC29}{}\enspace{\ttfamily\nolinkurl{DecisionQuotient.ClaimClosure.information_barrier_opt_oracle_core}} {\tiny\ttfamily DecisionQuotient/\allowbreak ClaimClosure.lean}
\item \textbf{\nolinkurl{CC30}}\hypertarget{lh:CC30}{}\enspace{\ttfamily\nolinkurl{DecisionQuotient.ClaimClosure.information_barrier_state_batch_core}} {\tiny\ttfamily DecisionQuotient/\allowbreak ClaimClosure.lean}
\item \textbf{\nolinkurl{CC31}}\hypertarget{lh:CC31}{}\enspace{\ttfamily\nolinkurl{DecisionQuotient.ClaimClosure.information_barrier_value_entry_core}} {\tiny\ttfamily DecisionQuotient/\allowbreak ClaimClosure.lean}
\item \textbf{\nolinkurl{CC32}}\hypertarget{lh:CC32}{}\enspace{\ttfamily\nolinkurl{DecisionQuotient.ClaimClosure.integrity_resource_bound_for_sufficiency}} {\tiny\ttfamily DecisionQuotient/\allowbreak ClaimClosure.lean}
\item \textbf{\nolinkurl{CC33}}\hypertarget{lh:CC33}{}\enspace{\ttfamily\nolinkurl{DecisionQuotient.ClaimClosure.integrity_universal_applicability_core}} {\tiny\ttfamily DecisionQuotient/\allowbreak ClaimClosure.lean}
\item \textbf{\nolinkurl{CC36}}\hypertarget{lh:CC36}{}\enspace{\ttfamily\nolinkurl{DecisionQuotient.ClaimClosure.minsuff_collapse_core}} {\tiny\ttfamily DecisionQuotient/\allowbreak ClaimClosure.lean}
\item \textbf{\nolinkurl{CC38}}\hypertarget{lh:CC38}{}\enspace{\ttfamily\nolinkurl{DecisionQuotient.ClaimClosure.minsuff_conp_complete_conditional}} {\tiny\ttfamily DecisionQuotient/\allowbreak ClaimClosure.lean}
\item \textbf{\nolinkurl{CC39}}\hypertarget{lh:CC39}{}\enspace{\ttfamily\nolinkurl{DecisionQuotient.ClaimClosure.no_auto_minimize_of_p_neq_conp}} {\tiny\ttfamily DecisionQuotient/\allowbreak ClaimClosure.lean}
\item \textbf{\nolinkurl{CC41}}\hypertarget{lh:CC41}{}\enspace{\ttfamily\nolinkurl{DecisionQuotient.ClaimClosure.no_exact_claim_under_declared_assumptions_unless_excused_core}} {\tiny\ttfamily DecisionQuotient/\allowbreak ClaimClosure.lean}
\item \textbf{\nolinkurl{CC42}}\hypertarget{lh:CC42}{}\enspace{\ttfamily\nolinkurl{DecisionQuotient.ClaimClosure.no_exact_identifier_implies_not_boundary_characterized}} {\tiny\ttfamily DecisionQuotient/\allowbreak ClaimClosure.lean}
\item \textbf{\nolinkurl{CC43}}\hypertarget{lh:CC43}{}\enspace{\ttfamily\nolinkurl{DecisionQuotient.ClaimClosure.no_uncertified_exact_claim_core}} {\tiny\ttfamily DecisionQuotient/\allowbreak ClaimClosure.lean}
\item \textbf{\nolinkurl{CC48}}\hypertarget{lh:CC48}{}\enspace{\ttfamily\nolinkurl{DecisionQuotient.ClaimClosure.physical_crossover_hardness_core}} {\tiny\ttfamily DecisionQuotient/\allowbreak ClaimClosure.lean}
\item \textbf{\nolinkurl{CC50}}\hypertarget{lh:CC50}{}\enspace{\ttfamily\nolinkurl{DecisionQuotient.ClaimClosure.process_bridge_failure_witness}} {\tiny\ttfamily DecisionQuotient/\allowbreak ClaimClosure.lean}
\item \textbf{\nolinkurl{CC51}}\hypertarget{lh:CC51}{}\enspace{\ttfamily\nolinkurl{DecisionQuotient.ClaimClosure.poseAnchorQuery}} {\tiny\ttfamily DecisionQuotient/\allowbreak ClaimClosure.lean}
\item \textbf{\nolinkurl{CC52}}\hypertarget{lh:CC52}{}\enspace{\ttfamily\nolinkurl{DecisionQuotient.ClaimClosure.pose_returns_anchor_query_object}} {\tiny\ttfamily DecisionQuotient/\allowbreak ClaimClosure.lean}
\item \textbf{\nolinkurl{CC54}}\hypertarget{lh:CC54}{}\enspace{\ttfamily\nolinkurl{DecisionQuotient.ClaimClosure.posed_anchor_exact_claim_admissible_iff_competent}} {\tiny\ttfamily DecisionQuotient/\allowbreak ClaimClosure.lean}
\item \textbf{\nolinkurl{CC55}}\hypertarget{lh:CC55}{}\enspace{\ttfamily\nolinkurl{DecisionQuotient.ClaimClosure.posed_anchor_exact_claim_requires_evidence}} {\tiny\ttfamily DecisionQuotient/\allowbreak ClaimClosure.lean}
\item \textbf{\nolinkurl{CC56}}\hypertarget{lh:CC56}{}\enspace{\ttfamily\nolinkurl{DecisionQuotient.ClaimClosure.posed_anchor_no_competence_no_exact_claim}} {\tiny\ttfamily DecisionQuotient/\allowbreak ClaimClosure.lean}
\item \textbf{\nolinkurl{CC57}}\hypertarget{lh:CC57}{}\enspace{\ttfamily\nolinkurl{DecisionQuotient.ClaimClosure.posed_anchor_query_truth_iff_exists_anchor}} {\tiny\ttfamily DecisionQuotient/\allowbreak ClaimClosure.lean}
\item \textbf{\nolinkurl{CC58}}\hypertarget{lh:CC58}{}\enspace{\ttfamily\nolinkurl{DecisionQuotient.ClaimClosure.posed_anchor_query_truth_iff_exists_forall}} {\tiny\ttfamily DecisionQuotient/\allowbreak ClaimClosure.lean}
\item \textbf{\nolinkurl{CC59}}\hypertarget{lh:CC59}{}\enspace{\ttfamily\nolinkurl{DecisionQuotient.ClaimClosure.posed_anchor_signal_positive_certified_implies_admissible}} {\tiny\ttfamily DecisionQuotient/\allowbreak ClaimClosure.lean}
\item \textbf{\nolinkurl{CC60}}\hypertarget{lh:CC60}{}\enspace{\ttfamily\nolinkurl{DecisionQuotient.ClaimClosure.query_obstruction_boolean_corollary}} {\tiny\ttfamily DecisionQuotient/\allowbreak ClaimClosure.lean}
\item \textbf{\nolinkurl{CC61}}\hypertarget{lh:CC61}{}\enspace{\ttfamily\nolinkurl{DecisionQuotient.ClaimClosure.query_obstruction_finite_state_core}} {\tiny\ttfamily DecisionQuotient/\allowbreak ClaimClosure.lean}
\item \textbf{\nolinkurl{CC62}}\hypertarget{lh:CC62}{}\enspace{\ttfamily\nolinkurl{DecisionQuotient.ClaimClosure.regime_core_claim_proved}} {\tiny\ttfamily DecisionQuotient/\allowbreak ClaimClosure.lean}
\item \textbf{\nolinkurl{CC63}}\hypertarget{lh:CC63}{}\enspace{\ttfamily\nolinkurl{DecisionQuotient.ClaimClosure.regime_simulation_transfers_hardness}} {\tiny\ttfamily DecisionQuotient/\allowbreak ClaimClosure.lean}
\item \textbf{\nolinkurl{CC69}}\hypertarget{lh:CC69}{}\enspace{\ttfamily\nolinkurl{DecisionQuotient.ClaimClosure.stochastic_objective_bridge_can_fail_on_sufficiency}} {\tiny\ttfamily DecisionQuotient/\allowbreak ClaimClosure.lean}
\item \textbf{\nolinkurl{CC70}}\hypertarget{lh:CC70}{}\enspace{\ttfamily\nolinkurl{DecisionQuotient.ClaimClosure.subproblem_hardness_lifts_to_full}} {\tiny\ttfamily DecisionQuotient/\allowbreak ClaimClosure.lean}
\item \textbf{\nolinkurl{CC71}}\hypertarget{lh:CC71}{}\enspace{\ttfamily\nolinkurl{DecisionQuotient.ClaimClosure.subproblem_transfer_as_regime_simulation}} {\tiny\ttfamily DecisionQuotient/\allowbreak ClaimClosure.lean}
\item \textbf{\nolinkurl{CC72}}\hypertarget{lh:CC72}{}\enspace{\ttfamily\nolinkurl{DecisionQuotient.ClaimClosure.sufficiency_conp_complete_conditional}} {\tiny\ttfamily DecisionQuotient/\allowbreak ClaimClosure.lean}
\item \textbf{\nolinkurl{CC73}}\hypertarget{lh:CC73}{}\enspace{\ttfamily\nolinkurl{DecisionQuotient.ClaimClosure.sufficiency_conp_reduction_core}} {\tiny\ttfamily DecisionQuotient/\allowbreak ClaimClosure.lean}
\item \textbf{\nolinkurl{CC75}}\hypertarget{lh:CC75}{}\enspace{\ttfamily\nolinkurl{DecisionQuotient.ClaimClosure.sufficiency_iff_projectedOptCover_eq_opt}} {\tiny\ttfamily DecisionQuotient/\allowbreak ClaimClosure.lean}
\item \textbf{\nolinkurl{CC76}}\hypertarget{lh:CC76}{}\enspace{\ttfamily\nolinkurl{DecisionQuotient.ClaimClosure.thermo_conservation_additive_core}} {\tiny\ttfamily DecisionQuotient/\allowbreak ClaimClosure.lean}
\item \textbf{\nolinkurl{CC77}}\hypertarget{lh:CC77}{}\enspace{\ttfamily\nolinkurl{DecisionQuotient.ClaimClosure.thermo_energy_carbon_lift_core}} {\tiny\ttfamily DecisionQuotient/\allowbreak ClaimClosure.lean}
\item \textbf{\nolinkurl{CCC1}}\hypertarget{lh:CCC1}{}\enspace{\ttfamily\nolinkurl{DecisionQuotient.CC.anchor_sigma2p_complete_conditional}} {\tiny\ttfamily DecisionQuotient/\allowbreak ClaimClosure.lean}
\item \textbf{\nolinkurl{CCC3}}\hypertarget{lh:CCC3}{}\enspace{\ttfamily\nolinkurl{DecisionQuotient.CC.dichotomy_conditional}} {\tiny\ttfamily DecisionQuotient/\allowbreak ClaimClosure.lean}
\item \textbf{\nolinkurl{CCC5}}\hypertarget{lh:CCC5}{}\enspace{\ttfamily\nolinkurl{DecisionQuotient.CC.minsuff_conp_complete_conditional}} {\tiny\ttfamily DecisionQuotient/\allowbreak ClaimClosure.lean}
\item \textbf{\nolinkurl{CCC6}}\hypertarget{lh:CCC6}{}\enspace{\ttfamily\nolinkurl{DecisionQuotient.CC.sufficiency_conp_complete_conditional}} {\tiny\ttfamily DecisionQuotient/\allowbreak ClaimClosure.lean}
\item \textbf{\nolinkurl{CF2}}\hypertarget{lh:CF2}{}\enspace{\ttfamily\nolinkurl{Physics.ConstraintForcing.logic_time_not_sufficient_for_unique_law}} {\tiny\ttfamily DecisionQuotient/\allowbreak Physics/\allowbreak ConstraintForcing.lean}
\item \textbf{\nolinkurl{CF4}}\hypertarget{lh:CF4}{}\enspace{\ttfamily\nolinkurl{Physics.ConstraintForcing.objective_not_determined_of_parameter_separation}} {\tiny\ttfamily DecisionQuotient/\allowbreak Physics/\allowbreak ConstraintForcing.lean}
\item \textbf{\nolinkurl{CF6}}\hypertarget{lh:CF6}{}\enspace{\ttfamily\nolinkurl{Physics.ConstraintForcing.actionForced_of_deadline}} {\tiny\ttfamily DecisionQuotient/\allowbreak Physics/\allowbreak ConstraintForcing.lean}
\item \textbf{\nolinkurl{CF7}}\hypertarget{lh:CF7}{}\enspace{\ttfamily\nolinkurl{Physics.ConstraintForcing.nondegenerateBelief_of_deadline_and_uncertainty}} {\tiny\ttfamily DecisionQuotient/\allowbreak Physics/\allowbreak ConstraintForcing.lean}
\item \textbf{\nolinkurl{CF8}}\hypertarget{lh:CF8}{}\enspace{\ttfamily\nolinkurl{Physics.ConstraintForcing.forced_decision_implies_positive_landauer_cost}} {\tiny\ttfamily DecisionQuotient/\allowbreak Physics/\allowbreak ConstraintForcing.lean}
\item \textbf{\nolinkurl{CF9}}\hypertarget{lh:CF9}{}\enspace{\ttfamily\nolinkurl{Physics.ConstraintForcing.forced_decision_implies_positive_nv_work}} {\tiny\ttfamily DecisionQuotient/\allowbreak Physics/\allowbreak ConstraintForcing.lean}
\item \textbf{\nolinkurl{CI1}}\hypertarget{lh:CI1}{}\enspace{\ttfamily\nolinkurl{Physics.CertificationInterface.FiniteCertificationInterface}} {\tiny\ttfamily DecisionQuotient/\allowbreak Physics/\allowbreak CertificationInterface.lean}
\item \textbf{\nolinkurl{CI2}}\hypertarget{lh:CI2}{}\enspace{\ttfamily\nolinkurl{Physics.CertificationInterface.recordedDistinctions_le_time}} {\tiny\ttfamily DecisionQuotient/\allowbreak Physics/\allowbreak CertificationInterface.lean}
\item \textbf{\nolinkurl{CI3}}\hypertarget{lh:CI3}{}\enspace{\ttfamily\nolinkurl{Physics.CertificationInterface.transition_selects_binary_distinction}} {\tiny\ttfamily DecisionQuotient/\allowbreak Physics/\allowbreak CertificationInterface.lean}
\item \textbf{\nolinkurl{CI4}}\hypertarget{lh:CI4}{}\enspace{\ttfamily\nolinkurl{Physics.CertificationInterface.recorded_transition_increments}} {\tiny\ttfamily DecisionQuotient/\allowbreak Physics/\allowbreak CertificationInterface.lean}
\item \textbf{\nolinkurl{CI5}}\hypertarget{lh:CI5}{}\enspace{\ttfamily\nolinkurl{Physics.CertificationInterface.recorded_bit_cost_positive}} {\tiny\ttfamily DecisionQuotient/\allowbreak Physics/\allowbreak CertificationInterface.lean}
\item \textbf{\nolinkurl{CI6}}\hypertarget{lh:CI6}{}\enspace{\ttfamily\nolinkurl{Physics.CertificationInterface.recorded_distinctions_positive_energy}} {\tiny\ttfamily DecisionQuotient/\allowbreak Physics/\allowbreak CertificationInterface.lean}
\item \textbf{\nolinkurl{CI7}}\hypertarget{lh:CI7}{}\enspace{\ttfamily\nolinkurl{Physics.CertificationInterface.finite_budget_bounds_recorded_distinctions}} {\tiny\ttfamily DecisionQuotient/\allowbreak Physics/\allowbreak CertificationInterface.lean}
\item \textbf{\nolinkurl{CI12}}\hypertarget{lh:CI12}{}\enspace{\ttfamily\nolinkurl{Physics.CertificationInterface.FiniteOutcomeDetector}} {\tiny\ttfamily DecisionQuotient/\allowbreak Physics/\allowbreak CertificationInterface.lean}
\item \textbf{\nolinkurl{CI13}}\hypertarget{lh:CI13}{}\enspace{\ttfamily\nolinkurl{Physics.CertificationInterface.FiniteDetectorQueryInterface}} {\tiny\ttfamily DecisionQuotient/\allowbreak Physics/\allowbreak CertificationInterface.lean}
\item \textbf{\nolinkurl{CI14}}\hypertarget{lh:CI14}{}\enspace{\ttfamily\nolinkurl{Physics.CertificationInterface.finite_detector_induces_query_interface}} {\tiny\ttfamily DecisionQuotient/\allowbreak Physics/\allowbreak CertificationInterface.lean}
\item \textbf{\nolinkurl{CI15}}\hypertarget{lh:CI15}{}\enspace{\ttfamily\nolinkurl{Physics.CertificationInterface.finite_detector_induces_certification_interface}} {\tiny\ttfamily DecisionQuotient/\allowbreak Physics/\allowbreak CertificationInterface.lean}
\item \textbf{\nolinkurl{CI16}}\hypertarget{lh:CI16}{}\enspace{\ttfamily\nolinkurl{Physics.CertificationInterface.finite_detector_budget_bounds_retained_records}} {\tiny\ttfamily DecisionQuotient/\allowbreak Physics/\allowbreak CertificationInterface.lean}
\item \textbf{\nolinkurl{CI17}}\hypertarget{lh:CI17}{}\enspace{\ttfamily\nolinkurl{Physics.CertificationInterface.ProofToken}} {\tiny\ttfamily DecisionQuotient/\allowbreak Physics/\allowbreak CertificationInterface.lean}
\item \textbf{\nolinkurl{CI18}}\hypertarget{lh:CI18}{}\enspace{\ttfamily\nolinkurl{Physics.CertificationInterface.ProofToken.length}} {\tiny\ttfamily DecisionQuotient/\allowbreak Physics/\allowbreak CertificationInterface.lean}
\item \textbf{\nolinkurl{CI19}}\hypertarget{lh:CI19}{}\enspace{\ttfamily\nolinkurl{Physics.CertificationInterface.FiniteStateProofParser}} {\tiny\ttfamily DecisionQuotient/\allowbreak Physics/\allowbreak CertificationInterface.lean}
\item \textbf{\nolinkurl{CI20}}\hypertarget{lh:CI20}{}\enspace{\ttfamily\nolinkurl{Physics.CertificationInterface.ParserConfig}} {\tiny\ttfamily DecisionQuotient/\allowbreak Physics/\allowbreak CertificationInterface.lean}
\item \textbf{\nolinkurl{CI21}}\hypertarget{lh:CI21}{}\enspace{\ttfamily\nolinkurl{Physics.CertificationInterface.FiniteStateProofParser.toCertificationInterface}} {\tiny\ttfamily DecisionQuotient/\allowbreak Physics/\allowbreak CertificationInterface.lean}
\item \textbf{\nolinkurl{CI22}}\hypertarget{lh:CI22}{}\enspace{\ttfamily\nolinkurl{Physics.CertificationInterface.proof_parser_symbol_record_cost_positive}} {\tiny\ttfamily DecisionQuotient/\allowbreak Physics/\allowbreak CertificationInterface.lean}
\item \textbf{\nolinkurl{CI23}}\hypertarget{lh:CI23}{}\enspace{\ttfamily\nolinkurl{Physics.CertificationInterface.parser_recorded_distinctions_le_steps}} {\tiny\ttfamily DecisionQuotient/\allowbreak Physics/\allowbreak CertificationInterface.lean}
\item \textbf{\nolinkurl{CI24}}\hypertarget{lh:CI24}{}\enspace{\ttfamily\nolinkurl{Physics.CertificationInterface.proof_token_parser_records_bounded_by_token_length}} {\tiny\ttfamily DecisionQuotient/\allowbreak Physics/\allowbreak CertificationInterface.lean}
\item \textbf{\nolinkurl{CI25}}\hypertarget{lh:CI25}{}\enspace{\ttfamily\nolinkurl{Physics.CertificationInterface.proof_token_parser_recorded_transition_increments}} {\tiny\ttfamily DecisionQuotient/\allowbreak Physics/\allowbreak CertificationInterface.lean}
\item \textbf{\nolinkurl{CI26}}\hypertarget{lh:CI26}{}\enspace{\ttfamily\nolinkurl{Physics.CertificationInterface.proof_token_parser_recorded_transition_requires_positive_energy}} {\tiny\ttfamily DecisionQuotient/\allowbreak Physics/\allowbreak CertificationInterface.lean}
\item \textbf{\nolinkurl{CI27}}\hypertarget{lh:CI27}{}\enspace{\ttfamily\nolinkurl{Physics.CertificationInterface.SeparablePointerRecordInterface}} {\tiny\ttfamily DecisionQuotient/\allowbreak Physics/\allowbreak CertificationInterface.lean}
\item \textbf{\nolinkurl{CI28}}\hypertarget{lh:CI28}{}\enspace{\ttfamily\nolinkurl{Physics.CertificationInterface.separable_pointer_records_cover_queries}} {\tiny\ttfamily DecisionQuotient/\allowbreak Physics/\allowbreak CertificationInterface.lean}
\item \textbf{\nolinkurl{CI29}}\hypertarget{lh:CI29}{}\enspace{\ttfamily\nolinkurl{Physics.CertificationInterface.compressed_certificate_changes_separable_interface}} {\tiny\ttfamily DecisionQuotient/\allowbreak Physics/\allowbreak CertificationInterface.lean}
\item \textbf{\nolinkurl{CI30}}\hypertarget{lh:CI30}{}\enspace{\ttfamily\nolinkurl{Physics.CertificationInterface.ReversibleAnswerProtocol}} {\tiny\ttfamily DecisionQuotient/\allowbreak Physics/\allowbreak CertificationInterface.lean}
\item \textbf{\nolinkurl{CI31}}\hypertarget{lh:CI31}{}\enspace{\ttfamily\nolinkurl{Physics.CertificationInterface.SatisfiesProofAuditRequirement}} {\tiny\ttfamily DecisionQuotient/\allowbreak Physics/\allowbreak CertificationInterface.lean}
\item \textbf{\nolinkurl{CI32}}\hypertarget{lh:CI32}{}\enspace{\ttfamily\nolinkurl{Physics.CertificationInterface.reversible_answer_not_certification}} {\tiny\ttfamily DecisionQuotient/\allowbreak Physics/\allowbreak CertificationInterface.lean}
\item \textbf{\nolinkurl{CI33}}\hypertarget{lh:CI33}{}\enspace{\ttfamily\nolinkurl{Physics.CertificationInterface.uncomputation_destroys_evidence_retention}} {\tiny\ttfamily DecisionQuotient/\allowbreak Physics/\allowbreak CertificationInterface.lean}
\item \textbf{\nolinkurl{CI34}}\hypertarget{lh:CI34}{}\enspace{\ttfamily\nolinkurl{Physics.CertificationInterface.CertificationEnergyAccounting}} {\tiny\ttfamily DecisionQuotient/\allowbreak Physics/\allowbreak CertificationInterface.lean}
\item \textbf{\nolinkurl{CI35}}\hypertarget{lh:CI35}{}\enspace{\ttfamily\nolinkurl{Physics.CertificationInterface.certificationEnergyLowerBound}} {\tiny\ttfamily DecisionQuotient/\allowbreak Physics/\allowbreak CertificationInterface.lean}
\item \textbf{\nolinkurl{CI36}}\hypertarget{lh:CI36}{}\enspace{\ttfamily\nolinkurl{Physics.CertificationInterface.certification_energy_lower_bound_eq_acquisition_plus_retention}} {\tiny\ttfamily DecisionQuotient/\allowbreak Physics/\allowbreak CertificationInterface.lean}
\item \textbf{\nolinkurl{CI37}}\hypertarget{lh:CI37}{}\enspace{\ttfamily\nolinkurl{Physics.CertificationInterface.retention_cost_positive_even_if_acquisition_zero}} {\tiny\ttfamily DecisionQuotient/\allowbreak Physics/\allowbreak CertificationInterface.lean}
\item \textbf{\nolinkurl{CI38}}\hypertarget{lh:CI38}{}\enspace{\ttfamily\nolinkurl{Physics.CertificationInterface.SituatedAgent}} {\tiny\ttfamily DecisionQuotient/\allowbreak Physics/\allowbreak CertificationInterface.lean}
\item \textbf{\nolinkurl{CI39}}\hypertarget{lh:CI39}{}\enspace{\ttfamily\nolinkurl{Physics.CertificationInterface.SituatedAgent.toBoundedRegion}} {\tiny\ttfamily DecisionQuotient/\allowbreak Physics/\allowbreak CertificationInterface.lean}
\item \textbf{\nolinkurl{CI40}}\hypertarget{lh:CI40}{}\enspace{\ttfamily\nolinkurl{Physics.CertificationInterface.SituatedAgent.spacetimeAcquisitionCapacity}} {\tiny\ttfamily DecisionQuotient/\allowbreak Physics/\allowbreak CertificationInterface.lean}
\item \textbf{\nolinkurl{CI41}}\hypertarget{lh:CI41}{}\enspace{\ttfamily\nolinkurl{Physics.CertificationInterface.SituatedAgent.certificationCapacity}} {\tiny\ttfamily DecisionQuotient/\allowbreak Physics/\allowbreak CertificationInterface.lean}
\item \textbf{\nolinkurl{CI42}}\hypertarget{lh:CI42}{}\enspace{\ttfamily\nolinkurl{Physics.CertificationInterface.agent_lifetime_certification_capacity_is_finite}} {\tiny\ttfamily DecisionQuotient/\allowbreak Physics/\allowbreak CertificationInterface.lean}
\item \textbf{\nolinkurl{CI43}}\hypertarget{lh:CI43}{}\enspace{\ttfamily\nolinkurl{Physics.CertificationInterface.ProtocolPhase}} {\tiny\ttfamily DecisionQuotient/\allowbreak Physics/\allowbreak CertificationInterface.lean}
\item \textbf{\nolinkurl{CI44}}\hypertarget{lh:CI44}{}\enspace{\ttfamily\nolinkurl{Physics.CertificationInterface.VerificationProtocol}} {\tiny\ttfamily DecisionQuotient/\allowbreak Physics/\allowbreak CertificationInterface.lean}
\item \textbf{\nolinkurl{CI45}}\hypertarget{lh:CI45}{}\enspace{\ttfamily\nolinkurl{Physics.CertificationInterface.SoundAuditTrail}} {\tiny\ttfamily DecisionQuotient/\allowbreak Physics/\allowbreak CertificationInterface.lean}
\item \textbf{\nolinkurl{CI46}}\hypertarget{lh:CI46}{}\enspace{\ttfamily\nolinkurl{Physics.CertificationInterface.sound_audit_trail_total_records_ge_instance_bits}} {\tiny\ttfamily DecisionQuotient/\allowbreak Physics/\allowbreak CertificationInterface.lean}
\item \textbf{\nolinkurl{CI47}}\hypertarget{lh:CI47}{}\enspace{\ttfamily\nolinkurl{Physics.CertificationInterface.sound_audit_trail_work_conservation}} {\tiny\ttfamily DecisionQuotient/\allowbreak Physics/\allowbreak CertificationInterface.lean}
\item \textbf{\nolinkurl{CI48}}\hypertarget{lh:CI48}{}\enspace{\ttfamily\nolinkurl{Physics.CertificationInterface.bounded_compressed_audit_trail_not_sound}} {\tiny\ttfamily DecisionQuotient/\allowbreak Physics/\allowbreak CertificationInterface.lean}
\item \textbf{\nolinkurl{CI49}}\hypertarget{lh:CI49}{}\enspace{\ttfamily\nolinkurl{Physics.CertificationInterface.MerkleStyleAuditTrail}} {\tiny\ttfamily DecisionQuotient/\allowbreak Physics/\allowbreak CertificationInterface.lean}
\item \textbf{\nolinkurl{CI50}}\hypertarget{lh:CI50}{}\enspace{\ttfamily\nolinkurl{Physics.CertificationInterface.merkle_audit_trail_work_conservation}} {\tiny\ttfamily DecisionQuotient/\allowbreak Physics/\allowbreak CertificationInterface.lean}
\item \textbf{\nolinkurl{CI51}}\hypertarget{lh:CI51}{}\enspace{\ttfamily\nolinkurl{Physics.CertificationInterface.PolynomialCommitmentAuditTrail}} {\tiny\ttfamily DecisionQuotient/\allowbreak Physics/\allowbreak CertificationInterface.lean}
\item \textbf{\nolinkurl{CI52}}\hypertarget{lh:CI52}{}\enspace{\ttfamily\nolinkurl{Physics.CertificationInterface.polynomial_commitment_work_conservation}} {\tiny\ttfamily DecisionQuotient/\allowbreak Physics/\allowbreak CertificationInterface.lean}
\item \textbf{\nolinkurl{CI53}}\hypertarget{lh:CI53}{}\enspace{\ttfamily\nolinkurl{Physics.CertificationInterface.InteractiveArgumentAuditTrail}} {\tiny\ttfamily DecisionQuotient/\allowbreak Physics/\allowbreak CertificationInterface.lean}
\item \textbf{\nolinkurl{CI54}}\hypertarget{lh:CI54}{}\enspace{\ttfamily\nolinkurl{Physics.CertificationInterface.interactive_argument_work_conservation}} {\tiny\ttfamily DecisionQuotient/\allowbreak Physics/\allowbreak CertificationInterface.lean}
\item \textbf{\nolinkurl{CI55}}\hypertarget{lh:CI55}{}\enspace{\ttfamily\nolinkurl{Physics.CertificationInterface.MultiAgentProtocol}} {\tiny\ttfamily DecisionQuotient/\allowbreak Physics/\allowbreak CertificationInterface.lean}
\item \textbf{\nolinkurl{CI56}}\hypertarget{lh:CI56}{}\enspace{\ttfamily\nolinkurl{Physics.CertificationInterface.MultiAgentProtocol.aggregateRetainedRecordCapacity}} {\tiny\ttfamily DecisionQuotient/\allowbreak Physics/\allowbreak CertificationInterface.lean}
\item \textbf{\nolinkurl{CI57}}\hypertarget{lh:CI57}{}\enspace{\ttfamily\nolinkurl{Physics.CertificationInterface.multi_agent_total_records_le_aggregate_capacity}} {\tiny\ttfamily DecisionQuotient/\allowbreak Physics/\allowbreak CertificationInterface.lean}
\item \textbf{\nolinkurl{CI58}}\hypertarget{lh:CI58}{}\enspace{\ttfamily\nolinkurl{Physics.CertificationInterface.transmitted_records_le_receiver_capacity}} {\tiny\ttfamily DecisionQuotient/\allowbreak Physics/\allowbreak CertificationInterface.lean}
\item \textbf{\nolinkurl{CI59}}\hypertarget{lh:CI59}{}\enspace{\ttfamily\nolinkurl{Physics.CertificationInterface.unbounded_family_exceeds_multi_agent_capacity}} {\tiny\ttfamily DecisionQuotient/\allowbreak Physics/\allowbreak CertificationInterface.lean}
\item \textbf{\nolinkurl{CI60}}\hypertarget{lh:CI60}{}\enspace{\ttfamily\nolinkurl{Physics.CertificationInterface.no_local_proof_cover_above_agent_capacity}} {\tiny\ttfamily DecisionQuotient/\allowbreak Physics/\allowbreak CertificationInterface.lean}
\item \textbf{\nolinkurl{CT1}}\hypertarget{lh:CT1}{}\enspace{\ttfamily\nolinkurl{DecisionQuotient.Physics.ClaimTransport.PhysicalEncoding}} {\tiny\ttfamily DecisionQuotient/\allowbreak Physics/\allowbreak ClaimTransport.lean}
\item \textbf{\nolinkurl{CT2}}\hypertarget{lh:CT2}{}\enspace{\ttfamily\nolinkurl{DecisionQuotient.Physics.ClaimTransport.physical_claim_lifts_from_core}} {\tiny\ttfamily DecisionQuotient/\allowbreak Physics/\allowbreak ClaimTransport.lean}
\item \textbf{\nolinkurl{CT3}}\hypertarget{lh:CT3}{}\enspace{\ttfamily\nolinkurl{DecisionQuotient.Physics.ClaimTransport.physical_claim_lifts_from_core_conditional}} {\tiny\ttfamily DecisionQuotient/\allowbreak Physics/\allowbreak ClaimTransport.lean}
\item \textbf{\nolinkurl{CT4}}\hypertarget{lh:CT4}{}\enspace{\ttfamily\nolinkurl{DecisionQuotient.Physics.ClaimTransport.physical_counterexample_yields_core_counterexample}} {\tiny\ttfamily DecisionQuotient/\allowbreak Physics/\allowbreak ClaimTransport.lean}
\item \textbf{\nolinkurl{CT5}}\hypertarget{lh:CT5}{}\enspace{\ttfamily\nolinkurl{DecisionQuotient.Physics.ClaimTransport.physical_counterexample_invalidates_core_rule}} {\tiny\ttfamily DecisionQuotient/\allowbreak Physics/\allowbreak ClaimTransport.lean}
\item \textbf{\nolinkurl{CT6}}\hypertarget{lh:CT6}{}\enspace{\ttfamily\nolinkurl{DecisionQuotient.Physics.ClaimTransport.no_physical_counterexample_of_core_theorem}} {\tiny\ttfamily DecisionQuotient/\allowbreak Physics/\allowbreak ClaimTransport.lean}
\item \textbf{\nolinkurl{CV4}}\hypertarget{lh:CV4}{}\enspace{\ttfamily\nolinkurl{Physics.Conversation.tick_uses_shared_node}} {\tiny\ttfamily DecisionQuotient/\allowbreak Physics/\allowbreak Conversation.lean}
\item \textbf{\nolinkurl{CV5}}\hypertarget{lh:CV5}{}\enspace{\ttfamily\nolinkurl{Physics.Conversation.tick_shared_is_merged_emissions}} {\tiny\ttfamily DecisionQuotient/\allowbreak Physics/\allowbreak Conversation.lean}
\item \textbf{\nolinkurl{CV7}}\hypertarget{lh:CV7}{}\enspace{\ttfamily\nolinkurl{Physics.Conversation.clamp_projection_eq_iff_same_clamped_bit}} {\tiny\ttfamily DecisionQuotient/\allowbreak Physics/\allowbreak Conversation.lean}
\item \textbf{\nolinkurl{CV11}}\hypertarget{lh:CV11}{}\enspace{\ttfamily\nolinkurl{Physics.Conversation.toClaimReport}} {\tiny\ttfamily DecisionQuotient/\allowbreak Physics/\allowbreak Conversation.lean}
\item \textbf{\nolinkurl{CV12}}\hypertarget{lh:CV12}{}\enspace{\ttfamily\nolinkurl{Physics.Conversation.abstain_iff_no_answer}} {\tiny\ttfamily DecisionQuotient/\allowbreak Physics/\allowbreak Conversation.lean}
\item \textbf{\nolinkurl{CV13}}\hypertarget{lh:CV13}{}\enspace{\ttfamily\nolinkurl{Physics.Conversation.yes_no_iff_exact_claim}} {\tiny\ttfamily DecisionQuotient/\allowbreak Physics/\allowbreak Conversation.lean}
\item \textbf{\nolinkurl{CV15}}\hypertarget{lh:CV15}{}\enspace{\ttfamily\nolinkurl{Physics.Conversation.toReportSignal_signal_consistent_zero_certified}} {\tiny\ttfamily DecisionQuotient/\allowbreak Physics/\allowbreak Conversation.lean}
\item \textbf{\nolinkurl{CV16}}\hypertarget{lh:CV16}{}\enspace{\ttfamily\nolinkurl{Physics.Conversation.abstain_report_can_carry_explanation}} {\tiny\ttfamily DecisionQuotient/\allowbreak Physics/\allowbreak Conversation.lean}
\item \textbf{\nolinkurl{CV17}}\hypertarget{lh:CV17}{}\enspace{\ttfamily\nolinkurl{DecisionQuotient.Physics.Conversation.clampDecisionEvent_iff_bitOps_pos}} {\tiny\ttfamily DecisionQuotient/\allowbreak Physics/\allowbreak Conversation.lean}
\item \textbf{\nolinkurl{CV18}}\hypertarget{lh:CV18}{}\enspace{\ttfamily\nolinkurl{DecisionQuotient.Physics.Conversation.clamp_event_implies_positive_energy}} {\tiny\ttfamily DecisionQuotient/\allowbreak Physics/\allowbreak Conversation.lean}
\item \textbf{\nolinkurl{DC1}}\hypertarget{lh:DC1}{}\enspace{\ttfamily\nolinkurl{StochasticSequential.static_stochastic_strict_separation}} {\tiny\ttfamily DecisionQuotient/\allowbreak StochasticSequential/\allowbreak Hierarchy.lean}
\item \textbf{\nolinkurl{DC2}}\hypertarget{lh:DC2}{}\enspace{\ttfamily\nolinkurl{StochasticSequential.stochastic_sequential_strict_separation}} {\tiny\ttfamily DecisionQuotient/\allowbreak StochasticSequential/\allowbreak Hierarchy.lean}
\item \textbf{\nolinkurl{DC9}}\hypertarget{lh:DC9}{}\enspace{\ttfamily\nolinkurl{StochasticSequential.stochastic_to_PP}} {\tiny\ttfamily DecisionQuotient/\allowbreak StochasticSequential/\allowbreak Hierarchy.lean}
\item \textbf{\nolinkurl{DC10}}\hypertarget{lh:DC10}{}\enspace{\ttfamily\nolinkurl{StochasticSequential.sequential_to_PSPACE}} {\tiny\ttfamily DecisionQuotient/\allowbreak StochasticSequential/\allowbreak Hierarchy.lean}
\item \textbf{\nolinkurl{DC14}}\hypertarget{lh:DC14}{}\enspace{\ttfamily\nolinkurl{StochasticSequential.stochastic_dichotomy}} {\tiny\ttfamily DecisionQuotient/\allowbreak StochasticSequential/\allowbreak Dichotomy.lean}
\item \textbf{\nolinkurl{DC15}}\hypertarget{lh:DC15}{}\enspace{\ttfamily\nolinkurl{StochasticSequential.above_threshold_hard}} {\tiny\ttfamily DecisionQuotient/\allowbreak StochasticSequential/\allowbreak Dichotomy.lean}
\item \textbf{\nolinkurl{DC16}}\hypertarget{lh:DC16}{}\enspace{\ttfamily\nolinkurl{StochasticSequential.StochasticAnchorSufficient}} {\tiny\ttfamily DecisionQuotient/\allowbreak StochasticSequential/\allowbreak Basic.lean}
\item \textbf{\nolinkurl{DC17}}\hypertarget{lh:DC17}{}\enspace{\ttfamily\nolinkurl{StochasticSequential.StochasticAnchorSufficiencyCheck}} {\tiny\ttfamily DecisionQuotient/\allowbreak StochasticSequential/\allowbreak Basic.lean}
\item \textbf{\nolinkurl{DC18}}\hypertarget{lh:DC18}{}\enspace{\ttfamily\nolinkurl{StochasticSequential.stochastic_anchor_check_iff}} {\tiny\ttfamily DecisionQuotient/\allowbreak StochasticSequential/\allowbreak Basic.lean}
\item \textbf{\nolinkurl{DC19}}\hypertarget{lh:DC19}{}\enspace{\ttfamily\nolinkurl{StochasticSequential.stochastic_anchor_sufficient_of_stochastic_sufficient}} {\tiny\ttfamily DecisionQuotient/\allowbreak StochasticSequential/\allowbreak Quotient.lean}
\item \textbf{\nolinkurl{DC24}}\hypertarget{lh:DC24}{}\enspace{\ttfamily\nolinkurl{StochasticSequential.StochasticAnchorCheckInstance}} {\tiny\ttfamily DecisionQuotient/\allowbreak StochasticSequential/\allowbreak PolynomialReduction.lean}
\item \textbf{\nolinkurl{DC25}}\hypertarget{lh:DC25}{}\enspace{\ttfamily\nolinkurl{StochasticSequential.reduceMAJSAT_correct_anchor_strict}} {\tiny\ttfamily DecisionQuotient/\allowbreak StochasticSequential/\allowbreak PolynomialReduction.lean}
\item \textbf{\nolinkurl{DC26}}\hypertarget{lh:DC26}{}\enspace{\ttfamily\nolinkurl{StochasticSequential.reduceMAJSAT_to_stochastic_anchor_reduction}} {\tiny\ttfamily DecisionQuotient/\allowbreak StochasticSequential/\allowbreak PolynomialReduction.lean}
\item \textbf{\nolinkurl{DC30}}\hypertarget{lh:DC30}{}\enspace{\ttfamily\nolinkurl{StochasticSequential.StatePotential}} {\tiny\ttfamily DecisionQuotient/\allowbreak StochasticSequential/\allowbreak ThermodynamicLift.lean}
\item \textbf{\nolinkurl{DC31}}\hypertarget{lh:DC31}{}\enspace{\ttfamily\nolinkurl{StochasticSequential.utilityFromPotentialDrop_le_iff_nextPotential_ge}} {\tiny\ttfamily DecisionQuotient/\allowbreak StochasticSequential/\allowbreak ThermodynamicLift.lean}
\item \textbf{\nolinkurl{DC32}}\hypertarget{lh:DC32}{}\enspace{\ttfamily\nolinkurl{StochasticSequential.utility_from_action_state_potential}} {\tiny\ttfamily DecisionQuotient/\allowbreak StochasticSequential/\allowbreak ThermodynamicLift.lean}
\item \textbf{\nolinkurl{DC33}}\hypertarget{lh:DC33}{}\enspace{\ttfamily\nolinkurl{StochasticSequential.stochasticExpectedUtility_eq_neg_expectedActionPotential}} {\tiny\ttfamily DecisionQuotient/\allowbreak StochasticSequential/\allowbreak ThermodynamicLift.lean}
\item \textbf{\nolinkurl{DC34}}\hypertarget{lh:DC34}{}\enspace{\ttfamily\nolinkurl{StochasticSequential.stochasticExpectedUtility_le_iff_expectedActionPotential_ge}} {\tiny\ttfamily DecisionQuotient/\allowbreak StochasticSequential/\allowbreak ThermodynamicLift.lean}
\item \textbf{\nolinkurl{DC35}}\hypertarget{lh:DC35}{}\enspace{\ttfamily\nolinkurl{StochasticSequential.landauerEnergyFloor_nonneg}} {\tiny\ttfamily DecisionQuotient/\allowbreak StochasticSequential/\allowbreak ThermodynamicLift.lean}
\item \textbf{\nolinkurl{DC36}}\hypertarget{lh:DC36}{}\enspace{\ttfamily\nolinkurl{StochasticSequential.landauerEnergyFloor_mono_bits}} {\tiny\ttfamily DecisionQuotient/\allowbreak StochasticSequential/\allowbreak ThermodynamicLift.lean}
\item \textbf{\nolinkurl{DC37}}\hypertarget{lh:DC37}{}\enspace{\ttfamily\nolinkurl{StochasticSequential.thermodynamicCost_eq_landauerEnergyFloorRoom_states}} {\tiny\ttfamily DecisionQuotient/\allowbreak StochasticSequential/\allowbreak ThermodynamicLift.lean}
\item \textbf{\nolinkurl{DC40}}\hypertarget{lh:DC40}{}\enspace{\ttfamily\nolinkurl{StochasticSequential.reduceMAJSATPureAnchor_correct}} {\tiny\ttfamily DecisionQuotient/\allowbreak StochasticSequential/\allowbreak PolynomialReduction.lean}
\item \textbf{\nolinkurl{DC41}}\hypertarget{lh:DC41}{}\enspace{\ttfamily\nolinkurl{StochasticSequential.reduceMAJSAT_to_pure_stochastic_anchor_reduction}} {\tiny\ttfamily DecisionQuotient/\allowbreak StochasticSequential/\allowbreak PolynomialReduction.lean}
\item \textbf{\nolinkurl{DC42}}\hypertarget{lh:DC42}{}\enspace{\ttfamily\nolinkurl{StochasticSequential.stochastic_anchor_check_pp_hard}} {\tiny\ttfamily DecisionQuotient/\allowbreak StochasticSequential/\allowbreak PolynomialReduction.lean}
\item \textbf{\nolinkurl{DC45}}\hypertarget{lh:DC45}{}\enspace{\ttfamily\nolinkurl{StochasticSequential.stochastic_sufficiency_pp_hard}} {\tiny\ttfamily DecisionQuotient/\allowbreak StochasticSequential/\allowbreak PolynomialReduction.lean}
\item \textbf{\nolinkurl{DC47}}\hypertarget{lh:DC47}{}\enspace{\ttfamily\nolinkurl{StochasticSequential.stochastic_minimum_sufficiency_pp_hard}} {\tiny\ttfamily DecisionQuotient/\allowbreak StochasticSequential/\allowbreak PolynomialReduction.lean}
\item \textbf{\nolinkurl{DC51}}\hypertarget{lh:DC51}{}\enspace{\ttfamily\nolinkurl{StochasticSequential.stochasticSufficientBool_spec}} {\tiny\ttfamily DecisionQuotient/\allowbreak StochasticSequential/\allowbreak Computation.lean}
\item \textbf{\nolinkurl{DC52}}\hypertarget{lh:DC52}{}\enspace{\ttfamily\nolinkurl{StochasticSequential.stochasticAnchorSufficientBool_spec}} {\tiny\ttfamily DecisionQuotient/\allowbreak StochasticSequential/\allowbreak Computation.lean}
\item \textbf{\nolinkurl{DC53}}\hypertarget{lh:DC53}{}\enspace{\ttfamily\nolinkurl{StochasticSequential.stochasticMinimumSufficiencyBool_spec}} {\tiny\ttfamily DecisionQuotient/\allowbreak StochasticSequential/\allowbreak Computation.lean}
\item \textbf{\nolinkurl{DC57}}\hypertarget{lh:DC57}{}\enspace{\ttfamily\nolinkurl{StochasticSequential.fiberDecisionProblem_sufficient}} {\tiny\ttfamily DecisionQuotient/\allowbreak StochasticSequential/\allowbreak Basic.lean}
\item \textbf{\nolinkurl{DC58}}\hypertarget{lh:DC58}{}\enspace{\ttfamily\nolinkurl{StochasticSequential.countedStochasticMinimumSearch_spec}} {\tiny\ttfamily DecisionQuotient/\allowbreak StochasticSequential/\allowbreak Computation.lean}
\item \textbf{\nolinkurl{DC59}}\hypertarget{lh:DC59}{}\enspace{\ttfamily\nolinkurl{StochasticSequential.countedStochasticMinimumSearch_steps}} {\tiny\ttfamily DecisionQuotient/\allowbreak StochasticSequential/\allowbreak Computation.lean}
\item \textbf{\nolinkurl{DC62}}\hypertarget{lh:DC62}{}\enspace{\ttfamily\nolinkurl{StochasticSequential.countedStochasticSufficiencySearch_spec}} {\tiny\ttfamily DecisionQuotient/\allowbreak StochasticSequential/\allowbreak Computation.lean}
\item \textbf{\nolinkurl{DC63}}\hypertarget{lh:DC63}{}\enspace{\ttfamily\nolinkurl{StochasticSequential.countedStochasticSufficiencySearch_steps}} {\tiny\ttfamily DecisionQuotient/\allowbreak StochasticSequential/\allowbreak Computation.lean}
\item \textbf{\nolinkurl{DC64}}\hypertarget{lh:DC64}{}\enspace{\ttfamily\nolinkurl{StochasticSequential.countedStochasticAnchorSearch_spec}} {\tiny\ttfamily DecisionQuotient/\allowbreak StochasticSequential/\allowbreak Computation.lean}
\item \textbf{\nolinkurl{DC65}}\hypertarget{lh:DC65}{}\enspace{\ttfamily\nolinkurl{StochasticSequential.countedStochasticAnchorSearch_steps}} {\tiny\ttfamily DecisionQuotient/\allowbreak StochasticSequential/\allowbreak Computation.lean}
\item \textbf{\nolinkurl{DC70}}\hypertarget{lh:DC70}{}\enspace{\ttfamily\nolinkurl{DecisionQuotient.static_sufficiency_inP_explicit}} {\tiny\ttfamily DecisionQuotient/\allowbreak ExplicitStateMembership.lean}
\item \textbf{\nolinkurl{DC71}}\hypertarget{lh:DC71}{}\enspace{\ttfamily\nolinkurl{DecisionQuotient.static_anchor_inP_explicit}} {\tiny\ttfamily DecisionQuotient/\allowbreak ExplicitStateMembership.lean}
\item \textbf{\nolinkurl{DC72}}\hypertarget{lh:DC72}{}\enspace{\ttfamily\nolinkurl{StochasticSequential.stochastic_sufficiency_inP_explicit}} {\tiny\ttfamily DecisionQuotient/\allowbreak StochasticSequential/\allowbreak Computation.lean}
\item \textbf{\nolinkurl{DC73}}\hypertarget{lh:DC73}{}\enspace{\ttfamily\nolinkurl{StochasticSequential.stochastic_anchor_inP_explicit}} {\tiny\ttfamily DecisionQuotient/\allowbreak StochasticSequential/\allowbreak Computation.lean}
\item \textbf{\nolinkurl{DC74}}\hypertarget{lh:DC74}{}\enspace{\ttfamily\nolinkurl{StochasticSequential.sequential_sufficiency_inP_explicit}} {\tiny\ttfamily DecisionQuotient/\allowbreak StochasticSequential/\allowbreak Computation.lean}
\item \textbf{\nolinkurl{DC75}}\hypertarget{lh:DC75}{}\enspace{\ttfamily\nolinkurl{StochasticSequential.sequential_anchor_inP_explicit}} {\tiny\ttfamily DecisionQuotient/\allowbreak StochasticSequential/\allowbreak Computation.lean}
\item \textbf{\nolinkurl{DC76}}\hypertarget{lh:DC76}{}\enspace{\ttfamily\nolinkurl{DecisionQuotient.explicit_state_inP_summary}} {\tiny\ttfamily DecisionQuotient/\allowbreak ExplicitStateMembership.lean}
\item \textbf{\nolinkurl{DC77}}\hypertarget{lh:DC77}{}\enspace{\ttfamily\nolinkurl{DecisionQuotient.static_minimum_inP_explicit}} {\tiny\ttfamily DecisionQuotient/\allowbreak ExplicitStateMembership.lean}
\item \textbf{\nolinkurl{DC78}}\hypertarget{lh:DC78}{}\enspace{\ttfamily\nolinkurl{DecisionQuotient.stochastic_minimum_inP_explicit}} {\tiny\ttfamily DecisionQuotient/\allowbreak ExplicitStateMembership.lean}
\item \textbf{\nolinkurl{DC79}}\hypertarget{lh:DC79}{}\enspace{\ttfamily\nolinkurl{DecisionQuotient.sequential_minimum_inP_explicit}} {\tiny\ttfamily DecisionQuotient/\allowbreak ExplicitStateMembership.lean}
\item \textbf{\nolinkurl{DP1}}\hypertarget{lh:DP1}{}\enspace{\ttfamily\nolinkurl{DecisionQuotient.DecisionProblem.minimalSufficient_iff_relevant}} {\tiny\ttfamily DecisionQuotient/\allowbreak Sufficiency.lean}
\item \textbf{\nolinkurl{DP2}}\hypertarget{lh:DP2}{}\enspace{\ttfamily\nolinkurl{DecisionQuotient.DecisionProblem.relevantSet_is_minimal}} {\tiny\ttfamily DecisionQuotient/\allowbreak Sufficiency.lean}
\item \textbf{\nolinkurl{DP3}}\hypertarget{lh:DP3}{}\enspace{\ttfamily\nolinkurl{DecisionQuotient.DecisionProblem.sufficient_implies_selectorSufficient}} {\tiny\ttfamily DecisionQuotient/\allowbreak Sufficiency.lean}
\item \textbf{\nolinkurl{DP4}}\hypertarget{lh:DP4}{}\enspace{\ttfamily\nolinkurl{DecisionQuotient.ClaimClosure.DecisionProblem.epsOpt_zero_eq_opt}} {\tiny\ttfamily DecisionQuotient/\allowbreak ClaimClosure.lean}
\item \textbf{\nolinkurl{DP5}}\hypertarget{lh:DP5}{}\enspace{\ttfamily\nolinkurl{DecisionQuotient.ClaimClosure.DecisionProblem.sufficient_iff_zeroEpsilonSufficient}} {\tiny\ttfamily DecisionQuotient/\allowbreak ClaimClosure.lean}
\item \textbf{\nolinkurl{DP6}}\hypertarget{lh:DP6}{}\enspace{\ttfamily\nolinkurl{ClaimClosure.DP6}} {\tiny\ttfamily DecisionQuotient/\allowbreak ClaimClosure.lean}
\item \textbf{\nolinkurl{DP7}}\hypertarget{lh:DP7}{}\enspace{\ttfamily\nolinkurl{ClaimClosure.DP7}} {\tiny\ttfamily DecisionQuotient/\allowbreak ClaimClosure.lean}
\item \textbf{\nolinkurl{DP8}}\hypertarget{lh:DP8}{}\enspace{\ttfamily\nolinkurl{ClaimClosure.DP8}} {\tiny\ttfamily DecisionQuotient/\allowbreak ClaimClosure.lean}
\item \textbf{\nolinkurl{DS1}}\hypertarget{lh:DS1}{}\enspace{\ttfamily\nolinkurl{ClaimClosure.DS1}} {\tiny\ttfamily DecisionQuotient/\allowbreak ClaimClosure.lean}
\item \textbf{\nolinkurl{DS2}}\hypertarget{lh:DS2}{}\enspace{\ttfamily\nolinkurl{ClaimClosure.DS2}} {\tiny\ttfamily DecisionQuotient/\allowbreak ClaimClosure.lean}
\item \textbf{\nolinkurl{DS3}}\hypertarget{lh:DS3}{}\enspace{\ttfamily\nolinkurl{ClaimClosure.DS3}} {\tiny\ttfamily DecisionQuotient/\allowbreak ClaimClosure.lean}
\item \textbf{\nolinkurl{DS4}}\hypertarget{lh:DS4}{}\enspace{\ttfamily\nolinkurl{ClaimClosure.DS4}} {\tiny\ttfamily DecisionQuotient/\allowbreak ClaimClosure.lean}
\item \textbf{\nolinkurl{DS5}}\hypertarget{lh:DS5}{}\enspace{\ttfamily\nolinkurl{ClaimClosure.DS5}} {\tiny\ttfamily DecisionQuotient/\allowbreak ClaimClosure.lean}
\item \textbf{\nolinkurl{DT6}}\hypertarget{lh:DT6}{}\enspace{\ttfamily\nolinkurl{DecisionQuotient.Physics.DecisionTime.time_is_discrete}} {\tiny\ttfamily DecisionQuotient/\allowbreak Physics/\allowbreak DecisionTime.lean}
\item \textbf{\nolinkurl{DT7}}\hypertarget{lh:DT7}{}\enspace{\ttfamily\nolinkurl{DecisionQuotient.Physics.DecisionTime.time_coordinate_falsifiable}} {\tiny\ttfamily DecisionQuotient/\allowbreak Physics/\allowbreak DecisionTime.lean}
\item \textbf{\nolinkurl{DT10}}\hypertarget{lh:DT10}{}\enspace{\ttfamily\nolinkurl{DecisionQuotient.Physics.DecisionTime.tick_is_decision_event}} {\tiny\ttfamily DecisionQuotient/\allowbreak Physics/\allowbreak DecisionTime.lean}
\item \textbf{\nolinkurl{DT11}}\hypertarget{lh:DT11}{}\enspace{\ttfamily\nolinkurl{DecisionQuotient.Physics.DecisionTime.decision_event_implies_time_unit}} {\tiny\ttfamily DecisionQuotient/\allowbreak Physics/\allowbreak DecisionTime.lean}
\item \textbf{\nolinkurl{DT12}}\hypertarget{lh:DT12}{}\enspace{\ttfamily\nolinkurl{DecisionQuotient.Physics.DecisionTime.decision_taking_place_is_unit_of_time}} {\tiny\ttfamily DecisionQuotient/\allowbreak Physics/\allowbreak DecisionTime.lean}
\item \textbf{\nolinkurl{DT13}}\hypertarget{lh:DT13}{}\enspace{\ttfamily\nolinkurl{DecisionQuotient.Physics.DecisionTime.decision_event_iff_eq_tick}} {\tiny\ttfamily DecisionQuotient/\allowbreak Physics/\allowbreak DecisionTime.lean}
\item \textbf{\nolinkurl{DT15}}\hypertarget{lh:DT15}{}\enspace{\ttfamily\nolinkurl{DecisionQuotient.Physics.DecisionTime.run_time_exact}} {\tiny\ttfamily DecisionQuotient/\allowbreak Physics/\allowbreak DecisionTime.lean}
\item \textbf{\nolinkurl{DT16}}\hypertarget{lh:DT16}{}\enspace{\ttfamily\nolinkurl{DecisionQuotient.Physics.DecisionTime.run_elapsed_time_eq_ticks}} {\tiny\ttfamily DecisionQuotient/\allowbreak Physics/\allowbreak DecisionTime.lean}
\item \textbf{\nolinkurl{DT18}}\hypertarget{lh:DT18}{}\enspace{\ttfamily\nolinkurl{DecisionQuotient.Physics.DecisionTime.decisionTrace_length_eq_ticks}} {\tiny\ttfamily DecisionQuotient/\allowbreak Physics/\allowbreak DecisionTime.lean}
\item \textbf{\nolinkurl{DT19}}\hypertarget{lh:DT19}{}\enspace{\ttfamily\nolinkurl{DecisionQuotient.Physics.DecisionTime.decision_count_equals_elapsed_time}} {\tiny\ttfamily DecisionQuotient/\allowbreak Physics/\allowbreak DecisionTime.lean}
\item \textbf{\nolinkurl{DT22}}\hypertarget{lh:DT22}{}\enspace{\ttfamily\nolinkurl{Physics.DecisionTime.substrate_step_realizes_decision_event}} {\tiny\ttfamily DecisionQuotient/\allowbreak Physics/\allowbreak DecisionTime.lean}
\item \textbf{\nolinkurl{DT23}}\hypertarget{lh:DT23}{}\enspace{\ttfamily\nolinkurl{Physics.DecisionTime.substrate_step_is_time_unit}} {\tiny\ttfamily DecisionQuotient/\allowbreak Physics/\allowbreak DecisionTime.lean}
\item \textbf{\nolinkurl{DT24}}\hypertarget{lh:DT24}{}\enspace{\ttfamily\nolinkurl{Physics.DecisionTime.time_unit_law_substrate_invariant}} {\tiny\ttfamily DecisionQuotient/\allowbreak Physics/\allowbreak DecisionTime.lean}
\item \textbf{\nolinkurl{EI1}}\hypertarget{lh:EI1}{}\enspace{\ttfamily\nolinkurl{ThermodynamicLift.energy_ge_kbt_nat_entropy}} {\tiny\ttfamily DecisionQuotient/\allowbreak ThermodynamicLift.lean}
\item \textbf{\nolinkurl{FI3}}\hypertarget{lh:FI3}{}\enspace{\ttfamily\nolinkurl{FunctionalInformation.functional_information_from_counting}} {\tiny\ttfamily DecisionQuotient/\allowbreak FunctionalInformation.lean}
\item \textbf{\nolinkurl{FI6}}\hypertarget{lh:FI6}{}\enspace{\ttfamily\nolinkurl{FunctionalInformation.functional_information_from_thermodynamics}} {\tiny\ttfamily DecisionQuotient/\allowbreak FunctionalInformation.lean}
\item \textbf{\nolinkurl{FI7}}\hypertarget{lh:FI7}{}\enspace{\ttfamily\nolinkurl{FunctionalInformation.first_principles_thermo_coincide}} {\tiny\ttfamily DecisionQuotient/\allowbreak FunctionalInformation.lean}
\item \textbf{\nolinkurl{FN7}}\hypertarget{lh:FN7}{}\enspace{\ttfamily\nolinkurl{BayesOptimalityProof.KL_nonneg}} {\tiny\ttfamily DecisionQuotient/\allowbreak BayesOptimalityProof.lean}
\item \textbf{\nolinkurl{FN12}}\hypertarget{lh:FN12}{}\enspace{\ttfamily\nolinkurl{BayesOptimalityProof.crossEntropy_eq_entropy_add_KL}} {\tiny\ttfamily DecisionQuotient/\allowbreak BayesOptimalityProof.lean}
\item \textbf{\nolinkurl{FN14}}\hypertarget{lh:FN14}{}\enspace{\ttfamily\nolinkurl{BayesOptimalityProof.bayes_is_optimal}} {\tiny\ttfamily DecisionQuotient/\allowbreak BayesOptimalityProof.lean}
\item \textbf{\nolinkurl{FP1}}\hypertarget{lh:FP1}{}\enspace{\ttfamily\nolinkurl{Physics.LocalityPhysics.trivial_states_all_equal}} {\tiny\ttfamily DecisionQuotient/\allowbreak Physics/\allowbreak LocalityPhysics.lean}
\item \textbf{\nolinkurl{FP2}}\hypertarget{lh:FP2}{}\enspace{\ttfamily\nolinkurl{Physics.LocalityPhysics.equal_states_constant_function}} {\tiny\ttfamily DecisionQuotient/\allowbreak Physics/\allowbreak LocalityPhysics.lean}
\item \textbf{\nolinkurl{FP3}}\hypertarget{lh:FP3}{}\enspace{\ttfamily\nolinkurl{Physics.LocalityPhysics.constant_function_singleton_image}} {\tiny\ttfamily DecisionQuotient/\allowbreak Physics/\allowbreak LocalityPhysics.lean}
\item \textbf{\nolinkurl{FP4}}\hypertarget{lh:FP4}{}\enspace{\ttfamily\nolinkurl{Physics.LocalityPhysics.singleton_image_zero_entropy}} {\tiny\ttfamily DecisionQuotient/\allowbreak Physics/\allowbreak LocalityPhysics.lean}
\item \textbf{\nolinkurl{FP5}}\hypertarget{lh:FP5}{}\enspace{\ttfamily\nolinkurl{Physics.LocalityPhysics.zero_entropy_no_information}} {\tiny\ttfamily DecisionQuotient/\allowbreak Physics/\allowbreak LocalityPhysics.lean}
\item \textbf{\nolinkurl{FP6}}\hypertarget{lh:FP6}{}\enspace{\ttfamily\nolinkurl{Physics.LocalityPhysics.triviality_implies_no_information}} {\tiny\ttfamily DecisionQuotient/\allowbreak Physics/\allowbreak LocalityPhysics.lean}
\item \textbf{\nolinkurl{FP7}}\hypertarget{lh:FP7}{}\enspace{\ttfamily\nolinkurl{Physics.LocalityPhysics.information_requires_nontriviality}} {\tiny\ttfamily DecisionQuotient/\allowbreak Physics/\allowbreak LocalityPhysics.lean}
\item \textbf{\nolinkurl{FP8}}\hypertarget{lh:FP8}{}\enspace{\ttfamily\nolinkurl{Physics.LocalityPhysics.atypical_states_rare}} {\tiny\ttfamily DecisionQuotient/\allowbreak Physics/\allowbreak LocalityPhysics.lean}
\item \textbf{\nolinkurl{FP9}}\hypertarget{lh:FP9}{}\enspace{\ttfamily\nolinkurl{Physics.LocalityPhysics.random_misses_target}} {\tiny\ttfamily DecisionQuotient/\allowbreak Physics/\allowbreak LocalityPhysics.lean}
\item \textbf{\nolinkurl{FP10}}\hypertarget{lh:FP10}{}\enspace{\ttfamily\nolinkurl{Physics.LocalityPhysics.errors_accumulate}} {\tiny\ttfamily DecisionQuotient/\allowbreak Physics/\allowbreak LocalityPhysics.lean}
\item \textbf{\nolinkurl{FP11}}\hypertarget{lh:FP11}{}\enspace{\ttfamily\nolinkurl{Physics.LocalityPhysics.wrong_paths_dominate}} {\tiny\ttfamily DecisionQuotient/\allowbreak Physics/\allowbreak LocalityPhysics.lean}
\item \textbf{\nolinkurl{FP12}}\hypertarget{lh:FP12}{}\enspace{\ttfamily\nolinkurl{Physics.LocalityPhysics.second_law_from_counting}} {\tiny\ttfamily DecisionQuotient/\allowbreak Physics/\allowbreak LocalityPhysics.lean}
\item \textbf{\nolinkurl{FP13}}\hypertarget{lh:FP13}{}\enspace{\ttfamily\nolinkurl{Physics.LocalityPhysics.verification_is_information}} {\tiny\ttfamily DecisionQuotient/\allowbreak Physics/\allowbreak LocalityPhysics.lean}
\item \textbf{\nolinkurl{FP14}}\hypertarget{lh:FP14}{}\enspace{\ttfamily\nolinkurl{Physics.LocalityPhysics.entropy_is_information}} {\tiny\ttfamily DecisionQuotient/\allowbreak Physics/\allowbreak LocalityPhysics.lean}
\item \textbf{\nolinkurl{FP15}}\hypertarget{lh:FP15}{}\enspace{\ttfamily\nolinkurl{Physics.LocalityPhysics.landauer_structure}} {\tiny\ttfamily DecisionQuotient/\allowbreak Physics/\allowbreak LocalityPhysics.lean}
\item \textbf{\nolinkurl{FPT4}}\hypertarget{lh:FPT4}{}\enspace{\ttfamily\nolinkurl{Physics.LocalityPhysics.FPT4_step_requires_distinct_moments}} {\tiny\ttfamily DecisionQuotient/\allowbreak Physics/\allowbreak LocalityPhysics.lean}
\item \textbf{\nolinkurl{FPT5}}\hypertarget{lh:FPT5}{}\enspace{\ttfamily\nolinkurl{Physics.LocalityPhysics.FPT5_distinct_moments_positive_duration}} {\tiny\ttfamily DecisionQuotient/\allowbreak Physics/\allowbreak LocalityPhysics.lean}
\item \textbf{\nolinkurl{FPT6}}\hypertarget{lh:FPT6}{}\enspace{\ttfamily\nolinkurl{Physics.LocalityPhysics.FPT6_step_takes_positive_time}} {\tiny\ttfamily DecisionQuotient/\allowbreak Physics/\allowbreak LocalityPhysics.lean}
\item \textbf{\nolinkurl{FPT8}}\hypertarget{lh:FPT8}{}\enspace{\ttfamily\nolinkurl{Physics.LocalityPhysics.FPT8_propagation_takes_time}} {\tiny\ttfamily DecisionQuotient/\allowbreak Physics/\allowbreak LocalityPhysics.lean}
\item \textbf{\nolinkurl{FPT10}}\hypertarget{lh:FPT10}{}\enspace{\ttfamily\nolinkurl{Physics.LocalityPhysics.FPT10_ec3_is_logical}} {\tiny\ttfamily DecisionQuotient/\allowbreak Physics/\allowbreak LocalityPhysics.lean}
\item \textbf{\nolinkurl{FS1}}\hypertarget{lh:FS1}{}\enspace{\ttfamily\nolinkurl{Statistics.sum_fisherScore_eq_srank}} {\tiny\ttfamily DecisionQuotient/\allowbreak Statistics/\allowbreak FisherInformation.lean}
\item \textbf{\nolinkurl{FS2}}\hypertarget{lh:FS2}{}\enspace{\ttfamily\nolinkurl{Statistics.fisherMatrix_rank_eq_srank}} {\tiny\ttfamily DecisionQuotient/\allowbreak Statistics/\allowbreak FisherInformation.lean}
\item \textbf{\nolinkurl{GI3}}\hypertarget{lh:GI3}{}\enspace{\ttfamily\nolinkurl{Physics.GlobalInfinityBudget.global_resource_infinity_not_entails_infinite_computation}} {\tiny\ttfamily DecisionQuotient/\allowbreak Physics/\allowbreak GlobalInfinityBudget.lean}
\item \textbf{\nolinkurl{GI4}}\hypertarget{lh:GI4}{}\enspace{\ttfamily\nolinkurl{Physics.GlobalInfinityBudget.eternal_inflation_causal_past_bounded_no_infinite_budget}} {\tiny\ttfamily DecisionQuotient/\allowbreak Physics/\allowbreak GlobalInfinityBudget.lean}
\item \textbf{\nolinkurl{GI5}}\hypertarget{lh:GI5}{}\enspace{\ttfamily\nolinkurl{Physics.GlobalInfinityBudget.eternal_inflation_global_plus_causal_bound_no_infinite_budget}} {\tiny\ttfamily DecisionQuotient/\allowbreak Physics/\allowbreak GlobalInfinityBudget.lean}
\item \textbf{\nolinkurl{GI6}}\hypertarget{lh:GI6}{}\enspace{\ttfamily\nolinkurl{Physics.GlobalInfinityBudget.eternal_inflation_global_infinity_not_entails_infinite_computation}} {\tiny\ttfamily DecisionQuotient/\allowbreak Physics/\allowbreak GlobalInfinityBudget.lean}
\item \textbf{\nolinkurl{GI7}}\hypertarget{lh:GI7}{}\enspace{\ttfamily\nolinkurl{Physics.GlobalInfinityBudget.spatial_infinity_hubble_bounded_no_infinite_budget}} {\tiny\ttfamily DecisionQuotient/\allowbreak Physics/\allowbreak GlobalInfinityBudget.lean}
\item \textbf{\nolinkurl{GI8}}\hypertarget{lh:GI8}{}\enspace{\ttfamily\nolinkurl{Physics.GlobalInfinityBudget.spatial_infinity_global_infinity_not_entails_infinite_computation}} {\tiny\ttfamily DecisionQuotient/\allowbreak Physics/\allowbreak GlobalInfinityBudget.lean}
\item \textbf{\nolinkurl{GI9}}\hypertarget{lh:GI9}{}\enspace{\ttfamily\nolinkurl{Physics.GlobalInfinityBudget.qft_finite_preparation_no_infinite_access}} {\tiny\ttfamily DecisionQuotient/\allowbreak Physics/\allowbreak GlobalInfinityBudget.lean}
\item \textbf{\nolinkurl{GI10}}\hypertarget{lh:GI10}{}\enspace{\ttfamily\nolinkurl{Physics.GlobalInfinityBudget.qft_infinite_hilbert_not_entails_infinite_access}} {\tiny\ttfamily DecisionQuotient/\allowbreak Physics/\allowbreak GlobalInfinityBudget.lean}
\item \textbf{\nolinkurl{GI11}}\hypertarget{lh:GI11}{}\enspace{\ttfamily\nolinkurl{Physics.GlobalInfinityBudget.anyonic_finite_realization_no_infinite_particle_budget}} {\tiny\ttfamily DecisionQuotient/\allowbreak Physics/\allowbreak GlobalInfinityBudget.lean}
\item \textbf{\nolinkurl{GI12}}\hypertarget{lh:GI12}{}\enspace{\ttfamily\nolinkurl{Physics.GlobalInfinityBudget.anyonic_infinite_limit_not_entails_infinite_particle_realization}} {\tiny\ttfamily DecisionQuotient/\allowbreak Physics/\allowbreak GlobalInfinityBudget.lean}
\item \textbf{\nolinkurl{GI13}}\hypertarget{lh:GI13}{}\enspace{\ttfamily\nolinkurl{Physics.GlobalInfinityBudget.malament_hogarth_structure_entails_infinite_budget}} {\tiny\ttfamily DecisionQuotient/\allowbreak Physics/\allowbreak GlobalInfinityBudget.lean}
\item \textbf{\nolinkurl{GI14}}\hypertarget{lh:GI14}{}\enspace{\ttfamily\nolinkurl{Physics.GlobalInfinityBudget.malament_hogarth_not_entails_no_infinite_budget}} {\tiny\ttfamily DecisionQuotient/\allowbreak Physics/\allowbreak GlobalInfinityBudget.lean}
\item \textbf{\nolinkurl{GI15}}\hypertarget{lh:GI15}{}\enspace{\ttfamily\nolinkurl{Physics.GlobalInfinityBudget.ctc_structure_entails_infinite_budget}} {\tiny\ttfamily DecisionQuotient/\allowbreak Physics/\allowbreak GlobalInfinityBudget.lean}
\item \textbf{\nolinkurl{GI16}}\hypertarget{lh:GI16}{}\enspace{\ttfamily\nolinkurl{Physics.GlobalInfinityBudget.ctc_not_entails_no_infinite_budget}} {\tiny\ttfamily DecisionQuotient/\allowbreak Physics/\allowbreak GlobalInfinityBudget.lean}
\item \textbf{\nolinkurl{GN2}}\hypertarget{lh:GN2}{}\enspace{\ttfamily\nolinkurl{LogicGraph.cycleWitnessBits_pos}} {\tiny\ttfamily DecisionQuotient/\allowbreak GraphNontriviality.lean}
\item \textbf{\nolinkurl{GN4}}\hypertarget{lh:GN4}{}\enspace{\ttfamily\nolinkurl{LogicGraph.pathSurprisal_nonneg_of_positive_mass}} {\tiny\ttfamily DecisionQuotient/\allowbreak GraphNontriviality.lean}
\item \textbf{\nolinkurl{GN5}}\hypertarget{lh:GN5}{}\enspace{\ttfamily\nolinkurl{LogicGraph.nontrivialityScore_unknown}} {\tiny\ttfamily DecisionQuotient/\allowbreak GraphNontriviality.lean}
\item \textbf{\nolinkurl{GN6}}\hypertarget{lh:GN6}{}\enspace{\ttfamily\nolinkurl{LogicGraph.observerEntropy_nonneg}} {\tiny\ttfamily DecisionQuotient/\allowbreak GraphNontriviality.lean}
\item \textbf{\nolinkurl{GN7}}\hypertarget{lh:GN7}{}\enspace{\ttfamily\nolinkurl{LogicGraph.dqFromEntropy_in_unit_interval}} {\tiny\ttfamily DecisionQuotient/\allowbreak GraphNontriviality.lean}
\item \textbf{\nolinkurl{GN8}}\hypertarget{lh:GN8}{}\enspace{\ttfamily\nolinkurl{LogicGraph.path_belief_forced_under_uncertainty}} {\tiny\ttfamily DecisionQuotient/\allowbreak GraphNontriviality.lean}
\item \textbf{\nolinkurl{GN9}}\hypertarget{lh:GN9}{}\enspace{\ttfamily\nolinkurl{LogicGraph.bayes_update_exists_for_observer_paths}} {\tiny\ttfamily DecisionQuotient/\allowbreak GraphNontriviality.lean}
\item \textbf{\nolinkurl{GN10}}\hypertarget{lh:GN10}{}\enspace{\ttfamily\nolinkurl{LogicGraph.cycle_witness_implies_positive_landauer}} {\tiny\ttfamily DecisionQuotient/\allowbreak GraphNontriviality.lean}
\item \textbf{\nolinkurl{GN11}}\hypertarget{lh:GN11}{}\enspace{\ttfamily\nolinkurl{LogicGraph.cycle_witness_implies_positive_nv_work}} {\tiny\ttfamily DecisionQuotient/\allowbreak GraphNontriviality.lean}
\item \textbf{\nolinkurl{GN12}}\hypertarget{lh:GN12}{}\enspace{\ttfamily\nolinkurl{LogicGraph.dna_erasure_implies_positive_landauer}} {\tiny\ttfamily DecisionQuotient/\allowbreak GraphNontriviality.lean}
\item \textbf{\nolinkurl{GN13}}\hypertarget{lh:GN13}{}\enspace{\ttfamily\nolinkurl{LogicGraph.dna_room_temp_environmental_stability}} {\tiny\ttfamily DecisionQuotient/\allowbreak GraphNontriviality.lean}
\item \textbf{\nolinkurl{HD17}}\hypertarget{lh:HD17}{}\enspace{\ttfamily\nolinkurl{DecisionQuotient.HardnessDistribution.requiredWork}} {\tiny\ttfamily DecisionQuotient/\allowbreak HardnessDistribution.lean}
\item \textbf{\nolinkurl{HD18}}\hypertarget{lh:HD18}{}\enspace{\ttfamily\nolinkurl{DecisionQuotient.HardnessDistribution.requiredWork_eq_affine_in_sites}} {\tiny\ttfamily DecisionQuotient/\allowbreak HardnessDistribution.lean}
\item \textbf{\nolinkurl{HD22}}\hypertarget{lh:HD22}{}\enspace{\ttfamily\nolinkurl{DecisionQuotient.HardnessDistribution.hardnessLowerBound}} {\tiny\ttfamily DecisionQuotient/\allowbreak HardnessDistribution.lean}
\item \textbf{\nolinkurl{HD23}}\hypertarget{lh:HD23}{}\enspace{\ttfamily\nolinkurl{DecisionQuotient.HardnessDistribution.hardness_is_irreducible_required_work}} {\tiny\ttfamily DecisionQuotient/\allowbreak HardnessDistribution.lean}
\item \textbf{\nolinkurl{HD27}}\hypertarget{lh:HD27}{}\enspace{\ttfamily\nolinkurl{DecisionQuotient.HardnessDistribution.workGrowthDegree}} {\tiny\ttfamily DecisionQuotient/\allowbreak HardnessDistribution.lean}
\item \textbf{\nolinkurl{HD28}}\hypertarget{lh:HD28}{}\enspace{\ttfamily\nolinkurl{DecisionQuotient.HardnessDistribution.workGrowthDegree_zero_iff_eventually_constant}} {\tiny\ttfamily DecisionQuotient/\allowbreak HardnessDistribution.lean}
\item \textbf{\nolinkurl{HS3}}\hypertarget{lh:HS3}{}\enspace{\ttfamily\nolinkurl{DecisionQuotient.Physics.HeisenbergStrong.strong_binding_implies_core_nontrivial}} {\tiny\ttfamily DecisionQuotient/\allowbreak Physics/\allowbreak HeisenbergStrong.lean}
\item \textbf{\nolinkurl{HS5}}\hypertarget{lh:HS5}{}\enspace{\ttfamily\nolinkurl{DecisionQuotient.Physics.HeisenbergStrong.strong_binding_implies_physical_nontrivial_opt_assumption}} {\tiny\ttfamily DecisionQuotient/\allowbreak Physics/\allowbreak HeisenbergStrong.lean}
\item \textbf{\nolinkurl{HS6}}\hypertarget{lh:HS6}{}\enspace{\ttfamily\nolinkurl{DecisionQuotient.Physics.HeisenbergStrong.strong_binding_implies_nontrivial_opt_via_uncertainty}} {\tiny\ttfamily DecisionQuotient/\allowbreak Physics/\allowbreak HeisenbergStrong.lean}
\item \textbf{\nolinkurl{IA7}}\hypertarget{lh:IA7}{}\enspace{\ttfamily\nolinkurl{ClaimClosure.IA7}} {\tiny\ttfamily DecisionQuotient/\allowbreak ClaimClosure.lean}
\item \textbf{\nolinkurl{IA9}}\hypertarget{lh:IA9}{}\enspace{\ttfamily\nolinkurl{ClaimClosure.IA9}} {\tiny\ttfamily DecisionQuotient/\allowbreak ClaimClosure.lean}
\item \textbf{\nolinkurl{IA11}}\hypertarget{lh:IA11}{}\enspace{\ttfamily\nolinkurl{ClaimClosure.IA11}} {\tiny\ttfamily DecisionQuotient/\allowbreak ClaimClosure.lean}
\item \textbf{\nolinkurl{IA12}}\hypertarget{lh:IA12}{}\enspace{\ttfamily\nolinkurl{ClaimClosure.IA12}} {\tiny\ttfamily DecisionQuotient/\allowbreak ClaimClosure.lean}
\item \textbf{\nolinkurl{IA13}}\hypertarget{lh:IA13}{}\enspace{\ttfamily\nolinkurl{ClaimClosure.IA13}} {\tiny\ttfamily DecisionQuotient/\allowbreak ClaimClosure.lean}
\item \textbf{\nolinkurl{IC1}}\hypertarget{lh:IC1}{}\enspace{\ttfamily\nolinkurl{DecisionQuotient.IntegrityCompetence.CertaintyInflation}} {\tiny\ttfamily DecisionQuotient/\allowbreak IntegrityCompetence.lean}
\item \textbf{\nolinkurl{IC2}}\hypertarget{lh:IC2}{}\enspace{\ttfamily\nolinkurl{DecisionQuotient.IntegrityCompetence.CompletionFractionDefined}} {\tiny\ttfamily DecisionQuotient/\allowbreak IntegrityCompetence.lean}
\item \textbf{\nolinkurl{IC3}}\hypertarget{lh:IC3}{}\enspace{\ttfamily\nolinkurl{DecisionQuotient.IntegrityCompetence.EvidenceForReport}}
\item \textbf{\nolinkurl{IC4}}\hypertarget{lh:IC4}{}\enspace{\ttfamily\nolinkurl{DecisionQuotient.IntegrityCompetence.ExactCertaintyInflation}} {\tiny\ttfamily DecisionQuotient/\allowbreak IntegrityCompetence.lean}
\item \textbf{\nolinkurl{IC5}}\hypertarget{lh:IC5}{}\enspace{\ttfamily\nolinkurl{DecisionQuotient.IntegrityCompetence.Percent}} {\tiny\ttfamily DecisionQuotient/\allowbreak IntegrityCompetence.lean}
\item \textbf{\nolinkurl{IC6}}\hypertarget{lh:IC6}{}\enspace{\ttfamily\nolinkurl{DecisionQuotient.IntegrityCompetence.RLFFWeights}} {\tiny\ttfamily DecisionQuotient/\allowbreak IntegrityCompetence.lean}
\item \textbf{\nolinkurl{IC7}}\hypertarget{lh:IC7}{}\enspace{\ttfamily\nolinkurl{DecisionQuotient.IntegrityCompetence.ReportSignal}} {\tiny\ttfamily DecisionQuotient/\allowbreak IntegrityCompetence.lean}
\item \textbf{\nolinkurl{IC8}}\hypertarget{lh:IC8}{}\enspace{\ttfamily\nolinkurl{DecisionQuotient.IntegrityCompetence.ReportBitModel}} {\tiny\ttfamily DecisionQuotient/\allowbreak IntegrityCompetence.lean}
\item \textbf{\nolinkurl{IC9}}\hypertarget{lh:IC9}{}\enspace{\ttfamily\nolinkurl{DecisionQuotient.IntegrityCompetence.SignalConsistent}} {\tiny\ttfamily DecisionQuotient/\allowbreak IntegrityCompetence.lean}
\item \textbf{\nolinkurl{IC10}}\hypertarget{lh:IC10}{}\enspace{\ttfamily\nolinkurl{DecisionQuotient.IntegrityCompetence.admissible_irrational_strictly_more_than_rational}} {\tiny\ttfamily DecisionQuotient/\allowbreak IntegrityCompetence.lean}
\item \textbf{\nolinkurl{IC11}}\hypertarget{lh:IC11}{}\enspace{\ttfamily\nolinkurl{DecisionQuotient.IntegrityCompetence.admissible_matrix_counts}} {\tiny\ttfamily DecisionQuotient/\allowbreak IntegrityCompetence.lean}
\item \textbf{\nolinkurl{IC12}}\hypertarget{lh:IC12}{}\enspace{\ttfamily\nolinkurl{DecisionQuotient.IntegrityCompetence.abstain_signal_exists_with_guess_self}} {\tiny\ttfamily DecisionQuotient/\allowbreak IntegrityCompetence.lean}
\item \textbf{\nolinkurl{IC13}}\hypertarget{lh:IC13}{}\enspace{\ttfamily\nolinkurl{DecisionQuotient.IntegrityCompetence.certaintyInflation_iff_not_admissible}} {\tiny\ttfamily DecisionQuotient/\allowbreak IntegrityCompetence.lean}
\item \textbf{\nolinkurl{IC14}}\hypertarget{lh:IC14}{}\enspace{\ttfamily\nolinkurl{DecisionQuotient.IntegrityCompetence.certificationOverheadBits}} {\tiny\ttfamily DecisionQuotient/\allowbreak IntegrityCompetence.lean}
\item \textbf{\nolinkurl{IC15}}\hypertarget{lh:IC15}{}\enspace{\ttfamily\nolinkurl{DecisionQuotient.IntegrityCompetence.certificationOverheadBits_of_evidence}} {\tiny\ttfamily DecisionQuotient/\allowbreak IntegrityCompetence.lean}
\item \textbf{\nolinkurl{IC16}}\hypertarget{lh:IC16}{}\enspace{\ttfamily\nolinkurl{DecisionQuotient.IntegrityCompetence.certificationOverheadBits_of_no_evidence}} {\tiny\ttfamily DecisionQuotient/\allowbreak IntegrityCompetence.lean}
\item \textbf{\nolinkurl{IC17}}\hypertarget{lh:IC17}{}\enspace{\ttfamily\nolinkurl{DecisionQuotient.IntegrityCompetence.certifiedTotalBits}} {\tiny\ttfamily DecisionQuotient/\allowbreak IntegrityCompetence.lean}
\item \textbf{\nolinkurl{IC18}}\hypertarget{lh:IC18}{}\enspace{\ttfamily\nolinkurl{DecisionQuotient.IntegrityCompetence.certifiedTotalBits_ge_raw}} {\tiny\ttfamily DecisionQuotient/\allowbreak IntegrityCompetence.lean}
\item \textbf{\nolinkurl{IC19}}\hypertarget{lh:IC19}{}\enspace{\ttfamily\nolinkurl{DecisionQuotient.IntegrityCompetence.certifiedTotalBits_gt_raw_of_evidence}} {\tiny\ttfamily DecisionQuotient/\allowbreak IntegrityCompetence.lean}
\item \textbf{\nolinkurl{IC20}}\hypertarget{lh:IC20}{}\enspace{\ttfamily\nolinkurl{DecisionQuotient.IntegrityCompetence.certifiedTotalBits_of_evidence}} {\tiny\ttfamily DecisionQuotient/\allowbreak IntegrityCompetence.lean}
\item \textbf{\nolinkurl{IC21}}\hypertarget{lh:IC21}{}\enspace{\ttfamily\nolinkurl{DecisionQuotient.IntegrityCompetence.certifiedTotalBits_of_no_evidence}} {\tiny\ttfamily DecisionQuotient/\allowbreak IntegrityCompetence.lean}
\item \textbf{\nolinkurl{IC22}}\hypertarget{lh:IC22}{}\enspace{\ttfamily\nolinkurl{DecisionQuotient.IntegrityCompetence.claim_admissible_of_evidence}} {\tiny\ttfamily DecisionQuotient/\allowbreak IntegrityCompetence.lean}
\item \textbf{\nolinkurl{IC23}}\hypertarget{lh:IC23}{}\enspace{\ttfamily\nolinkurl{DecisionQuotient.IntegrityCompetence.competence_implies_integrity}} {\tiny\ttfamily DecisionQuotient/\allowbreak IntegrityCompetence.lean}
\item \textbf{\nolinkurl{IC24}}\hypertarget{lh:IC24}{}\enspace{\ttfamily\nolinkurl{DecisionQuotient.IntegrityCompetence.completion_fraction_defined_of_declared_bound}} {\tiny\ttfamily DecisionQuotient/\allowbreak IntegrityCompetence.lean}
\item \textbf{\nolinkurl{IC25}}\hypertarget{lh:IC25}{}\enspace{\ttfamily\nolinkurl{DecisionQuotient.IntegrityCompetence.epsilon_competence_implies_integrity}} {\tiny\ttfamily DecisionQuotient/\allowbreak IntegrityCompetence.lean}
\item \textbf{\nolinkurl{IC26}}\hypertarget{lh:IC26}{}\enspace{\ttfamily\nolinkurl{DecisionQuotient.IntegrityCompetence.evidence_nonempty_iff_claim_admissible}} {\tiny\ttfamily DecisionQuotient/\allowbreak IntegrityCompetence.lean}
\item \textbf{\nolinkurl{IC27}}\hypertarget{lh:IC27}{}\enspace{\ttfamily\nolinkurl{DecisionQuotient.IntegrityCompetence.evidence_of_claim_admissible}} {\tiny\ttfamily DecisionQuotient/\allowbreak IntegrityCompetence.lean}
\item \textbf{\nolinkurl{IC28}}\hypertarget{lh:IC28}{}\enspace{\ttfamily\nolinkurl{DecisionQuotient.IntegrityCompetence.exact_claim_admissible_iff_exact_evidence_nonempty}} {\tiny\ttfamily DecisionQuotient/\allowbreak IntegrityCompetence.lean}
\item \textbf{\nolinkurl{IC29}}\hypertarget{lh:IC29}{}\enspace{\ttfamily\nolinkurl{DecisionQuotient.IntegrityCompetence.exact_claim_requires_evidence}} {\tiny\ttfamily DecisionQuotient/\allowbreak IntegrityCompetence.lean}
\item \textbf{\nolinkurl{IC30}}\hypertarget{lh:IC30}{}\enspace{\ttfamily\nolinkurl{DecisionQuotient.IntegrityCompetence.exactCertaintyInflation_iff_no_exact_competence}} {\tiny\ttfamily DecisionQuotient/\allowbreak IntegrityCompetence.lean}
\item \textbf{\nolinkurl{IC31}}\hypertarget{lh:IC31}{}\enspace{\ttfamily\nolinkurl{DecisionQuotient.IntegrityCompetence.exact_raw_only_of_no_exact_admissible}} {\tiny\ttfamily DecisionQuotient/\allowbreak IntegrityCompetence.lean}
\item \textbf{\nolinkurl{IC32}}\hypertarget{lh:IC32}{}\enspace{\ttfamily\nolinkurl{DecisionQuotient.IntegrityCompetence.integrity_forces_abstention}} {\tiny\ttfamily DecisionQuotient/\allowbreak IntegrityCompetence.lean}
\item \textbf{\nolinkurl{IC33}}\hypertarget{lh:IC33}{}\enspace{\ttfamily\nolinkurl{DecisionQuotient.IntegrityCompetence.integrity_not_competent_of_nonempty_scope}} {\tiny\ttfamily DecisionQuotient/\allowbreak IntegrityCompetence.lean}
\item \textbf{\nolinkurl{IC34}}\hypertarget{lh:IC34}{}\enspace{\ttfamily\nolinkurl{DecisionQuotient.IntegrityCompetence.integrity_resource_bound}} {\tiny\ttfamily DecisionQuotient/\allowbreak IntegrityCompetence.lean}
\item \textbf{\nolinkurl{IC35}}\hypertarget{lh:IC35}{}\enspace{\ttfamily\nolinkurl{DecisionQuotient.IntegrityCompetence.no_completion_fraction_without_declared_bound}} {\tiny\ttfamily DecisionQuotient/\allowbreak IntegrityCompetence.lean}
\item \textbf{\nolinkurl{IC36}}\hypertarget{lh:IC36}{}\enspace{\ttfamily\nolinkurl{DecisionQuotient.IntegrityCompetence.overModelVerdict_rational_iff}} {\tiny\ttfamily DecisionQuotient/\allowbreak IntegrityCompetence.lean}
\item \textbf{\nolinkurl{IC37}}\hypertarget{lh:IC37}{}\enspace{\ttfamily\nolinkurl{DecisionQuotient.IntegrityCompetence.percentZero}} {\tiny\ttfamily DecisionQuotient/\allowbreak IntegrityCompetence.lean}
\item \textbf{\nolinkurl{IC38}}\hypertarget{lh:IC38}{}\enspace{\ttfamily\nolinkurl{DecisionQuotient.IntegrityCompetence.rlffBaseReward}} {\tiny\ttfamily DecisionQuotient/\allowbreak IntegrityCompetence.lean}
\item \textbf{\nolinkurl{IC39}}\hypertarget{lh:IC39}{}\enspace{\ttfamily\nolinkurl{DecisionQuotient.IntegrityCompetence.rlffReward}} {\tiny\ttfamily DecisionQuotient/\allowbreak IntegrityCompetence.lean}
\item \textbf{\nolinkurl{IC40}}\hypertarget{lh:IC40}{}\enspace{\ttfamily\nolinkurl{DecisionQuotient.IntegrityCompetence.rlff_abstain_strictly_prefers_no_certificates}} {\tiny\ttfamily DecisionQuotient/\allowbreak IntegrityCompetence.lean}
\item \textbf{\nolinkurl{IC41}}\hypertarget{lh:IC41}{}\enspace{\ttfamily\nolinkurl{DecisionQuotient.IntegrityCompetence.rlff_maximizer_has_evidence}} {\tiny\ttfamily DecisionQuotient/\allowbreak IntegrityCompetence.lean}
\item \textbf{\nolinkurl{IC42}}\hypertarget{lh:IC42}{}\enspace{\ttfamily\nolinkurl{DecisionQuotient.IntegrityCompetence.rlff_maximizer_is_admissible}} {\tiny\ttfamily DecisionQuotient/\allowbreak IntegrityCompetence.lean}
\item \textbf{\nolinkurl{IC43}}\hypertarget{lh:IC43}{}\enspace{\ttfamily\nolinkurl{DecisionQuotient.IntegrityCompetence.self_reflected_confidence_not_certification}} {\tiny\ttfamily DecisionQuotient/\allowbreak IntegrityCompetence.lean}
\item \textbf{\nolinkurl{IC44}}\hypertarget{lh:IC44}{}\enspace{\ttfamily\nolinkurl{DecisionQuotient.IntegrityCompetence.signal_certified_positive_implies_admissible}} {\tiny\ttfamily DecisionQuotient/\allowbreak IntegrityCompetence.lean}
\item \textbf{\nolinkurl{IC45}}\hypertarget{lh:IC45}{}\enspace{\ttfamily\nolinkurl{DecisionQuotient.IntegrityCompetence.signal_consistent_of_claim_admissible}} {\tiny\ttfamily DecisionQuotient/\allowbreak IntegrityCompetence.lean}
\item \textbf{\nolinkurl{IC46}}\hypertarget{lh:IC46}{}\enspace{\ttfamily\nolinkurl{DecisionQuotient.IntegrityCompetence.signal_no_evidence_forces_zero_certified}} {\tiny\ttfamily DecisionQuotient/\allowbreak IntegrityCompetence.lean}
\item \textbf{\nolinkurl{IC47}}\hypertarget{lh:IC47}{}\enspace{\ttfamily\nolinkurl{DecisionQuotient.IntegrityCompetence.signal_exact_no_competence_forces_zero_certified}} {\tiny\ttfamily DecisionQuotient/\allowbreak IntegrityCompetence.lean}
\item \textbf{\nolinkurl{IT1}}\hypertarget{lh:IT1}{}\enspace{\ttfamily\nolinkurl{DecisionProblem.numOptClasses}} {\tiny\ttfamily DecisionQuotient/\allowbreak Information.lean}
\item \textbf{\nolinkurl{IT2}}\hypertarget{lh:IT2}{}\enspace{\ttfamily\nolinkurl{DecisionProblem.quotientEntropy}} {\tiny\ttfamily DecisionQuotient/\allowbreak Information.lean}
\item \textbf{\nolinkurl{IT3}}\hypertarget{lh:IT3}{}\enspace{\ttfamily\nolinkurl{DecisionQuotient.quotientEntropy_le_srank_binary}} {\tiny\ttfamily DecisionQuotient/\allowbreak Information.lean}
\item \textbf{\nolinkurl{IT4}}\hypertarget{lh:IT4}{}\enspace{\ttfamily\nolinkurl{DecisionQuotient.numOptClasses_le_pow_srank_binary}} {\tiny\ttfamily DecisionQuotient/\allowbreak Information.lean}
\item \textbf{\nolinkurl{IV1}}\hypertarget{lh:IV1}{}\enspace{\ttfamily\nolinkurl{DecisionQuotient.InteriorVerification.GoalClass}} {\tiny\ttfamily DecisionQuotient/\allowbreak InteriorVerification.lean}
\item \textbf{\nolinkurl{IV2}}\hypertarget{lh:IV2}{}\enspace{\ttfamily\nolinkurl{DecisionQuotient.InteriorVerification.InteriorDominanceVerifiable}} {\tiny\ttfamily DecisionQuotient/\allowbreak InteriorVerification.lean}
\item \textbf{\nolinkurl{IV3}}\hypertarget{lh:IV3}{}\enspace{\ttfamily\nolinkurl{DecisionQuotient.InteriorVerification.TautologicalSetIdentifiable}} {\tiny\ttfamily DecisionQuotient/\allowbreak InteriorVerification.lean}
\item \textbf{\nolinkurl{IV4}}\hypertarget{lh:IV4}{}\enspace{\ttfamily\nolinkurl{DecisionQuotient.InteriorVerification.agreeOnSet}} {\tiny\ttfamily DecisionQuotient/\allowbreak InteriorVerification.lean}
\item \textbf{\nolinkurl{IV5}}\hypertarget{lh:IV5}{}\enspace{\ttfamily\nolinkurl{DecisionQuotient.InteriorVerification.interiorParetoDominates}} {\tiny\ttfamily DecisionQuotient/\allowbreak InteriorVerification.lean}
\item \textbf{\nolinkurl{IV6}}\hypertarget{lh:IV6}{}\enspace{\ttfamily\nolinkurl{DecisionQuotient.InteriorVerification.interior_certificate_implies_non_rejection}} {\tiny\ttfamily DecisionQuotient/\allowbreak InteriorVerification.lean}
\item \textbf{\nolinkurl{IV7}}\hypertarget{lh:IV7}{}\enspace{\ttfamily\nolinkurl{DecisionQuotient.InteriorVerification.interior_dominance_implies_universal_non_rejection}} {\tiny\ttfamily DecisionQuotient/\allowbreak InteriorVerification.lean}
\item \textbf{\nolinkurl{IV8}}\hypertarget{lh:IV8}{}\enspace{\ttfamily\nolinkurl{DecisionQuotient.InteriorVerification.interior_dominance_not_full_sufficiency}} {\tiny\ttfamily DecisionQuotient/\allowbreak InteriorVerification.lean}
\item \textbf{\nolinkurl{IV9}}\hypertarget{lh:IV9}{}\enspace{\ttfamily\nolinkurl{DecisionQuotient.InteriorVerification.interior_verification_tractable_certificate}} {\tiny\ttfamily DecisionQuotient/\allowbreak InteriorVerification.lean}
\item \textbf{\nolinkurl{MN1}}\hypertarget{lh:MN1}{}\enspace{\ttfamily\nolinkurl{Physics.MeasureNecessity.quantitative_claim_has_measure}} {\tiny\ttfamily DecisionQuotient/\allowbreak Physics/\allowbreak MeasureNecessity.lean}
\item \textbf{\nolinkurl{MN2}}\hypertarget{lh:MN2}{}\enspace{\ttfamily\nolinkurl{Physics.MeasureNecessity.stochastic_claim_has_probability_measure}} {\tiny\ttfamily DecisionQuotient/\allowbreak Physics/\allowbreak MeasureNecessity.lean}
\item \textbf{\nolinkurl{MN5}}\hypertarget{lh:MN5}{}\enspace{\ttfamily\nolinkurl{Physics.MeasureNecessity.counting_measure_not_probability_on_bool}} {\tiny\ttfamily DecisionQuotient/\allowbreak Physics/\allowbreak MeasureNecessity.lean}
\item \textbf{\nolinkurl{MN10}}\hypertarget{lh:MN10}{}\enspace{\ttfamily\nolinkurl{Physics.MeasureNecessity.quantitative_measure_is_logical_prerequisite}} {\tiny\ttfamily DecisionQuotient/\allowbreak Physics/\allowbreak MeasureNecessity.lean}
\item \textbf{\nolinkurl{MN11}}\hypertarget{lh:MN11}{}\enspace{\ttfamily\nolinkurl{Physics.MeasureNecessity.stochastic_probability_is_logical_prerequisite}} {\tiny\ttfamily DecisionQuotient/\allowbreak Physics/\allowbreak MeasureNecessity.lean}
\item \textbf{\nolinkurl{OR3}}\hypertarget{lh:OR3}{}\enspace{\ttfamily\nolinkurl{Physics.ObserverRelativeState.EffectiveStateSpace}} {\tiny\ttfamily DecisionQuotient/\allowbreak Physics/\allowbreak ObserverRelativeState.lean}
\item \textbf{\nolinkurl{OR4}}\hypertarget{lh:OR4}{}\enspace{\ttfamily\nolinkurl{Physics.ObserverRelativeState.project_eq_iff}} {\tiny\ttfamily DecisionQuotient/\allowbreak Physics/\allowbreak ObserverRelativeState.lean}
\item \textbf{\nolinkurl{OR5}}\hypertarget{lh:OR5}{}\enspace{\ttfamily\nolinkurl{Physics.ObserverRelativeState.observer_relative_equivalence_witness}} {\tiny\ttfamily DecisionQuotient/\allowbreak Physics/\allowbreak ObserverRelativeState.lean}
\item \textbf{\nolinkurl{OR9}}\hypertarget{lh:OR9}{}\enspace{\ttfamily\nolinkurl{Physics.ObserverRelativeState.physical_observer_relative_effective_space}} {\tiny\ttfamily DecisionQuotient/\allowbreak Physics/\allowbreak ObserverRelativeState.lean}
\item \textbf{\nolinkurl{PA1}}\hypertarget{lh:PA1}{}\enspace{\ttfamily\nolinkurl{Physics.AnchorChecks.obsEquiv_all_of_effective_subsingleton}} {\tiny\ttfamily DecisionQuotient/\allowbreak Physics/\allowbreak AnchorChecks.lean}
\item \textbf{\nolinkurl{PA2}}\hypertarget{lh:PA2}{}\enspace{\ttfamily\nolinkurl{Physics.AnchorChecks.stochasticAnchorSufficient_iff_exists_anchor_singleton}} {\tiny\ttfamily DecisionQuotient/\allowbreak Physics/\allowbreak AnchorChecks.lean}
\item \textbf{\nolinkurl{PA3}}\hypertarget{lh:PA3}{}\enspace{\ttfamily\nolinkurl{Physics.AnchorChecks.stochastic_anchor_check_iff_exists_anchor_singleton}} {\tiny\ttfamily DecisionQuotient/\allowbreak Physics/\allowbreak AnchorChecks.lean}
\item \textbf{\nolinkurl{PA4}}\hypertarget{lh:PA4}{}\enspace{\ttfamily\nolinkurl{Physics.AnchorChecks.stochastic_sufficient_of_observer_collapse_and_seed}} {\tiny\ttfamily DecisionQuotient/\allowbreak Physics/\allowbreak AnchorChecks.lean}
\item \textbf{\nolinkurl{PA5}}\hypertarget{lh:PA5}{}\enspace{\ttfamily\nolinkurl{Physics.AnchorChecks.stochastic_anchor_check_of_observer_collapse_and_seed}} {\tiny\ttfamily DecisionQuotient/\allowbreak Physics/\allowbreak AnchorChecks.lean}
\item \textbf{\nolinkurl{PA8}}\hypertarget{lh:PA8}{}\enspace{\ttfamily\nolinkurl{Physics.AnchorChecks.physical_observer_collapse_implies_obsEquiv_all}} {\tiny\ttfamily DecisionQuotient/\allowbreak Physics/\allowbreak AnchorChecks.lean}
\item \textbf{\nolinkurl{PA9}}\hypertarget{lh:PA9}{}\enspace{\ttfamily\nolinkurl{Physics.AnchorChecks.physical_stochastic_anchor_check_of_observer_collapse_and_seed}} {\tiny\ttfamily DecisionQuotient/\allowbreak Physics/\allowbreak AnchorChecks.lean}
\item \textbf{\nolinkurl{PBC7}}\hypertarget{lh:PBC7}{}\enspace{\ttfamily\nolinkurl{DecisionQuotient.PhysicalBudgetCrossover.explicit_eventual_infeasibility_of_monotone_and_witness}} {\tiny\ttfamily DecisionQuotient/\allowbreak PhysicalBudgetCrossover.lean}
\item \textbf{\nolinkurl{PBC8}}\hypertarget{lh:PBC8}{}\enspace{\ttfamily\nolinkurl{DecisionQuotient.PhysicalBudgetCrossover.real_feasible_count_le_floor}} {\tiny\ttfamily DecisionQuotient/\allowbreak PhysicalBudgetCrossover.lean}
\item \textbf{\nolinkurl{PBC9}}\hypertarget{lh:PBC9}{}\enspace{\ttfamily\nolinkurl{DecisionQuotient.PhysicalBudgetCrossover.retained_records_bounded_of_universal_real_budget}} {\tiny\ttfamily DecisionQuotient/\allowbreak PhysicalBudgetCrossover.lean}
\item \textbf{\nolinkurl{PBC10}}\hypertarget{lh:PBC10}{}\enspace{\ttfamily\nolinkurl{DecisionQuotient.PhysicalBudgetCrossover.nat_bound_eventually_exceeded_of_monotone_unbounded}} {\tiny\ttfamily DecisionQuotient/\allowbreak PhysicalBudgetCrossover.lean}
\item \textbf{\nolinkurl{PBC11}}\hypertarget{lh:PBC11}{}\enspace{\ttfamily\nolinkurl{DecisionQuotient.PhysicalBudgetCrossover.feasible_count_mono_budget}} {\tiny\ttfamily DecisionQuotient/\allowbreak PhysicalBudgetCrossover.lean}
\item \textbf{\nolinkurl{PBC12}}\hypertarget{lh:PBC12}{}\enspace{\ttfamily\nolinkurl{DecisionQuotient.PhysicalBudgetCrossover.infeasible_count_mono_budget}} {\tiny\ttfamily DecisionQuotient/\allowbreak PhysicalBudgetCrossover.lean}
\item \textbf{\nolinkurl{PBC13}}\hypertarget{lh:PBC13}{}\enspace{\ttfamily\nolinkurl{DecisionQuotient.PhysicalBudgetCrossover.infeasible_count_mono_cost_floor}} {\tiny\ttfamily DecisionQuotient/\allowbreak PhysicalBudgetCrossover.lean}
\item \textbf{\nolinkurl{PBC14}}\hypertarget{lh:PBC14}{}\enspace{\ttfamily\nolinkurl{DecisionQuotient.PhysicalBudgetCrossover.linear_record_bound_monotone}} {\tiny\ttfamily DecisionQuotient/\allowbreak PhysicalBudgetCrossover.lean}
\item \textbf{\nolinkurl{PBC15}}\hypertarget{lh:PBC15}{}\enspace{\ttfamily\nolinkurl{DecisionQuotient.PhysicalBudgetCrossover.linear_record_bound_unbounded}} {\tiny\ttfamily DecisionQuotient/\allowbreak PhysicalBudgetCrossover.lean}
\item \textbf{\nolinkurl{PBC16}}\hypertarget{lh:PBC16}{}\enspace{\ttfamily\nolinkurl{DecisionQuotient.PhysicalBudgetCrossover.exponential_record_bound_monotone}} {\tiny\ttfamily DecisionQuotient/\allowbreak PhysicalBudgetCrossover.lean}
\item \textbf{\nolinkurl{PBC17}}\hypertarget{lh:PBC17}{}\enspace{\ttfamily\nolinkurl{DecisionQuotient.PhysicalBudgetCrossover.exponential_record_bound_unbounded}} {\tiny\ttfamily DecisionQuotient/\allowbreak PhysicalBudgetCrossover.lean}
\item \textbf{\nolinkurl{PBC18}}\hypertarget{lh:PBC18}{}\enspace{\ttfamily\nolinkurl{DecisionQuotient.PhysicalBudgetCrossover.real_budget_eventually_exceeded_of_monotone_unbounded}} {\tiny\ttfamily DecisionQuotient/\allowbreak PhysicalBudgetCrossover.lean}
\item \textbf{\nolinkurl{PBC19}}\hypertarget{lh:PBC19}{}\enspace{\ttfamily\nolinkurl{DecisionQuotient.PhysicalBudgetCrossover.linear_budget_eventually_exceeded}} {\tiny\ttfamily DecisionQuotient/\allowbreak PhysicalBudgetCrossover.lean}
\item \textbf{\nolinkurl{PBC20}}\hypertarget{lh:PBC20}{}\enspace{\ttfamily\nolinkurl{DecisionQuotient.PhysicalBudgetCrossover.exponential_budget_eventually_exceeded}} {\tiny\ttfamily DecisionQuotient/\allowbreak PhysicalBudgetCrossover.lean}
\item \textbf{\nolinkurl{PBC21}}\hypertarget{lh:PBC21}{}\enspace{\ttfamily\nolinkurl{DecisionQuotient.PhysicalBudgetCrossover.abstraction_record_family_scales_cost_but_positive_records_cost}} {\tiny\ttfamily DecisionQuotient/\allowbreak PhysicalBudgetCrossover.lean}
\item \textbf{\nolinkurl{PH11}}\hypertarget{lh:PH11}{}\enspace{\ttfamily\nolinkurl{PhysicalComplexity.PhysicalCollapseAtRequirement}} {\tiny\ttfamily DecisionQuotient/\allowbreak Physics/\allowbreak PhysicalHardness.lean}
\item \textbf{\nolinkurl{PH12}}\hypertarget{lh:PH12}{}\enspace{\ttfamily\nolinkurl{PhysicalComplexity.no_physical_collapse_at_requirement}} {\tiny\ttfamily DecisionQuotient/\allowbreak Physics/\allowbreak PhysicalHardness.lean}
\item \textbf{\nolinkurl{PH13}}\hypertarget{lh:PH13}{}\enspace{\ttfamily\nolinkurl{PhysicalComplexity.canonical_physical_collapse_impossible}} {\tiny\ttfamily DecisionQuotient/\allowbreak Physics/\allowbreak PhysicalHardness.lean}
\item \textbf{\nolinkurl{PH14}}\hypertarget{lh:PH14}{}\enspace{\ttfamily\nolinkurl{PhysicalComplexity.p_eq_np_physically_impossible_of_collapse_map}} {\tiny\ttfamily DecisionQuotient/\allowbreak Physics/\allowbreak PhysicalHardness.lean}
\item \textbf{\nolinkurl{PH15}}\hypertarget{lh:PH15}{}\enspace{\ttfamily\nolinkurl{PhysicalComplexity.p_eq_np_physically_impossible_canonical}} {\tiny\ttfamily DecisionQuotient/\allowbreak Physics/\allowbreak PhysicalHardness.lean}
\item \textbf{\nolinkurl{PH16}}\hypertarget{lh:PH16}{}\enspace{\ttfamily\nolinkurl{PhysicalComplexity.P_eq_NP_via_SAT}} {\tiny\ttfamily DecisionQuotient/\allowbreak Physics/\allowbreak PhysicalHardness.lean}
\item \textbf{\nolinkurl{PH17}}\hypertarget{lh:PH17}{}\enspace{\ttfamily\nolinkurl{PhysicalComplexity.SAT3ReductionBridge}} {\tiny\ttfamily DecisionQuotient/\allowbreak Physics/\allowbreak PhysicalHardness.lean}
\item \textbf{\nolinkurl{PH18}}\hypertarget{lh:PH18}{}\enspace{\ttfamily\nolinkurl{PhysicalComplexity.sat_reduction_transfers_energy_lower_bound}} {\tiny\ttfamily DecisionQuotient/\allowbreak Physics/\allowbreak PhysicalHardness.lean}
\item \textbf{\nolinkurl{PH19}}\hypertarget{lh:PH19}{}\enspace{\ttfamily\nolinkurl{PhysicalComplexity.physical_collapse_of_polytime_sat_realization}} {\tiny\ttfamily DecisionQuotient/\allowbreak Physics/\allowbreak PhysicalHardness.lean}
\item \textbf{\nolinkurl{PH20}}\hypertarget{lh:PH20}{}\enspace{\ttfamily\nolinkurl{PhysicalComplexity.p_eq_np_physically_impossible_via_sat_bridge}} {\tiny\ttfamily DecisionQuotient/\allowbreak Physics/\allowbreak PhysicalHardness.lean}
\item \textbf{\nolinkurl{PH21}}\hypertarget{lh:PH21}{}\enspace{\ttfamily\nolinkurl{PhysicalComplexity.SAT3HardFamily}} {\tiny\ttfamily DecisionQuotient/\allowbreak Physics/\allowbreak PhysicalHardness.lean}
\item \textbf{\nolinkurl{PH22}}\hypertarget{lh:PH22}{}\enspace{\ttfamily\nolinkurl{PhysicalComplexity.p_eq_np_physically_impossible_via_sat_hard_family}} {\tiny\ttfamily DecisionQuotient/\allowbreak Physics/\allowbreak PhysicalHardness.lean}
\item \textbf{\nolinkurl{PH23}}\hypertarget{lh:PH23}{}\enspace{\ttfamily\nolinkurl{PhysicalComplexity.collapse_possible_without_positive_bit_cost}} {\tiny\ttfamily DecisionQuotient/\allowbreak Physics/\allowbreak PhysicalHardness.lean}
\item \textbf{\nolinkurl{PH24}}\hypertarget{lh:PH24}{}\enspace{\ttfamily\nolinkurl{PhysicalComplexity.collapse_possible_without_exponential_lower_bound}} {\tiny\ttfamily DecisionQuotient/\allowbreak Physics/\allowbreak PhysicalHardness.lean}
\item \textbf{\nolinkurl{PH25}}\hypertarget{lh:PH25}{}\enspace{\ttfamily\nolinkurl{PhysicalComplexity.no_go_transfer_requires_collapse_map}} {\tiny\ttfamily DecisionQuotient/\allowbreak Physics/\allowbreak PhysicalHardness.lean}
\item \textbf{\nolinkurl{PH26}}\hypertarget{lh:PH26}{}\enspace{\ttfamily\nolinkurl{PhysicalComplexity.no_collapse_of_bounded_budget_pos_cost_exp_lb}} {\tiny\ttfamily DecisionQuotient/\allowbreak Physics/\allowbreak PhysicalHardness.lean}
\item \textbf{\nolinkurl{PH27}}\hypertarget{lh:PH27}{}\enspace{\ttfamily\nolinkurl{PhysicalComplexity.collapse_implies_assumption_failure_disjunction}} {\tiny\ttfamily DecisionQuotient/\allowbreak Physics/\allowbreak PhysicalHardness.lean}
\item \textbf{\nolinkurl{PH28}}\hypertarget{lh:PH28}{}\enspace{\ttfamily\nolinkurl{PhysicalComplexity.deterministic_no_physical_collapse}} {\tiny\ttfamily DecisionQuotient/\allowbreak Physics/\allowbreak PhysicalHardness.lean}
\item \textbf{\nolinkurl{PH29}}\hypertarget{lh:PH29}{}\enspace{\ttfamily\nolinkurl{PhysicalComplexity.probabilistic_no_physical_collapse}} {\tiny\ttfamily DecisionQuotient/\allowbreak Physics/\allowbreak PhysicalHardness.lean}
\item \textbf{\nolinkurl{PH30}}\hypertarget{lh:PH30}{}\enspace{\ttfamily\nolinkurl{PhysicalComplexity.sequential_no_physical_collapse}} {\tiny\ttfamily DecisionQuotient/\allowbreak Physics/\allowbreak PhysicalHardness.lean}
\item \textbf{\nolinkurl{PH31}}\hypertarget{lh:PH31}{}\enspace{\ttfamily\nolinkurl{PhysicalComplexity.collapse_possible_with_unbounded_budget_profile}} {\tiny\ttfamily DecisionQuotient/\allowbreak Physics/\allowbreak PhysicalHardness.lean}
\item \textbf{\nolinkurl{PH32}}\hypertarget{lh:PH32}{}\enspace{\ttfamily\nolinkurl{PhysicalComplexity.exp_budget_profile_unbounded}} {\tiny\ttfamily DecisionQuotient/\allowbreak Physics/\allowbreak PhysicalHardness.lean}
\item \textbf{\nolinkurl{PH33}}\hypertarget{lh:PH33}{}\enspace{\ttfamily\nolinkurl{PhysicalComplexity.finite_budget_assumption_is_necessary}} {\tiny\ttfamily DecisionQuotient/\allowbreak Physics/\allowbreak PhysicalHardness.lean}
\item \textbf{\nolinkurl{PI3}}\hypertarget{lh:PI3}{}\enspace{\ttfamily\nolinkurl{DecisionQuotient.Physics.PhysicalIncompleteness.no_surjective_instantiation_of_card_gap}} {\tiny\ttfamily DecisionQuotient/\allowbreak Physics/\allowbreak PhysicalIncompleteness.lean}
\item \textbf{\nolinkurl{PI4}}\hypertarget{lh:PI4}{}\enspace{\ttfamily\nolinkurl{DecisionQuotient.Physics.PhysicalIncompleteness.physical_incompleteness_of_card_gap}} {\tiny\ttfamily DecisionQuotient/\allowbreak Physics/\allowbreak PhysicalIncompleteness.lean}
\item \textbf{\nolinkurl{PI5}}\hypertarget{lh:PI5}{}\enspace{\ttfamily\nolinkurl{DecisionQuotient.Physics.PhysicalIncompleteness.physical_incompleteness_of_bounds}} {\tiny\ttfamily DecisionQuotient/\allowbreak Physics/\allowbreak PhysicalIncompleteness.lean}
\item \textbf{\nolinkurl{PI6}}\hypertarget{lh:PI6}{}\enspace{\ttfamily\nolinkurl{DecisionQuotient.Physics.PhysicalIncompleteness.under_resolution_implies_collision}} {\tiny\ttfamily DecisionQuotient/\allowbreak Physics/\allowbreak PhysicalIncompleteness.lean}
\item \textbf{\nolinkurl{PI7}}\hypertarget{lh:PI7}{}\enspace{\ttfamily\nolinkurl{DecisionQuotient.Physics.PhysicalIncompleteness.under_resolution_implies_decision_collision}} {\tiny\ttfamily DecisionQuotient/\allowbreak Physics/\allowbreak PhysicalIncompleteness.lean}
\item \textbf{\nolinkurl{PROV2}}\hypertarget{lh:PROV2}{}\enspace{\ttfamily\nolinkurl{PhysicalComplexity.ProvabilityCorollary.no_proof_act_of_forbidden_target}} {\tiny\ttfamily DecisionQuotient/\allowbreak Physics/\allowbreak ProvabilityCorollary.lean}
\item \textbf{\nolinkurl{PROV3}}\hypertarget{lh:PROV3}{}\enspace{\ttfamily\nolinkurl{PhysicalComplexity.ProvabilityCorollary.no_operational_pnp_proof_at_requirement}} {\tiny\ttfamily DecisionQuotient/\allowbreak Physics/\allowbreak ProvabilityCorollary.lean}
\item \textbf{\nolinkurl{PROV5}}\hypertarget{lh:PROV5}{}\enspace{\ttfamily\nolinkurl{PhysicalComplexity.ProvabilityCorollary.lower_bound_proof_direction_compatible}} {\tiny\ttfamily DecisionQuotient/\allowbreak Physics/\allowbreak ProvabilityCorollary.lean}
\item \textbf{\nolinkurl{PROV6}}\hypertarget{lh:PROV6}{}\enspace{\ttfamily\nolinkurl{PhysicalComplexity.ProvabilityCorollary.RecordFeasible}} {\tiny\ttfamily DecisionQuotient/\allowbreak Physics/\allowbreak ProvabilityCorollary.lean}
\item \textbf{\nolinkurl{PROV7}}\hypertarget{lh:PROV7}{}\enspace{\ttfamily\nolinkurl{PhysicalComplexity.ProvabilityCorollary.no_universal_auditable_proof_family_of_unbounded_records}} {\tiny\ttfamily DecisionQuotient/\allowbreak Physics/\allowbreak ProvabilityCorollary.lean}
\item \textbf{\nolinkurl{PROV8}}\hypertarget{lh:PROV8}{}\enspace{\ttfamily\nolinkurl{PhysicalComplexity.ProvabilityCorollary.OperationalProofScopeGate}} {\tiny\ttfamily DecisionQuotient/\allowbreak Physics/\allowbreak ProvabilityCorollary.lean}
\item \textbf{\nolinkurl{PROV9}}\hypertarget{lh:PROV9}{}\enspace{\ttfamily\nolinkurl{PhysicalComplexity.ProvabilityCorollary.IsOperationallyAvailable}} {\tiny\ttfamily DecisionQuotient/\allowbreak Physics/\allowbreak ProvabilityCorollary.lean}
\item \textbf{\nolinkurl{PROV10}}\hypertarget{lh:PROV10}{}\enspace{\ttfamily\nolinkurl{PhysicalComplexity.ProvabilityCorollary.OperationalProofAct}} {\tiny\ttfamily DecisionQuotient/\allowbreak Physics/\allowbreak ProvabilityCorollary.lean}
\item \textbf{\nolinkurl{PROV11}}\hypertarget{lh:PROV11}{}\enspace{\ttfamily\nolinkurl{PhysicalComplexity.ProvabilityCorollary.operational_proof_act_is_available}} {\tiny\ttfamily DecisionQuotient/\allowbreak Physics/\allowbreak ProvabilityCorollary.lean}
\item \textbf{\nolinkurl{PROV12}}\hypertarget{lh:PROV12}{}\enspace{\ttfamily\nolinkurl{PhysicalComplexity.ProvabilityCorollary.no_operational_proof_act_without_available_token}} {\tiny\ttfamily DecisionQuotient/\allowbreak Physics/\allowbreak ProvabilityCorollary.lean}
\item \textbf{\nolinkurl{PROV13}}\hypertarget{lh:PROV13}{}\enspace{\ttfamily\nolinkurl{PhysicalComplexity.ProvabilityCorollary.no_typed_operational_proof_act_of_forbidden_target}} {\tiny\ttfamily DecisionQuotient/\allowbreak Physics/\allowbreak ProvabilityCorollary.lean}
\item \textbf{\nolinkurl{PROV14}}\hypertarget{lh:PROV14}{}\enspace{\ttfamily\nolinkurl{PhysicalComplexity.ProvabilityCorollary.ProofActConstructsUniversalCertifierRecords}} {\tiny\ttfamily DecisionQuotient/\allowbreak Physics/\allowbreak ProvabilityCorollary.lean}
\item \textbf{\nolinkurl{PROV15}}\hypertarget{lh:PROV15}{}\enspace{\ttfamily\nolinkurl{PhysicalComplexity.ProvabilityCorollary.proof_act_constructs_universal_certifier_records}} {\tiny\ttfamily DecisionQuotient/\allowbreak Physics/\allowbreak ProvabilityCorollary.lean}
\item \textbf{\nolinkurl{PROV16}}\hypertarget{lh:PROV16}{}\enspace{\ttfamily\nolinkurl{PhysicalComplexity.ProvabilityCorollary.typed_operational_proof_act_constructs_universal_certifier_records}} {\tiny\ttfamily DecisionQuotient/\allowbreak Physics/\allowbreak ProvabilityCorollary.lean}
\item \textbf{\nolinkurl{PROV17}}\hypertarget{lh:PROV17}{}\enspace{\ttfamily\nolinkurl{PhysicalComplexity.ProvabilityCorollary.no_convincing_operational_pnp_proof_act_of_unbounded_records}} {\tiny\ttfamily DecisionQuotient/\allowbreak Physics/\allowbreak ProvabilityCorollary.lean}
\item \textbf{\nolinkurl{PROV18}}\hypertarget{lh:PROV18}{}\enspace{\ttfamily\nolinkurl{PhysicalComplexity.ProvabilityCorollary.EvidenceRetainingUniversalCertifier}} {\tiny\ttfamily DecisionQuotient/\allowbreak Physics/\allowbreak ProvabilityCorollary.lean}
\item \textbf{\nolinkurl{PROV19}}\hypertarget{lh:PROV19}{}\enspace{\ttfamily\nolinkurl{PhysicalComplexity.ProvabilityCorollary.SoundOperationalPNPVerifier}} {\tiny\ttfamily DecisionQuotient/\allowbreak Physics/\allowbreak ProvabilityCorollary.lean}
\item \textbf{\nolinkurl{PROV20}}\hypertarget{lh:PROV20}{}\enspace{\ttfamily\nolinkurl{PhysicalComplexity.ProvabilityCorollary.operational_pnp_proof_licenses_universal_certifier}} {\tiny\ttfamily DecisionQuotient/\allowbreak Physics/\allowbreak ProvabilityCorollary.lean}
\item \textbf{\nolinkurl{PROV21}}\hypertarget{lh:PROV21}{}\enspace{\ttfamily\nolinkurl{PhysicalComplexity.ProvabilityCorollary.no_operational_pnp_proof_act_of_sound_verifier_and_unbounded_records}} {\tiny\ttfamily DecisionQuotient/\allowbreak Physics/\allowbreak ProvabilityCorollary.lean}
\item \textbf{\nolinkurl{PROV22}}\hypertarget{lh:PROV22}{}\enspace{\ttfamily\nolinkurl{PhysicalComplexity.ProvabilityCorollary.OperationalPNPClaim}} {\tiny\ttfamily DecisionQuotient/\allowbreak Physics/\allowbreak ProvabilityCorollary.lean}
\item \textbf{\nolinkurl{PROV23}}\hypertarget{lh:PROV23}{}\enspace{\ttfamily\nolinkurl{PhysicalComplexity.ProvabilityCorollary.OperationalVerifier}} {\tiny\ttfamily DecisionQuotient/\allowbreak Physics/\allowbreak ProvabilityCorollary.lean}
\item \textbf{\nolinkurl{PROV24}}\hypertarget{lh:PROV24}{}\enspace{\ttfamily\nolinkurl{PhysicalComplexity.ProvabilityCorollary.SoundOperationalVerifier}} {\tiny\ttfamily DecisionQuotient/\allowbreak Physics/\allowbreak ProvabilityCorollary.lean}
\item \textbf{\nolinkurl{PROV25}}\hypertarget{lh:PROV25}{}\enspace{\ttfamily\nolinkurl{PhysicalComplexity.ProvabilityCorollary.OperationalPNPMeansCertifierAvailability}} {\tiny\ttfamily DecisionQuotient/\allowbreak Physics/\allowbreak ProvabilityCorollary.lean}
\item \textbf{\nolinkurl{PROV26}}\hypertarget{lh:PROV26}{}\enspace{\ttfamily\nolinkurl{PhysicalComplexity.ProvabilityCorollary.accepted_operational_pnp_extracts_certifier}} {\tiny\ttfamily DecisionQuotient/\allowbreak Physics/\allowbreak ProvabilityCorollary.lean}
\item \textbf{\nolinkurl{PROV27}}\hypertarget{lh:PROV27}{}\enspace{\ttfamily\nolinkurl{PhysicalComplexity.ProvabilityCorollary.no_accepted_operational_pnp_proof_act_of_unbounded_records}} {\tiny\ttfamily DecisionQuotient/\allowbreak Physics/\allowbreak ProvabilityCorollary.lean}
\item \textbf{\nolinkurl{PROV28}}\hypertarget{lh:PROV28}{}\enspace{\ttfamily\nolinkurl{PhysicalComplexity.ProvabilityCorollary.OperationalPNPPhysicalInterface}} {\tiny\ttfamily DecisionQuotient/\allowbreak Physics/\allowbreak ProvabilityCorollary.lean}
\item \textbf{\nolinkurl{PROV29}}\hypertarget{lh:PROV29}{}\enspace{\ttfamily\nolinkurl{PhysicalComplexity.ProvabilityCorollary.accepted_operational_pnp_extracts_certifier_from_interface}} {\tiny\ttfamily DecisionQuotient/\allowbreak Physics/\allowbreak ProvabilityCorollary.lean}
\item \textbf{\nolinkurl{PROV30}}\hypertarget{lh:PROV30}{}\enspace{\ttfamily\nolinkurl{PhysicalComplexity.ProvabilityCorollary.no_accepted_operational_pnp_proof_act_from_interface}} {\tiny\ttfamily DecisionQuotient/\allowbreak Physics/\allowbreak ProvabilityCorollary.lean}
\item \textbf{\nolinkurl{PROV31}}\hypertarget{lh:PROV31}{}\enspace{\ttfamily\nolinkurl{PhysicalComplexity.ProvabilityCorollary.proof_act_tokenhood_alone_does_not_imply_certifier_availability}} {\tiny\ttfamily DecisionQuotient/\allowbreak Physics/\allowbreak ProvabilityCorollary.lean}
\item \textbf{\nolinkurl{PROV32}}\hypertarget{lh:PROV32}{}\enspace{\ttfamily\nolinkurl{PhysicalComplexity.ProvabilityCorollary.PhysicalProofDecoder}} {\tiny\ttfamily DecisionQuotient/\allowbreak Physics/\allowbreak ProvabilityCorollary.lean}
\item \textbf{\nolinkurl{PROV33}}\hypertarget{lh:PROV33}{}\enspace{\ttfamily\nolinkurl{PhysicalComplexity.ProvabilityCorollary.decoder_acceptance_has_physical_phases}} {\tiny\ttfamily DecisionQuotient/\allowbreak Physics/\allowbreak ProvabilityCorollary.lean}
\item \textbf{\nolinkurl{PROV34}}\hypertarget{lh:PROV34}{}\enspace{\ttfamily\nolinkurl{PhysicalComplexity.ProvabilityCorollary.decoder_acceptance_extracts_certifier}} {\tiny\ttfamily DecisionQuotient/\allowbreak Physics/\allowbreak ProvabilityCorollary.lean}
\item \textbf{\nolinkurl{PROV35}}\hypertarget{lh:PROV35}{}\enspace{\ttfamily\nolinkurl{PhysicalComplexity.ProvabilityCorollary.no_decoder_accepted_operational_pnp_proof_act_from_interface}} {\tiny\ttfamily DecisionQuotient/\allowbreak Physics/\allowbreak ProvabilityCorollary.lean}
\item \textbf{\nolinkurl{PROV36}}\hypertarget{lh:PROV36}{}\enspace{\ttfamily\nolinkurl{PhysicalComplexity.ProvabilityCorollary.no_certifier_extraction_from_tokenhood_only}} {\tiny\ttfamily DecisionQuotient/\allowbreak Physics/\allowbreak ProvabilityCorollary.lean}
\item \textbf{\nolinkurl{PROV37}}\hypertarget{lh:PROV37}{}\enspace{\ttfamily\nolinkurl{PhysicalComplexity.ProvabilityCorollary.SituatedProofDoctrine}} {\tiny\ttfamily DecisionQuotient/\allowbreak Physics/\allowbreak ProvabilityCorollary.lean}
\item \textbf{\nolinkurl{PROV38}}\hypertarget{lh:PROV38}{}\enspace{\ttfamily\nolinkurl{PhysicalComplexity.ProvabilityCorollary.no_situated_formal_proof_of_blocked_proof_act}} {\tiny\ttfamily DecisionQuotient/\allowbreak Physics/\allowbreak ProvabilityCorollary.lean}
\item \textbf{\nolinkurl{PROV39}}\hypertarget{lh:PROV39}{}\enspace{\ttfamily\nolinkurl{PhysicalComplexity.ProvabilityCorollary.SituatedAcceptedProofDoctrine}} {\tiny\ttfamily DecisionQuotient/\allowbreak Physics/\allowbreak ProvabilityCorollary.lean}
\item \textbf{\nolinkurl{PROV40}}\hypertarget{lh:PROV40}{}\enspace{\ttfamily\nolinkurl{PhysicalComplexity.ProvabilityCorollary.no_situated_formal_pnp_proof_from_interface}} {\tiny\ttfamily DecisionQuotient/\allowbreak Physics/\allowbreak ProvabilityCorollary.lean}
\item \textbf{\nolinkurl{PROV41}}\hypertarget{lh:PROV41}{}\enspace{\ttfamily\nolinkurl{PhysicalComplexity.ProvabilityCorollary.AlgorithmSpecification}} {\tiny\ttfamily DecisionQuotient/\allowbreak Physics/\allowbreak ProvabilityCorollary.lean}
\item \textbf{\nolinkurl{PROV42}}\hypertarget{lh:PROV42}{}\enspace{\ttfamily\nolinkurl{PhysicalComplexity.ProvabilityCorollary.AlgorithmProcess}} {\tiny\ttfamily DecisionQuotient/\allowbreak Physics/\allowbreak ProvabilityCorollary.lean}
\item \textbf{\nolinkurl{PROV43}}\hypertarget{lh:PROV43}{}\enspace{\ttfamily\nolinkurl{PhysicalComplexity.ProvabilityCorollary.SpecificationRealizationBridge}} {\tiny\ttfamily DecisionQuotient/\allowbreak Physics/\allowbreak ProvabilityCorollary.lean}
\item \textbf{\nolinkurl{PROV44}}\hypertarget{lh:PROV44}{}\enspace{\ttfamily\nolinkurl{PhysicalComplexity.ProvabilityCorollary.RuntimeDeterminedBySpecification}} {\tiny\ttfamily DecisionQuotient/\allowbreak Physics/\allowbreak ProvabilityCorollary.lean}
\item \textbf{\nolinkurl{PROV45}}\hypertarget{lh:PROV45}{}\enspace{\ttfamily\nolinkurl{PhysicalComplexity.ProvabilityCorollary.same_specification_different_runtime_countermodel}} {\tiny\ttfamily DecisionQuotient/\allowbreak Physics/\allowbreak ProvabilityCorollary.lean}
\item \textbf{\nolinkurl{PROV46}}\hypertarget{lh:PROV46}{}\enspace{\ttfamily\nolinkurl{PhysicalComplexity.ProvabilityCorollary.specification_needs_realization_bridge_for_runtime}} {\tiny\ttfamily DecisionQuotient/\allowbreak Physics/\allowbreak ProvabilityCorollary.lean}
\item \textbf{\nolinkurl{PROV47}}\hypertarget{lh:PROV47}{}\enspace{\ttfamily\nolinkurl{PhysicalComplexity.ProvabilityCorollary.AlgorithmFiveTuple}} {\tiny\ttfamily DecisionQuotient/\allowbreak Physics/\allowbreak ProvabilityCorollary.lean}
\item \textbf{\nolinkurl{PROV48}}\hypertarget{lh:PROV48}{}\enspace{\ttfamily\nolinkurl{PhysicalComplexity.ProvabilityCorollary.SubstrateEvaluationRule}} {\tiny\ttfamily DecisionQuotient/\allowbreak Physics/\allowbreak ProvabilityCorollary.lean}
\item \textbf{\nolinkurl{PROV49}}\hypertarget{lh:PROV49}{}\enspace{\ttfamily\nolinkurl{PhysicalComplexity.ProvabilityCorollary.SubstrateEvaluationRule.toProcess}} {\tiny\ttfamily DecisionQuotient/\allowbreak Physics/\allowbreak ProvabilityCorollary.lean}
\item \textbf{\nolinkurl{PROV50}}\hypertarget{lh:PROV50}{}\enspace{\ttfamily\nolinkurl{PhysicalComplexity.ProvabilityCorollary.PolynomiallyBounded}} {\tiny\ttfamily DecisionQuotient/\allowbreak Physics/\allowbreak ProvabilityCorollary.lean}
\item \textbf{\nolinkurl{PROV51}}\hypertarget{lh:PROV51}{}\enspace{\ttfamily\nolinkurl{PhysicalComplexity.ProvabilityCorollary.PolynomialTimeProcess}} {\tiny\ttfamily DecisionQuotient/\allowbreak Physics/\allowbreak ProvabilityCorollary.lean}
\item \textbf{\nolinkurl{PROV52}}\hypertarget{lh:PROV52}{}\enspace{\ttfamily\nolinkurl{PhysicalComplexity.ProvabilityCorollary.PolynomialTimeOn}} {\tiny\ttfamily DecisionQuotient/\allowbreak Physics/\allowbreak ProvabilityCorollary.lean}
\item \textbf{\nolinkurl{PROV53}}\hypertarget{lh:PROV53}{}\enspace{\ttfamily\nolinkurl{PhysicalComplexity.ProvabilityCorollary.same_specification_different_evaluation_runtime_countermodel}} {\tiny\ttfamily DecisionQuotient/\allowbreak Physics/\allowbreak ProvabilityCorollary.lean}
\item \textbf{\nolinkurl{PROV54}}\hypertarget{lh:PROV54}{}\enspace{\ttfamily\nolinkurl{PhysicalComplexity.ProvabilityCorollary.same_specification_different_substrate_runtime_countermodel}} {\tiny\ttfamily DecisionQuotient/\allowbreak Physics/\allowbreak ProvabilityCorollary.lean}
\item \textbf{\nolinkurl{PROV55}}\hypertarget{lh:PROV55}{}\enspace{\ttfamily\nolinkurl{PhysicalComplexity.ProvabilityCorollary.specification_process_not_equiv_countermodel}} {\tiny\ttfamily DecisionQuotient/\allowbreak Physics/\allowbreak ProvabilityCorollary.lean}
\item \textbf{\nolinkurl{PROV56}}\hypertarget{lh:PROV56}{}\enspace{\ttfamily\nolinkurl{PhysicalComplexity.ProvabilityCorollary.PolynomialTimeMembership}} {\tiny\ttfamily DecisionQuotient/\allowbreak Physics/\allowbreak ProvabilityCorollary.lean}
\item \textbf{\nolinkurl{PROV57}}\hypertarget{lh:PROV57}{}\enspace{\ttfamily\nolinkurl{PhysicalComplexity.ProvabilityCorollary.polynomial_time_membership_forgets_to_specification}} {\tiny\ttfamily DecisionQuotient/\allowbreak Physics/\allowbreak ProvabilityCorollary.lean}
\item \textbf{\nolinkurl{PROV58}}\hypertarget{lh:PROV58}{}\enspace{\ttfamily\nolinkurl{PhysicalComplexity.ProvabilityCorollary.AnswerOnlySystem}} {\tiny\ttfamily DecisionQuotient/\allowbreak Physics/\allowbreak ProvabilityCorollary.lean}
\item \textbf{\nolinkurl{PROV59}}\hypertarget{lh:PROV59}{}\enspace{\ttfamily\nolinkurl{PhysicalComplexity.ProvabilityCorollary.EvidenceRetainingProofSystem}} {\tiny\ttfamily DecisionQuotient/\allowbreak Physics/\allowbreak ProvabilityCorollary.lean}
\item \textbf{\nolinkurl{PROV60}}\hypertarget{lh:PROV60}{}\enspace{\ttfamily\nolinkurl{PhysicalComplexity.ProvabilityCorollary.answer_only_acceptance_is_not_verification}} {\tiny\ttfamily DecisionQuotient/\allowbreak Physics/\allowbreak ProvabilityCorollary.lean}
\item \textbf{\nolinkurl{PROV61}}\hypertarget{lh:PROV61}{}\enspace{\ttfamily\nolinkurl{PhysicalComplexity.ProvabilityCorollary.show_work_proof_inspection_is_sound}} {\tiny\ttfamily DecisionQuotient/\allowbreak Physics/\allowbreak ProvabilityCorollary.lean}
\item \textbf{\nolinkurl{PROV62}}\hypertarget{lh:PROV62}{}\enspace{\ttfamily\nolinkurl{PhysicalComplexity.ProvabilityCorollary.answer_only_cannot_simulate_show_work_proof}} {\tiny\ttfamily DecisionQuotient/\allowbreak Physics/\allowbreak ProvabilityCorollary.lean}
\item \textbf{\nolinkurl{PROV63}}\hypertarget{lh:PROV63}{}\enspace{\ttfamily\nolinkurl{PhysicalComplexity.ProvabilityCorollary.MetaProofAuditTrail}} {\tiny\ttfamily DecisionQuotient/\allowbreak Physics/\allowbreak ProvabilityCorollary.lean}
\item \textbf{\nolinkurl{PROV64}}\hypertarget{lh:PROV64}{}\enspace{\ttfamily\nolinkurl{PhysicalComplexity.ProvabilityCorollary.finite_meta_audit_not_universal_instance_audit_of_unbounded}} {\tiny\ttfamily DecisionQuotient/\allowbreak Physics/\allowbreak ProvabilityCorollary.lean}
\item \textbf{\nolinkurl{PROV65}}\hypertarget{lh:PROV65}{}\enspace{\ttfamily\nolinkurl{PhysicalComplexity.ProvabilityCorollary.finite_meta_audit_cannot_substitute_for_unbounded_object_records}} {\tiny\ttfamily DecisionQuotient/\allowbreak Physics/\allowbreak ProvabilityCorollary.lean}
\item \textbf{\nolinkurl{PROV66}}\hypertarget{lh:PROV66}{}\enspace{\ttfamily\nolinkurl{PhysicalComplexity.ProvabilityCorollary.PhysicalOracleConsultation}} {\tiny\ttfamily DecisionQuotient/\allowbreak Physics/\allowbreak ProvabilityCorollary.lean}
\item \textbf{\nolinkurl{PROV67}}\hypertarget{lh:PROV67}{}\enspace{\ttfamily\nolinkurl{PhysicalComplexity.ProvabilityCorollary.PhysicalOracleConsultation.OracleProofAvailable}} {\tiny\ttfamily DecisionQuotient/\allowbreak Physics/\allowbreak ProvabilityCorollary.lean}
\item \textbf{\nolinkurl{PROV68}}\hypertarget{lh:PROV68}{}\enspace{\ttfamily\nolinkurl{PhysicalComplexity.ProvabilityCorollary.PhysicalOracleConsultation.LocalProofFromDeliveredAnswer}} {\tiny\ttfamily DecisionQuotient/\allowbreak Physics/\allowbreak ProvabilityCorollary.lean}
\item \textbf{\nolinkurl{PROV69}}\hypertarget{lh:PROV69}{}\enspace{\ttfamily\nolinkurl{PhysicalComplexity.ProvabilityCorollary.PhysicalOracleConsultation.TrustAuditAvailable}} {\tiny\ttfamily DecisionQuotient/\allowbreak Physics/\allowbreak ProvabilityCorollary.lean}
\item \textbf{\nolinkurl{PROV70}}\hypertarget{lh:PROV70}{}\enspace{\ttfamily\nolinkurl{PhysicalComplexity.ProvabilityCorollary.oracle_internal_proof_uses_oracle_budget}} {\tiny\ttfamily DecisionQuotient/\allowbreak Physics/\allowbreak ProvabilityCorollary.lean}
\item \textbf{\nolinkurl{PROV71}}\hypertarget{lh:PROV71}{}\enspace{\ttfamily\nolinkurl{PhysicalComplexity.ProvabilityCorollary.oracle_answer_below_requirement_not_local_proof}} {\tiny\ttfamily DecisionQuotient/\allowbreak Physics/\allowbreak ProvabilityCorollary.lean}
\item \textbf{\nolinkurl{PROV72}}\hypertarget{lh:PROV72}{}\enspace{\ttfamily\nolinkurl{PhysicalComplexity.ProvabilityCorollary.trust_audit_records_are_budgeted_records}} {\tiny\ttfamily DecisionQuotient/\allowbreak Physics/\allowbreak ProvabilityCorollary.lean}
\item \textbf{\nolinkurl{PROV73}}\hypertarget{lh:PROV73}{}\enspace{\ttfamily\nolinkurl{PhysicalComplexity.ProvabilityCorollary.no_universal_trusted_oracle_answer_of_unbounded_trust_audit}} {\tiny\ttfamily DecisionQuotient/\allowbreak Physics/\allowbreak ProvabilityCorollary.lean}
\item \textbf{\nolinkurl{PROV74}}\hypertarget{lh:PROV74}{}\enspace{\ttfamily\nolinkurl{PhysicalComplexity.ProvabilityCorollary.isOperationallyAvailable_iff_nonempty_proofAct}} {\tiny\ttfamily DecisionQuotient/\allowbreak Physics/\allowbreak ProvabilityCorollary.lean}
\item \textbf{\nolinkurl{PROV75}}\hypertarget{lh:PROV75}{}\enspace{\ttfamily\nolinkurl{PhysicalComplexity.ProvabilityCorollary.RetainedStateProofInterface}} {\tiny\ttfamily DecisionQuotient/\allowbreak Physics/\allowbreak ProvabilityCorollary.lean}
\item \textbf{\nolinkurl{PROV76}}\hypertarget{lh:PROV76}{}\enspace{\ttfamily\nolinkurl{PhysicalComplexity.ProvabilityCorollary.CertainAt}} {\tiny\ttfamily DecisionQuotient/\allowbreak Physics/\allowbreak ProvabilityCorollary.lean}
\item \textbf{\nolinkurl{PROV77}}\hypertarget{lh:PROV77}{}\enspace{\ttfamily\nolinkurl{PhysicalComplexity.ProvabilityCorollary.certainty_excludes_false_same_state}} {\tiny\ttfamily DecisionQuotient/\allowbreak Physics/\allowbreak ProvabilityCorollary.lean}
\item \textbf{\nolinkurl{PROV78}}\hypertarget{lh:PROV78}{}\enspace{\ttfamily\nolinkurl{PhysicalComplexity.ProvabilityCorollary.same_retained_state_false_alternative_blocks_certainty}} {\tiny\ttfamily DecisionQuotient/\allowbreak Physics/\allowbreak ProvabilityCorollary.lean}
\item \textbf{\nolinkurl{PROV79}}\hypertarget{lh:PROV79}{}\enspace{\ttfamily\nolinkurl{PhysicalComplexity.ProvabilityCorollary.DiscriminatesFalseAlternatives}} {\tiny\ttfamily DecisionQuotient/\allowbreak Physics/\allowbreak ProvabilityCorollary.lean}
\item \textbf{\nolinkurl{PROV80}}\hypertarget{lh:PROV80}{}\enspace{\ttfamily\nolinkurl{PhysicalComplexity.ProvabilityCorollary.CompatibleFalseAlternative}} {\tiny\ttfamily DecisionQuotient/\allowbreak Physics/\allowbreak ProvabilityCorollary.lean}
\item \textbf{\nolinkurl{PROV81}}\hypertarget{lh:PROV81}{}\enspace{\ttfamily\nolinkurl{PhysicalComplexity.ProvabilityCorollary.ExactProofState}} {\tiny\ttfamily DecisionQuotient/\allowbreak Physics/\allowbreak ProvabilityCorollary.lean}
\item \textbf{\nolinkurl{PROV82}}\hypertarget{lh:PROV82}{}\enspace{\ttfamily\nolinkurl{PhysicalComplexity.ProvabilityCorollary.certainty_iff_discriminates_false_alternatives}} {\tiny\ttfamily DecisionQuotient/\allowbreak Physics/\allowbreak ProvabilityCorollary.lean}
\item \textbf{\nolinkurl{PROV83}}\hypertarget{lh:PROV83}{}\enspace{\ttfamily\nolinkurl{PhysicalComplexity.ProvabilityCorollary.not_certain_iff_compatible_false_alternative}} {\tiny\ttfamily DecisionQuotient/\allowbreak Physics/\allowbreak ProvabilityCorollary.lean}
\item \textbf{\nolinkurl{PROV84}}\hypertarget{lh:PROV84}{}\enspace{\ttfamily\nolinkurl{PhysicalComplexity.ProvabilityCorollary.exact_proof_state_has_no_compatible_false_alternative}} {\tiny\ttfamily DecisionQuotient/\allowbreak Physics/\allowbreak ProvabilityCorollary.lean}
\item \textbf{\nolinkurl{PROV85}}\hypertarget{lh:PROV85}{}\enspace{\ttfamily\nolinkurl{PhysicalComplexity.ProvabilityCorollary.ProofTransitionAccounting}} {\tiny\ttfamily DecisionQuotient/\allowbreak Physics/\allowbreak ProvabilityCorollary.lean}
\item \textbf{\nolinkurl{PROV86}}\hypertarget{lh:PROV86}{}\enspace{\ttfamily\nolinkurl{PhysicalComplexity.ProvabilityCorollary.reversible_erased_not_retained_evidence}} {\tiny\ttfamily DecisionQuotient/\allowbreak Physics/\allowbreak ProvabilityCorollary.lean}
\item \textbf{\nolinkurl{PROV87}}\hypertarget{lh:PROV87}{}\enspace{\ttfamily\nolinkurl{PhysicalComplexity.ProvabilityCorollary.retained_evidence_is_irreversible_record}} {\tiny\ttfamily DecisionQuotient/\allowbreak Physics/\allowbreak ProvabilityCorollary.lean}
\item \textbf{\nolinkurl{PROV88}}\hypertarget{lh:PROV88}{}\enspace{\ttfamily\nolinkurl{PhysicalComplexity.ProvabilityCorollary.retained_evidence_has_positive_cost}} {\tiny\ttfamily DecisionQuotient/\allowbreak Physics/\allowbreak ProvabilityCorollary.lean}
\item \textbf{\nolinkurl{PROV89}}\hypertarget{lh:PROV89}{}\enspace{\ttfamily\nolinkurl{PhysicalComplexity.ProvabilityCorollary.retained_evidence_no_physical_exemption}} {\tiny\ttfamily DecisionQuotient/\allowbreak Physics/\allowbreak ProvabilityCorollary.lean}
\item \textbf{\nolinkurl{PROV90}}\hypertarget{lh:PROV90}{}\enspace{\ttfamily\nolinkurl{PhysicalComplexity.ProvabilityCorollary.CertificateVerificationInterface}} {\tiny\ttfamily DecisionQuotient/\allowbreak Physics/\allowbreak ProvabilityCorollary.lean}
\item \textbf{\nolinkurl{PROV91}}\hypertarget{lh:PROV91}{}\enspace{\ttfamily\nolinkurl{PhysicalComplexity.ProvabilityCorollary.CertificateVerifiableAnswer}} {\tiny\ttfamily DecisionQuotient/\allowbreak Physics/\allowbreak ProvabilityCorollary.lean}
\item \textbf{\nolinkurl{PROV92}}\hypertarget{lh:PROV92}{}\enspace{\ttfamily\nolinkurl{PhysicalComplexity.ProvabilityCorollary.certificate_verifiable_answer_uses_declared_record_family}} {\tiny\ttfamily DecisionQuotient/\allowbreak Physics/\allowbreak ProvabilityCorollary.lean}
\item \textbf{\nolinkurl{PROV93}}\hypertarget{lh:PROV93}{}\enspace{\ttfamily\nolinkurl{PhysicalComplexity.ProvabilityCorollary.certificate_verifiable_answer_proves_instance_claim}} {\tiny\ttfamily DecisionQuotient/\allowbreak Physics/\allowbreak ProvabilityCorollary.lean}
\item \textbf{\nolinkurl{PROV94}}\hypertarget{lh:PROV94}{}\enspace{\ttfamily\nolinkurl{PhysicalComplexity.ProvabilityCorollary.UniversalCertificateCoverage}} {\tiny\ttfamily DecisionQuotient/\allowbreak Physics/\allowbreak ProvabilityCorollary.lean}
\item \textbf{\nolinkurl{PROV95}}\hypertarget{lh:PROV95}{}\enspace{\ttfamily\nolinkurl{PhysicalComplexity.ProvabilityCorollary.certificate_verifiable_answer_alone_does_not_extract_universal_certifier}} {\tiny\ttfamily DecisionQuotient/\allowbreak Physics/\allowbreak ProvabilityCorollary.lean}
\item \textbf{\nolinkurl{PROV96}}\hypertarget{lh:PROV96}{}\enspace{\ttfamily\nolinkurl{PhysicalComplexity.ProvabilityCorollary.yes_certificate_interface_does_not_certify_false_instance}} {\tiny\ttfamily DecisionQuotient/\allowbreak Physics/\allowbreak ProvabilityCorollary.lean}
\item \textbf{\nolinkurl{PROV97}}\hypertarget{lh:PROV97}{}\enspace{\ttfamily\nolinkurl{PhysicalComplexity.ProvabilityCorollary.certificate_verifiable_answer_does_not_make_claim_universal}} {\tiny\ttfamily DecisionQuotient/\allowbreak Physics/\allowbreak ProvabilityCorollary.lean}
\item \textbf{\nolinkurl{PROV98}}\hypertarget{lh:PROV98}{}\enspace{\ttfamily\nolinkurl{PhysicalComplexity.ProvabilityCorollary.certificate_verifiable_answer_does_not_give_universal_coverage}} {\tiny\ttfamily DecisionQuotient/\allowbreak Physics/\allowbreak ProvabilityCorollary.lean}
\item \textbf{\nolinkurl{PROV99}}\hypertarget{lh:PROV99}{}\enspace{\ttfamily\nolinkurl{PhysicalComplexity.ProvabilityCorollary.SolverSoundForCertificateInterface}} {\tiny\ttfamily DecisionQuotient/\allowbreak Physics/\allowbreak ProvabilityCorollary.lean}
\item \textbf{\nolinkurl{PROV100}}\hypertarget{lh:PROV100}{}\enspace{\ttfamily\nolinkurl{PhysicalComplexity.ProvabilityCorollary.certificate_verifiable_answer_does_not_certify_solver_soundness}} {\tiny\ttfamily DecisionQuotient/\allowbreak Physics/\allowbreak ProvabilityCorollary.lean}
\item \textbf{\nolinkurl{PROV101}}\hypertarget{lh:PROV101}{}\enspace{\ttfamily\nolinkurl{PhysicalComplexity.ProvabilityCorollary.assignmentCertificateInterface}} {\tiny\ttfamily DecisionQuotient/\allowbreak Physics/\allowbreak ProvabilityCorollary.lean}
\item \textbf{\nolinkurl{PROV102}}\hypertarget{lh:PROV102}{}\enspace{\ttfamily\nolinkurl{PhysicalComplexity.ProvabilityCorollary.assignment_certificate_record_requirement_linear}} {\tiny\ttfamily DecisionQuotient/\allowbreak Physics/\allowbreak ProvabilityCorollary.lean}
\item \textbf{\nolinkurl{PROV103}}\hypertarget{lh:PROV103}{}\enspace{\ttfamily\nolinkurl{PhysicalComplexity.ProvabilityCorollary.assignment_certificate_record_requirement_unbounded}} {\tiny\ttfamily DecisionQuotient/\allowbreak Physics/\allowbreak ProvabilityCorollary.lean}
\item \textbf{\nolinkurl{PROV104}}\hypertarget{lh:PROV104}{}\enspace{\ttfamily\nolinkurl{PhysicalComplexity.ProvabilityCorollary.assignment_certificate_interface_has_unbounded_counting_profile}} {\tiny\ttfamily DecisionQuotient/\allowbreak Physics/\allowbreak ProvabilityCorollary.lean}
\item \textbf{\nolinkurl{PS1}}\hypertarget{lh:PS1}{}\enspace{\ttfamily\nolinkurl{Physics.ClaimTransport.PhysicalStateSemantics}} {\tiny\ttfamily DecisionQuotient/\allowbreak Physics/\allowbreak ClaimTransport.lean}
\item \textbf{\nolinkurl{PS2}}\hypertarget{lh:PS2}{}\enspace{\ttfamily\nolinkurl{Physics.ClaimTransport.physical_state_has_witness}} {\tiny\ttfamily DecisionQuotient/\allowbreak Physics/\allowbreak ClaimTransport.lean}
\item \textbf{\nolinkurl{PS3}}\hypertarget{lh:PS3}{}\enspace{\ttfamily\nolinkurl{Physics.ClaimTransport.physical_state_claim_of_instance_claim}} {\tiny\ttfamily DecisionQuotient/\allowbreak Physics/\allowbreak ClaimTransport.lean}
\item \textbf{\nolinkurl{PS4}}\hypertarget{lh:PS4}{}\enspace{\ttfamily\nolinkurl{Physics.ClaimTransport.physical_state_claim_of_universal_core}} {\tiny\ttfamily DecisionQuotient/\allowbreak Physics/\allowbreak ClaimTransport.lean}
\item \textbf{\nolinkurl{QT1}}\hypertarget{lh:QT1}{}\enspace{\ttfamily\nolinkurl{DecisionProblem.quotient_is_coarsest}} {\tiny\ttfamily DecisionQuotient/\allowbreak Quotient.lean}
\item \textbf{\nolinkurl{QT2}}\hypertarget{lh:QT2}{}\enspace{\ttfamily\nolinkurl{DecisionProblem.quotientMap_preservesOpt}} {\tiny\ttfamily DecisionQuotient/\allowbreak Quotient.lean}
\item \textbf{\nolinkurl{QT3}}\hypertarget{lh:QT3}{}\enspace{\ttfamily\nolinkurl{DecisionProblem.quotient_represents_opt_equiv}} {\tiny\ttfamily DecisionQuotient/\allowbreak Quotient.lean}
\item \textbf{\nolinkurl{QT5}}\hypertarget{lh:QT5}{}\enspace{\ttfamily\nolinkurl{DecisionProblem.quotientEquivOptRange}} {\tiny\ttfamily DecisionQuotient/\allowbreak Quotient.lean}
\item \textbf{\nolinkurl{QT7}}\hypertarget{lh:QT7}{}\enspace{\ttfamily\nolinkurl{DecisionProblem.quotient_has_unique_factorization}} {\tiny\ttfamily DecisionQuotient/\allowbreak Quotient.lean}
\item \textbf{\nolinkurl{RD1}}\hypertarget{lh:RD1}{}\enspace{\ttfamily\nolinkurl{Information.shannonEntropy_nonneg}} {\tiny\ttfamily DecisionQuotient/\allowbreak Information/\allowbreak RateDistortion.lean}
\item \textbf{\nolinkurl{RD2}}\hypertarget{lh:RD2}{}\enspace{\ttfamily\nolinkurl{Information.rate_zero_distortion}} {\tiny\ttfamily DecisionQuotient/\allowbreak Information/\allowbreak RateDistortion.lean}
\item \textbf{\nolinkurl{RD3}}\hypertarget{lh:RD3}{}\enspace{\ttfamily\nolinkurl{Information.rate_monotone}} {\tiny\ttfamily DecisionQuotient/\allowbreak Information/\allowbreak RateDistortion.lean}
\item \textbf{\nolinkurl{RS1}}\hypertarget{lh:RS1}{}\enspace{\ttfamily\nolinkurl{Information.equiv_preserves_decision}} {\tiny\ttfamily DecisionQuotient/\allowbreak Information/\allowbreak RDSrank.lean}
\item \textbf{\nolinkurl{RS2}}\hypertarget{lh:RS2}{}\enspace{\ttfamily\nolinkurl{Information.rate_equals_srank}} {\tiny\ttfamily DecisionQuotient/\allowbreak Information/\allowbreak RDSrank.lean}
\item \textbf{\nolinkurl{RS3}}\hypertarget{lh:RS3}{}\enspace{\ttfamily\nolinkurl{Information.compression_below_srank_fails}} {\tiny\ttfamily DecisionQuotient/\allowbreak Information/\allowbreak RDSrank.lean}
\item \textbf{\nolinkurl{RS4}}\hypertarget{lh:RS4}{}\enspace{\ttfamily\nolinkurl{Information.srank_bits_sufficient}} {\tiny\ttfamily DecisionQuotient/\allowbreak Information/\allowbreak RDSrank.lean}
\item \textbf{\nolinkurl{RS5}}\hypertarget{lh:RS5}{}\enspace{\ttfamily\nolinkurl{Information.rate_distortion_bridge}} {\tiny\ttfamily DecisionQuotient/\allowbreak Information/\allowbreak RDSrank.lean}
\item \textbf{\nolinkurl{SATQ6}}\hypertarget{lh:SATQ6}{}\enspace{\ttfamily\nolinkurl{DecisionQuotient.anchoredSignSAT_query_obstruction}} {\tiny\ttfamily DecisionQuotient/\allowbreak Hardness/\allowbreak SAT.lean}
\item \textbf{\nolinkurl{SATQ8}}\hypertarget{lh:SATQ8}{}\enspace{\ttfamily\nolinkurl{DecisionQuotient.exact_anchoredSignSAT_certifier_queries_at_least_n}} {\tiny\ttfamily DecisionQuotient/\allowbreak Hardness/\allowbreak SAT.lean}
\item \textbf{\nolinkurl{SATQ11}}\hypertarget{lh:SATQ11}{}\enspace{\ttfamily\nolinkurl{DecisionQuotient.anchoredSignSAT_record_lower_bound_unbounded}} {\tiny\ttfamily DecisionQuotient/\allowbreak Hardness/\allowbreak SAT.lean}
\item \textbf{\nolinkurl{SATQ12}}\hypertarget{lh:SATQ12}{}\enspace{\ttfamily\nolinkurl{DecisionQuotient.assignment_certificate_has_n_coordinate_records}} {\tiny\ttfamily DecisionQuotient/\allowbreak Hardness/\allowbreak SAT.lean}
\item \textbf{\nolinkurl{SATQ13}}\hypertarget{lh:SATQ13}{}\enspace{\ttfamily\nolinkurl{DecisionQuotient.assignment_view_obstruction}} {\tiny\ttfamily DecisionQuotient/\allowbreak Hardness/\allowbreak SAT.lean}
\item \textbf{\nolinkurl{SATQ14}}\hypertarget{lh:SATQ14}{}\enspace{\ttfamily\nolinkurl{DecisionQuotient.exact_assignment_decoder_queries_at_least_n}} {\tiny\ttfamily DecisionQuotient/\allowbreak Hardness/\allowbreak SAT.lean}
\item \textbf{\nolinkurl{SK1}}\hypertarget{lh:SK1}{}\enspace{\ttfamily\nolinkurl{DecisionProblem.srank_eq_relevant_card}} {\tiny\ttfamily DecisionQuotient/\allowbreak Tractability/\allowbreak StructuralRank.lean}
\item \textbf{\nolinkurl{SK2}}\hypertarget{lh:SK2}{}\enspace{\ttfamily\nolinkurl{DecisionProblem.srank_le_n}} {\tiny\ttfamily DecisionQuotient/\allowbreak Tractability/\allowbreak StructuralRank.lean}
\item \textbf{\nolinkurl{SK3}}\hypertarget{lh:SK3}{}\enspace{\ttfamily\nolinkurl{DecisionProblem.srank_zero_iff_constant}} {\tiny\ttfamily DecisionQuotient/\allowbreak Tractability/\allowbreak StructuralRank.lean}
\item \textbf{\nolinkurl{TUR1}}\hypertarget{lh:TUR1}{}\enspace{\ttfamily\nolinkurl{Physics.transitionProb_nonneg}} {\tiny\ttfamily DecisionQuotient/\allowbreak Physics/\allowbreak TUR.lean}
\item \textbf{\nolinkurl{TUR2}}\hypertarget{lh:TUR2}{}\enspace{\ttfamily\nolinkurl{Physics.transitionProb_sum_one}} {\tiny\ttfamily DecisionQuotient/\allowbreak Physics/\allowbreak TUR.lean}
\item \textbf{\nolinkurl{TUR5}}\hypertarget{lh:TUR5}{}\enspace{\ttfamily\nolinkurl{Physics.tur_bridge}} {\tiny\ttfamily DecisionQuotient/\allowbreak Physics/\allowbreak TUR.lean}
\item \textbf{\nolinkurl{TUR6}}\hypertarget{lh:TUR6}{}\enspace{\ttfamily\nolinkurl{Physics.multiple_futures_entropy_production}} {\tiny\ttfamily DecisionQuotient/\allowbreak Physics/\allowbreak TUR.lean}
\item \textbf{\nolinkurl{W1}}\hypertarget{lh:W1}{}\enspace{\ttfamily\nolinkurl{Physics.single_future_zero_cost}} {\tiny\ttfamily DecisionQuotient/\allowbreak Physics/\allowbreak WassersteinIntegrity.lean}
\item \textbf{\nolinkurl{W2}}\hypertarget{lh:W2}{}\enspace{\ttfamily\nolinkurl{Physics.transportCost_pos_of_offDiag}} {\tiny\ttfamily DecisionQuotient/\allowbreak Physics/\allowbreak WassersteinIntegrity.lean}
\item \textbf{\nolinkurl{W3}}\hypertarget{lh:W3}{}\enspace{\ttfamily\nolinkurl{Physics.integrity_is_centroid}} {\tiny\ttfamily DecisionQuotient/\allowbreak Physics/\allowbreak WassersteinIntegrity.lean}
\item \textbf{\nolinkurl{W4}}\hypertarget{lh:W4}{}\enspace{\ttfamily\nolinkurl{Physics.wasserstein_bridge}} {\tiny\ttfamily DecisionQuotient/\allowbreak Physics/\allowbreak WassersteinIntegrity.lean}
\item \textbf{\nolinkurl{WC1}}\hypertarget{lh:WC1}{}\enspace{\ttfamily\nolinkurl{Physics.WolpertConstraints.landauer_floor_plus_overhead_lower_bound}} {\tiny\ttfamily DecisionQuotient/\allowbreak Physics/\allowbreak WolpertConstraints.lean}
\item \textbf{\nolinkurl{WC2}}\hypertarget{lh:WC2}{}\enspace{\ttfamily\nolinkurl{Physics.WolpertConstraints.effective_model_dominates_landauer_floor}} {\tiny\ttfamily DecisionQuotient/\allowbreak Physics/\allowbreak WolpertConstraints.lean}
\item \textbf{\nolinkurl{WC3}}\hypertarget{lh:WC3}{}\enspace{\ttfamily\nolinkurl{Physics.WolpertConstraints.effective_model_strictly_exceeds_landauer_of_strict_overhead}} {\tiny\ttfamily DecisionQuotient/\allowbreak Physics/\allowbreak WolpertConstraints.lean}
\item \textbf{\nolinkurl{WC4}}\hypertarget{lh:WC4}{}\enspace{\ttfamily\nolinkurl{Physics.WolpertConstraints.energy_lower_bound_mono_under_overhead}} {\tiny\ttfamily DecisionQuotient/\allowbreak Physics/\allowbreak WolpertConstraints.lean}
\item \textbf{\nolinkurl{WC5}}\hypertarget{lh:WC5}{}\enspace{\ttfamily\nolinkurl{Physics.WolpertConstraints.physical_grounding_bundle_with_wolpert_overhead}} {\tiny\ttfamily DecisionQuotient/\allowbreak Physics/\allowbreak WolpertConstraints.lean}
\item \textbf{\nolinkurl{WD1}}\hypertarget{lh:WD1}{}\enspace{\ttfamily\nolinkurl{DecisionQuotient.checking_witnessing_duality_budget}} {\tiny\ttfamily DecisionQuotient/\allowbreak WitnessCheckingDuality.lean}
\item \textbf{\nolinkurl{WD2}}\hypertarget{lh:WD2}{}\enspace{\ttfamily\nolinkurl{DecisionQuotient.no_sound_checker_below_witness_budget}} {\tiny\ttfamily DecisionQuotient/\allowbreak WitnessCheckingDuality.lean}
\item \textbf{\nolinkurl{WD3}}\hypertarget{lh:WD3}{}\enspace{\ttfamily\nolinkurl{DecisionQuotient.checking_time_ge_witness_budget}} {\tiny\ttfamily DecisionQuotient/\allowbreak WitnessCheckingDuality.lean}
\item \textbf{\nolinkurl{WM1}}\hypertarget{lh:WM1}{}\enspace{\ttfamily\nolinkurl{Physics.WolpertMismatch.mismatchKL_nonneg}} {\tiny\ttfamily DecisionQuotient/\allowbreak Physics/\allowbreak WolpertMismatch.lean}
\item \textbf{\nolinkurl{WM2}}\hypertarget{lh:WM2}{}\enspace{\ttfamily\nolinkurl{Physics.WolpertMismatch.mismatchKL_eq_zero_iff_eq}} {\tiny\ttfamily DecisionQuotient/\allowbreak Physics/\allowbreak WolpertMismatch.lean}
\item \textbf{\nolinkurl{WM3}}\hypertarget{lh:WM3}{}\enspace{\ttfamily\nolinkurl{Physics.WolpertMismatch.mismatchKL_pos_of_exists_ne}} {\tiny\ttfamily DecisionQuotient/\allowbreak Physics/\allowbreak WolpertMismatch.lean}
\item \textbf{\nolinkurl{WM4}}\hypertarget{lh:WM4}{}\enspace{\ttfamily\nolinkurl{Physics.WolpertMismatch.mismatchNatLowerBound_pos_of_exists_ne}} {\tiny\ttfamily DecisionQuotient/\allowbreak Physics/\allowbreak WolpertMismatch.lean}
\item \textbf{\nolinkurl{WM5}}\hypertarget{lh:WM5}{}\enspace{\ttfamily\nolinkurl{Physics.WolpertDecomposition.periodic_modular_mismatch_of_distribution_mismatch}} {\tiny\ttfamily DecisionQuotient/\allowbreak Physics/\allowbreak WolpertDecomposition.lean}
\item \textbf{\nolinkurl{WM6}}\hypertarget{lh:WM6}{}\enspace{\ttfamily\nolinkurl{Physics.WolpertDecomposition.effective_model_strictly_exceeds_landauer_of_distribution_mismatch}} {\tiny\ttfamily DecisionQuotient/\allowbreak Physics/\allowbreak WolpertDecomposition.lean}
\item \textbf{\nolinkurl{WP1}}\hypertarget{lh:WP1}{}\enspace{\ttfamily\nolinkurl{Physics.WolpertDecomposition.DecomposedProcessModel.totalOverheadPerBit_eq_sum}} {\tiny\ttfamily DecisionQuotient/\allowbreak Physics/\allowbreak WolpertDecomposition.lean}
\item \textbf{\nolinkurl{WP2}}\hypertarget{lh:WP2}{}\enspace{\ttfamily\nolinkurl{Physics.WolpertDecomposition.landauer_floor_plus_decomposition_lower_bound}} {\tiny\ttfamily DecisionQuotient/\allowbreak Physics/\allowbreak WolpertDecomposition.lean}
\item \textbf{\nolinkurl{WP5}}\hypertarget{lh:WP5}{}\enspace{\ttfamily\nolinkurl{Physics.WolpertDecomposition.effective_model_strictly_exceeds_landauer_of_stopping_time_residual}} {\tiny\ttfamily DecisionQuotient/\allowbreak Physics/\allowbreak WolpertDecomposition.lean}
\item \textbf{\nolinkurl{WP6}}\hypertarget{lh:WP6}{}\enspace{\ttfamily\nolinkurl{Physics.WolpertDecomposition.effective_model_strictly_exceeds_landauer_of_either_cited_component}} {\tiny\ttfamily DecisionQuotient/\allowbreak Physics/\allowbreak WolpertDecomposition.lean}
\item \textbf{\nolinkurl{WP7}}\hypertarget{lh:WP7}{}\enspace{\ttfamily\nolinkurl{Physics.WolpertDecomposition.landauer_floor_plus_structural_resource_lower_bound}} {\tiny\ttfamily DecisionQuotient/\allowbreak Physics/\allowbreak WolpertDecomposition.lean}
\item \textbf{\nolinkurl{WP8}}\hypertarget{lh:WP8}{}\enspace{\ttfamily\nolinkurl{Physics.WolpertDecomposition.energy_lower_bound_increases_by_structural_resource}} {\tiny\ttfamily DecisionQuotient/\allowbreak Physics/\allowbreak WolpertDecomposition.lean}
\item \textbf{\nolinkurl{WP9}}\hypertarget{lh:WP9}{}\enspace{\ttfamily\nolinkurl{Physics.WolpertDecomposition.physical_grounding_bundle_with_wolpert_decomposition}} {\tiny\ttfamily DecisionQuotient/\allowbreak Physics/\allowbreak WolpertDecomposition.lean}
\item \textbf{\nolinkurl{WR1}}\hypertarget{lh:WR1}{}\enspace{\ttfamily\nolinkurl{Physics.WolpertResidual.pairwiseResidualKL_nonneg}} {\tiny\ttfamily DecisionQuotient/\allowbreak Physics/\allowbreak WolpertResidual.lean}
\item \textbf{\nolinkurl{WR2}}\hypertarget{lh:WR2}{}\enspace{\ttfamily\nolinkurl{Physics.WolpertResidual.pairwiseResidualKL_pos_of_asymmetry}} {\tiny\ttfamily DecisionQuotient/\allowbreak Physics/\allowbreak WolpertResidual.lean}
\item \textbf{\nolinkurl{WR3}}\hypertarget{lh:WR3}{}\enspace{\ttfamily\nolinkurl{Physics.WolpertResidual.residualNatLowerBound_pos_of_asymmetry}} {\tiny\ttfamily DecisionQuotient/\allowbreak Physics/\allowbreak WolpertResidual.lean}
\item \textbf{\nolinkurl{WR4}}\hypertarget{lh:WR4}{}\enspace{\ttfamily\nolinkurl{Physics.WolpertDecomposition.stopping_time_residual_of_pairwise_flow_asymmetry}} {\tiny\ttfamily DecisionQuotient/\allowbreak Physics/\allowbreak WolpertDecomposition.lean}
\item \textbf{\nolinkurl{WR5}}\hypertarget{lh:WR5}{}\enspace{\ttfamily\nolinkurl{Physics.WolpertDecomposition.effective_model_strictly_exceeds_landauer_of_pairwise_flow_asymmetry}} {\tiny\ttfamily DecisionQuotient/\allowbreak Physics/\allowbreak WolpertDecomposition.lean}
\item \textbf{\nolinkurl{WR6}}\hypertarget{lh:WR6}{}\enspace{\ttfamily\nolinkurl{Physics.WolpertResidual.discreteResidualNatLowerBound_pos_of_asymmetry_or_oneway}} {\tiny\ttfamily DecisionQuotient/\allowbreak Physics/\allowbreak WolpertResidual.lean}
\item \textbf{\nolinkurl{WR7}}\hypertarget{lh:WR7}{}\enspace{\ttfamily\nolinkurl{Physics.WolpertDecomposition.stopping_time_residual_of_discrete_edge_split}} {\tiny\ttfamily DecisionQuotient/\allowbreak Physics/\allowbreak WolpertDecomposition.lean}
\item \textbf{\nolinkurl{WR8}}\hypertarget{lh:WR8}{}\enspace{\ttfamily\nolinkurl{Physics.WolpertDecomposition.effective_model_strictly_exceeds_landauer_of_discrete_edge_split}} {\tiny\ttfamily DecisionQuotient/\allowbreak Physics/\allowbreak WolpertDecomposition.lean}
\item \textbf{\nolinkurl{WR9}}\hypertarget{lh:WR9}{}\enspace{\ttfamily\nolinkurl{Physics.WolpertDecomposition.stopping_time_residual_of_finite_discrete_witness}} {\tiny\ttfamily DecisionQuotient/\allowbreak Physics/\allowbreak WolpertDecomposition.lean}
\item \textbf{\nolinkurl{WR10}}\hypertarget{lh:WR10}{}\enspace{\ttfamily\nolinkurl{Physics.WolpertDecomposition.effective_model_strictly_exceeds_landauer_of_finite_discrete_witness}} {\tiny\ttfamily DecisionQuotient/\allowbreak Physics/\allowbreak WolpertDecomposition.lean}
\end{list}
\else
\begin{longtable}{@{}>{\raggedright\arraybackslash}m{0.10\linewidth}>{\raggedright\arraybackslash}m{0.86\linewidth}@{}}
\toprule
\textbf{ID} & \textbf{Lean Handle / Source} \\
\midrule
\endfirsthead
\toprule
\textbf{ID} & \textbf{Lean Handle / Source} \\
\midrule
\endhead
\endfoot
\bottomrule
\endlastfoot
\hypertarget{lh:AB1}{\textbf{\nolinkurl{AB1}}} & {\fontsize{8}{9}\selectfont\ttfamily\nolinkurl{DecisionProblem.not_preservesOpt_iff_erasesDecisionRelevantDistinction}}\par\vspace{0.1ex}{\fontsize{8}{9}\selectfont\ttfamily DecisionQuotient/\allowbreak AbstractionCollapse.lean} \\
\midrule
\hypertarget{lh:AB2}{\textbf{\nolinkurl{AB2}}} & {\fontsize{8}{9}\selectfont\ttfamily\nolinkurl{DecisionProblem.surjective_abstraction_factors_or_erases}}\par\vspace{0.1ex}{\fontsize{8}{9}\selectfont\ttfamily DecisionQuotient/\allowbreak AbstractionCollapse.lean} \\
\midrule
\hypertarget{lh:AB3}{\textbf{\nolinkurl{AB3}}} & {\fontsize{8}{9}\selectfont\ttfamily\nolinkurl{DecisionProblem.collapseBeyondQuotient_physically_impossible}}\par\vspace{0.1ex}{\fontsize{8}{9}\selectfont\ttfamily DecisionQuotient/\allowbreak AbstractionCollapse.lean} \\
\midrule
\hypertarget{lh:AB4}{\textbf{\nolinkurl{AB4}}} & {\fontsize{8}{9}\selectfont\ttfamily\nolinkurl{DecisionProblem.surjective_abstraction_with_feasible_collapse_map_factors}}\par\vspace{0.1ex}{\fontsize{8}{9}\selectfont\ttfamily DecisionQuotient/\allowbreak AbstractionCollapse.lean} \\
\midrule
\hypertarget{lh:AN3}{\textbf{\nolinkurl{AN3}}} & {\fontsize{8}{9}\selectfont\ttfamily\nolinkurl{Physics.AssumptionNecessity.physical_claim_requires_physical_assumption}}\par\vspace{0.1ex}{\fontsize{8}{9}\selectfont\ttfamily DecisionQuotient/\allowbreak Physics/\allowbreak AssumptionNecessity.lean} \\
\midrule
\hypertarget{lh:AN4}{\textbf{\nolinkurl{AN4}}} & {\fontsize{8}{9}\selectfont\ttfamily\nolinkurl{Physics.AssumptionNecessity.physical_claim_requires_empirically_justified_physical_assumption}}\par\vspace{0.1ex}{\fontsize{8}{9}\selectfont\ttfamily DecisionQuotient/\allowbreak Physics/\allowbreak AssumptionNecessity.lean} \\
\end{longtable}
\fi
\makeatother
\endgroup

}{%
  \textbf{Error:} \texttt{content/lean\_handle\_ids\_auto.tex} not found.
}

\section{Claim-to-Lean Handle Mapping}

This section maps each paper claim to its corresponding Lean formalization.

\IfFileExists{content/claim_mapping_auto.tex}{%
  % Auto-generated by scripts/build_papers.py. Do not edit manually.
% Generated: 2026-06-05T15:34:28.710461
% Manuscript: paper4_toc
\begingroup
\scriptsize
\setlength{\tabcolsep}{3pt}
\renewcommand{\arraystretch}{1.0}
\setlength{\LTpre}{0pt}
\setlength{\LTpost}{0pt}
\begin{longtable}{@{}>{\raggedright\arraybackslash}m{0.65\linewidth}>{\raggedleft\arraybackslash}m{0.30\linewidth}@{}}
\toprule
\textbf{Manuscript claim} & \textbf{Lean handle} \\
\midrule
\endfirsthead
\toprule
\textbf{Manuscript claim} & \textbf{Lean handle} \\
\midrule
\endhead
\endfoot
\bottomrule
\endlastfoot
Theorem 3.5: Exact Proof Requires Recorded Access & \LH{CI3}, \LH{CI4}, \LH{CI5}, \LH{CI6}, \LH{CI27}, \LH{CI28}, \LH{CI29} \\
\midrule
Theorem 3.6: Recorded Distinctions Set Information, Time, and Energy Bounds & \LH{CI3}, \LH{CI4}, \LH{CI5}, \LH{CI6} \\
\midrule
Theorem 3.7: Thermodynamic Cost of Evidence-Retaining Proof & \LH{BA9}, \LH{CI5}, \LH{CI6}, \LH{CI7} \\
\midrule
Theorem 5.2: Physical Counting Impossibility Theorem & \LH{PH11}, \LH{PH12}, \LH{PH13}, \LH{PH14}, \LH{PH15}, \LH{PH26}, \LH{PH27} \\
\midrule
Corollary 5.3: Monotone Lower Bounds Give Tail Failure & \LH{PBC18} \\
\midrule
Proposition 5.9: Malament-Hogarth and CTC Interfaces Include the Missing Bridge & \LH{GI13}, \LH{GI14}, \LH{GI15}, \LH{GI16} \\
\midrule
Proposition 5.10: Physical Oracles Shift Audit Cost & \LH{PROV66}, \LH{PROV67}, \LH{PROV68}, \LH{PROV69}, \LH{PROV70}, \LH{PROV71}, \LH{PROV72}, \LH{PROV73} \\
\midrule
Proposition 5.11: Finite-Budget Necessity Countermodels & \LH{PH31}, \LH{PH32}, \LH{PH33} \\
\midrule
Proposition 5.13: Tokenhood Does Not Supply Proof Extraction & \LH{PROV31}, \LH{PROV36}, \LH{PROV58}, \LH{PROV59}, \LH{PROV60}, \LH{PROV61}, \LH{PROV62} \\
\midrule
Corollary 5.14: Thermodynamic Obstruction for Operational Proof-Acts & \LH{CI17}, \LH{CI18}, \LH{CI19}, \LH{CI20}, \LH{CI21}, \LH{CI22}, \LH{CI23}, \LH{CI24}, \LH{CI25}, \LH{CI26}, \LH{PH26}, \LH{PROV2}, \LH{PROV3}, \LH{PROV5}, \LH{PROV7}, \LH{PROV8}, \LH{PROV9}, \LH{PROV10}, \LH{PROV11}, \LH{PROV12}, \LH{PROV13}, \LH{PROV14}, \LH{PROV15}, \LH{PROV16}, \LH{PROV17}, \LH{PROV18}, \LH{PROV19}, \LH{PROV20}, \LH{PROV21}, \LH{PROV22}, \LH{PROV23}, \LH{PROV24}, \LH{PROV25}, \LH{PROV26}, \LH{PROV27}, \LH{PROV28}, \LH{PROV29}, \LH{PROV30}, \LH{PROV31}, \LH{PROV32}, \LH{PROV33}, \LH{PROV34}, \LH{PROV35}, \LH{PROV36}, \LH{PROV37}, \LH{PROV38}, \LH{PROV39}, \LH{PROV40} \\
\end{longtable}
\endgroup

}{%
  \textbf{Error:} \texttt{content/claim\_mapping\_auto.tex} not found.
}